\documentclass[twocolumn,floatfix,amsmath,amssymb,prb,showpacs,superscriptaddress,hyperref]{revtex4-2}
\usepackage{graphicx}
\usepackage{color}
\usepackage{bbold}
\usepackage{tikz}
\usepackage{dcolumn}
\usepackage{bm,listings}
\usepackage{dsfont}
\usepackage{ulem}

\usepackage{amsfonts}
\usepackage{comment}

\begin{document}

\title{Intensity landscapes in elliptical and oval billiards with a circular absorbing region}
%\date{\today}
\author{Katherine Holmes}\author{Joseph Hall}\author{Eva-Maria Graefe}
\address{Department of Mathematics, Imperial College London, London, SW7 2AZ, United Kingdom}
\begin{abstract}
Billiard models of single particles moving freely in two-dimensional regions enclosed by hard walls, have long provided ideal toy models for the investigation of dynamical systems and chaos. Recently, billiards with (semi-)permeable walls and internal holes have been used to study open systems. Here we introduce a billiard model containing an internal region with partial absorption. The absorption does not change the trajectories, but instead reduces an intensity variable associated with each trajectory.  The value of the intensity can be tracked as a function of the initial configuration and the number of reflections from the wall and depicted in intensity landscapes over the Poincar\'e phase space.  This is similar in spirit to escape time diagrams that are often considered in dynamical systems with holes. We analyse the resulting intensity landscapes for three different geometries; a circular, elliptic, and oval billiard, respectively, all with a centrally placed circular absorbing region. The intensity landscapes feature increasingly more complex structures, organised around  the sets of points in phase space that intersect the absorbing region in a given iteration, which we study in some detail. On top of these, the intensity landscapes are enriched by effects arising from  multiple absorption events for a given trajectory. 
\end{abstract}

\maketitle

\section{Introduction}
Billiard models play a prominent role in our understanding of chaos and dynamical systems. In their simplest form they consist of two-dimensional regions with hard walls at the boundary, where free particles or rays may traverse without friction and reflect from the hard walls. While the corresponding phase space is four-dimensional, the simple behaviour of the trajectories between reflections from the walls, and the scaling invariance with respect to the total momentum of a trajectory, allows for the analysis of the dynamical behaviour on a two-dimensional Poincar\'e section $\Omega=S\times p$, spanned by Poincar\'e-Birkhoff coordinates $(S,p)$, characterising the points of reflection. \\
The billiard dynamics can thus be described in the form of a symplectic map $T:(S_n,p_n)\to(S_{n+1},p_{n+1})$ on the Poincar\'e section, the dynamical behaviour of which is determined by the shape of the billiard boundary \cite{berry1981regularity,tabachnikov2005geometry,bunimovich1979ergodic}. Integrable (regular) dynamics can be studied through e.g., circular and elliptical boundaries; and globally chaotic dynamics through e.g. a stadium boundary or Sinai geometries. The generic case of mixed dynamics arises in many perturbations of these shapes, such as in ovals or in billiards with an internal boundary/scatterer such as annulus or iris billiards \cite{page2020iris,da2015dynamics}.  In particular the investigation of mixed and chaotic billiards and their classical maps, has played an important role in the development of the theory of quantum and wave chaos, where quantum counterparts of these systems provide important theoretical examples \cite{sieber1993non,backer1995spectral,waalkens1997elliptic,jain2017nodal,batistic2019statistical}. Experimentally these systems can be implemented for example as quasi-two-dimensional cavities 
for microwaves \cite{Stoeckmann_1999,dietz2015quantum} or for light 
\cite{nockel1997ray,hentschel2002quantum,shinohara2010chaos,wiersig2006unidirectional,cao2015dielectric}. Recently there has also been intense interest in relativistic billiards \cite{berry1987neutrino,silvestrov2007quantum,dietz2020semiclassical,dietz2015spectral}.

While classical dissipative chaos is a well developed field, the study of wave chaos in the presence of loss is still in the early stages and many phenomena such as multifractal distributions of eigenstates and the fractal Weyl law are still only partially understood \cite{lu2003fractal,Keating_2006,Hen_2010,Hen_2012_weyl_law_open_sys,Alttman_2013_Chaotic_Systems_Absorbtion,Ketz_2018,Clauss2019,Ketz_2021_intensity_stats,Roland_three_disk_scattering_2023,Joe_2023,montes2024average}. 
This has motivated various studies of modified classical billiard dynamics, which include a loss mechanism. In particular, leaking  billiards - billiards with partial or total escape through a region in the boundary - have been studied in some detail \cite{Altmann_2013_Leaking,altmann2013chaotic,hansen2016influence,carlo2022lagrangian}. A hole placed at the boundary corresponds to escape through a simple rectangular area in the Poincar\'e section, and does not affect the mapping $T$, but stops the iteration of the map when the hole is first encountered. The study of escape time diagrams, where initial points on the Poincar\'e section are colour-coded in correspondence with the number of iterations by which they are separated from the hole, has uncovered rich dynamical structures and provided a new tool for understanding the corresponding wave systems \cite{Roland_three_disk_scattering_2023}. 
In \cite{Nagler_2007}  leaking billiards with a hole in the inside of the billiard, rather than on its boundary, were studied. The dynamics again terminate on first encounter of a trajectory with the hole, and escape time diagrams and average escape times can be studied. In contrast to a hole in the boundary a hole in the interior is not connected to a static area in the Poincar\'e section. The behaviour of the escape times was found to depend crucially on the dynamical nature of the billiard, with characteristic features associated to chaotic versus regular cases. 

The concept of a partial leak, on the other hand does not terminate trajectories on encounter with the leaking region, but can be described by a loss of \textit{intensity} - an additional coordinate attached to the trajectories that changes due to the leak \cite{Altmann_2013_Leaking,altmann2013chaotic,clauss2022local,Joe_2023,holmes2023husimi}.  Hitherto partial leaks have only been investigated on the boundary of the system, corresponding to losses from a static region in the Poincar\'e phase space. In \cite{altmann2015chaotic} a fully chaotic billiard with an internal amplifying region has been considered. In the present paper we consider a loss region in the interior of the billiard. The loss region itself is considered spatially homogeneous and does not change the dynamics encoded in $T$. However, due to the finite size of the loss region, different trajectories spend different amounts of time within the loss region between iterations, leading to a nontrivial change in intensity  that cannot be connected to a static area of loss in the two-dimensional Poincar\'e phase space. The intensity dynamics can be visualised as false colour plots on the Poincar\'e phase space, displaying the value of the intensity after a given number of iterations in dependence of the initial phase-space point. The resulting \textit{intensity landscapes} have been found to unravel classical structures underlying quantum systems with decay in simpler model systems \cite{Joe_2023,holmes2023husimi}. 

Here we investigate billiards with elliptic and oval boundaries, (corresponding to regular and mixed chaotic-regular maps) with a partial leak from a central circular region within the billiard.
We observe that the intensity landscapes over time develop characteristics of the Poincar\'e section for the closed billiard, and the interplay with the absorbing region. We show that the intensity landscapes are organised around a union of sets related to the infinite absorption limit (i.e. the limit in which the absorbing region acts as a hole), and are modulated with intricate intensity variations arising from a finite absorption strength.

The paper is organised as follows. We begin with a brief summary of the background and notation of the billiard dynamics expressed as a symplectic map in Poincar\'e-Birkhoff coordinates and introduce the concept of the intensity landscape in section \ref{sec-II}. Then we discuss the intensity landscapes in detail for three examples (circle, ellipse, and oval) in sections \ref{sec-circle}, \ref{sec-ellipse}, and \ref{sec-oval}. We finish with a short summary and outlook in section \ref{sec-sumout}. The influence of a displaced absorbing region, breaking the symmetry of the systems, is explored in an Appendix.

\section{Billiard Map and intensity landscapes}\label{sec-II}
The classical dynamics of an absorption-free two-dimensional billiard system are conveniently analysed via the symplectic map $T:(S_n,p_n)\to(S_{n+1},p_{n+1})$ on the Poincar\'e section $\Omega$, spanned by the Poincar\'e-Birkhoff coordinates $(S,p)$. These coordinates characterise the points of impact for each collision with the boundary, with $(S,p) = (S(\phi),\cos(\alpha))$, where $S(\phi)\in[0,1)$ denotes the normalised arc-length along the boundary to the of reflection with polar coordinate \(\phi\), and $p=\cos(\alpha)$, where $\alpha\in(0,\pi)$ is the angle of reflection. 

\begin{figure}[t]
\begin{center}
\includegraphics[width=0.75\columnwidth]{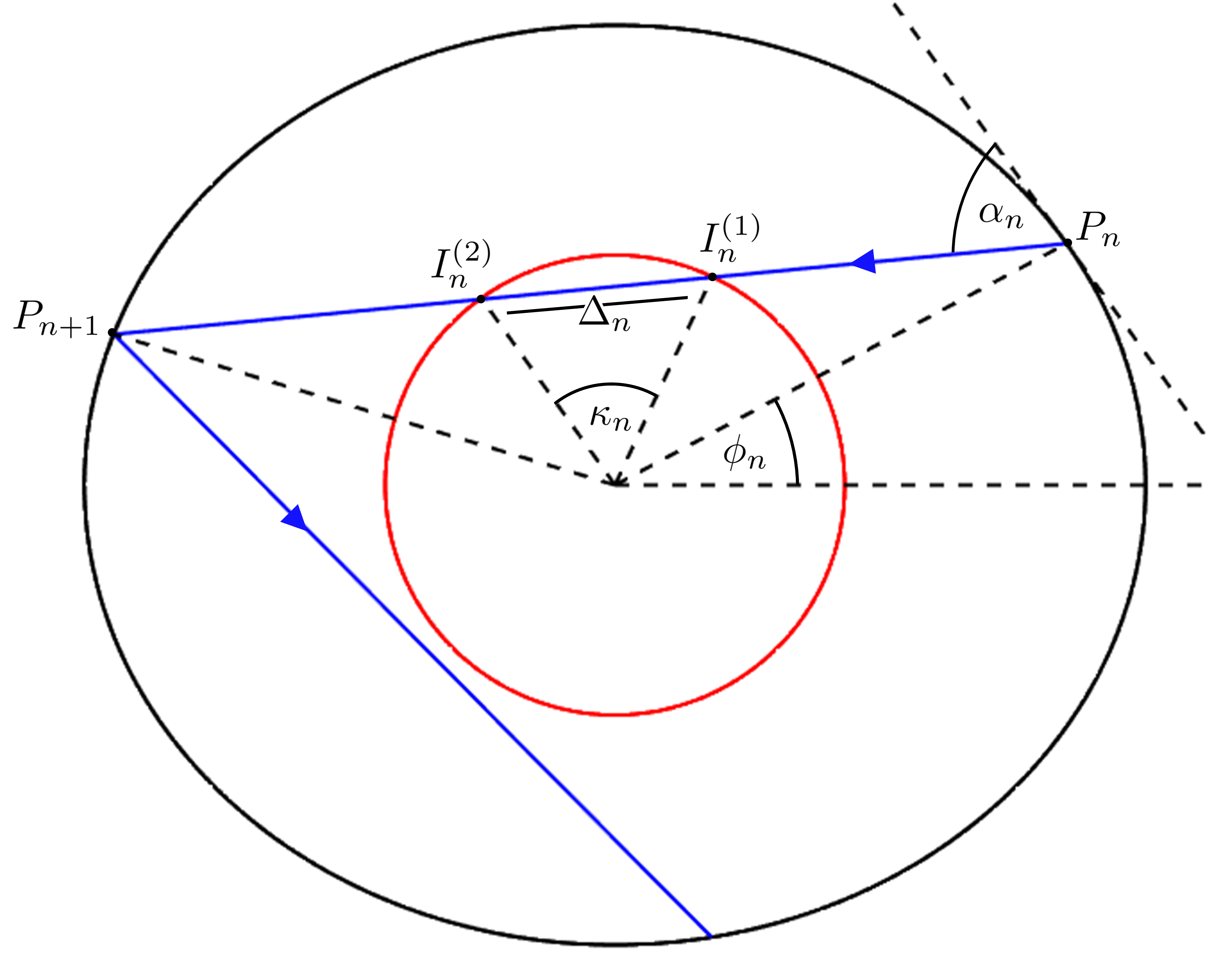}
\end{center}
\caption{A convex billiard with an absorbing region. The solid black line depicts the boundary of an elliptic billiard, the red circle indicates the boundary of an internal absorbing region. An example of the first two iterations of a trajectory is depicted in blue.}
\label{diagram:trajectory_with_patch}
\end{figure}

To model absorption in the system, we introduce an additional coordinate $w$ describing an \textit{intensity} or amplitude, with $w_0=1$ for all trajectories. We consider a circular region with radius $R$, centred at the origin of the billiard as illustrated in Figure \ref{diagram:trajectory_with_patch}, in which we assume a constant absorption, leading to an exponential decay of the intensity with rate $\gamma\in\mathbb{R_{0}^{+}}$, when a trajectory passes through it. The billiard map $T$ of the original system is unchanged by the presence of this absorbing region. The influence of the absorption is captured solely in the change of the intensity coordinate of a given trajectory. If, between two collisions with the boundary, a trajectory enters the absorbing region  it experiences an intensity change 
\begin{equation}
\label{eqn:norm_evolution}
     w_{n+1} = e^{-\gamma \Delta_{n}}w_{n},
\end{equation}
with $w_0=1$, and
where $\Delta_n$ denotes the length of the section of the trajectory that passes through the absorbing region. 
Geometrically, the length $\Delta_n$ can be found as   
\begin{equation}
\label{eqn:delta_n}
    \Delta_{n} = R\sqrt{2(1-\cos(\kappa_{n}))},
\end{equation}
where \(\kappa_{n}\) is the angle between the origin and the radial connections to the two points of intersection with the absorbing region, $I_{n}^{(1)}$ and $I_{n}^{(2)}$, as illustrated in Figure~\ref{diagram:trajectory_with_patch}.

We now introduce the main object of our study - the intensity landscapes $w_n(S,p,\gamma)$. They are constructed by associating to each initial configuration $(S_0,p_0)=(S,p)\in\Omega$ the resulting  intensity $w_n$ after $n$ iterations of the map. The intensity landscapes typically change from iteration to iteration, and can evolve highly non-trivial structures. At each iteration they are divided into two regions - the set of initial conditions corresponding to trajectories that do not intersect the absorbing region up to the $n$-th iteration, forming a plateau where $w_n=1$, and the complement of this set, which is affected by the absorption and on which $w_n$ has a valley structure with $w_n<1$. The plateau region is the intersection of the sets that do not enter the absorbing region in the $n$-th iteration. Let us denote the set of initial points corresponding to trajectories that intersect the absorbing region in the $n$-th iteration by ${\cal S}_n$. These sets are related to each other by the recursion 
\begin{equation}
\label{eqn:evolution_point_set}
{\cal S}_{n+1}=T^{-1}{\cal S}_{n}=T^{-n}{\cal S}_1,  
\end{equation}
where $T^{-1}$ is the inverse of the billiard map. Thus, the sets ${\cal S}_n$ can be found from the set ${\cal S}_1$ (that intersects the absorbing region in the first iteration) and the billiard map of the absorption-free system. The boundary of the set ${\cal S}_1$ has been considered in \cite{da2015dynamics}, in the context of absorption-free billiards with a circular scatterer. The boundary is described by two curves in the Poincar\'e section, $p^{(+)}(S)$ and $p^{(-)}(S)$. For a given initial $S(\phi_{0})$ the values  $(p^{(+)},p^{(-)})$ for which the corresponding trajectory is tangential to the loss region are given by 
\begin{align}
   p^{(\pm)} = \cos\left(\phi \pm \arcsin\left(\frac{R}{r(\phi)}\right) - \zeta(\phi) \right) \label{eqn:red_lines},
\end{align}
where
\begin{equation}
\zeta(\phi) = \arctan\left(\frac{\left(\frac{d}{d\phi}r(\phi)\right)\sin(\phi)+r(\phi)\cos(\phi)}{\left(\frac{d}{d\phi}r(\phi)\right)\cos(\phi)-r(\phi)\sin(\phi)}\right)\label{theta_n},
\end{equation}
and \(r(\phi)\) is the radius function for the billiard boundary.

In the limit of total escape ($\gamma\to\infty$), the absorbing region acts as a hole in the configuration space. In this limit, 
the intensity only takes the discrete values $w_n\in\{0,1\}$, depending on whether an initial point corresponds to a trajectory that encounters the hole before the $n$-th iteration or not. That is, the intensity landscape is given by 
\begin{equation}
\label{eqn:w-escape}
w^{\rm esc}_n(S,p)=\mathbb{1}_{\Omega\setminus U}(S,p),
\end{equation}
where  $U = \cup_n{\cal S}_n$ is the union of the sets ${\cal S}_n$, and where $\mathbb{1}_A$ denotes the indicator function of the set $A$. In this limit, the sets ${\cal S}_{n}$ make up the components of what is known as \textit{escape time diagrams}, where the ${\cal S}_n$ for different values of $n$ are represented in different colours (see, e.g., \cite{Altmann_2013_Leaking} and references therein).  

For a finite value of $\gamma$, the values of $w_n$ away from the plateau region range between zero and one, and can show intricate patterns. Since the underlying dynamics are independent of the  absorption strength $\gamma$, so are the length of the trajectory segments in the absorbing region, $\Delta_n$. That is, changing the value of the absorption $\gamma$, amounts to a simple scaling of the intensity; the intensity landscape for a given \(\gamma_c\in\mathbb{R}^{+}_{0}\), $w_n(S_0,p_0,\gamma_c)$, is related to the landscape for arbitrary $\gamma$ by 
\begin{equation}
\label{eqn:scaling_property}
w_n(S_0,p_0,\gamma)=w_n(S_0,p_0,\gamma_c)^{\frac{\gamma}{\gamma_c}}.
\end{equation}

In what follows we will analyse the intensity landscapes for small and intermediate iteration numbers for three types of convex Birkhoff billiards, a circle, an ellipse and an oval, with a circular uniformly absorbing region at their centres. 

\begin{figure}[t]
\begin{center}
\includegraphics[height=0.44\columnwidth]{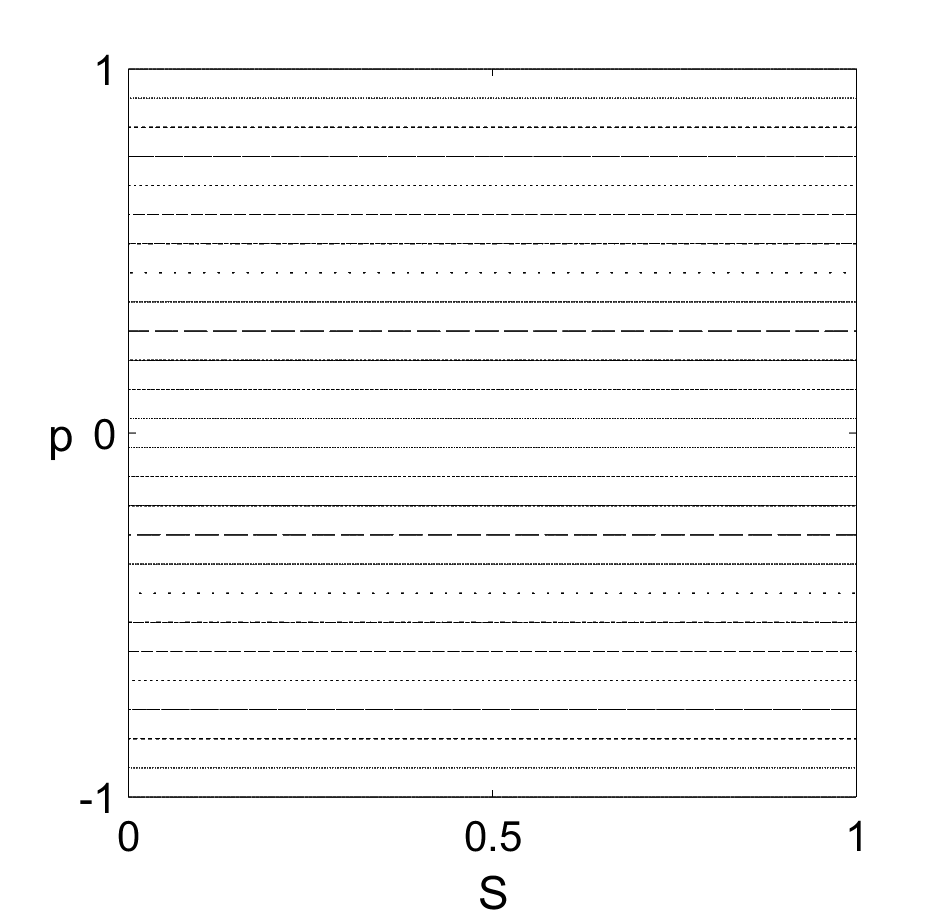}
\includegraphics[height=0.44\columnwidth]{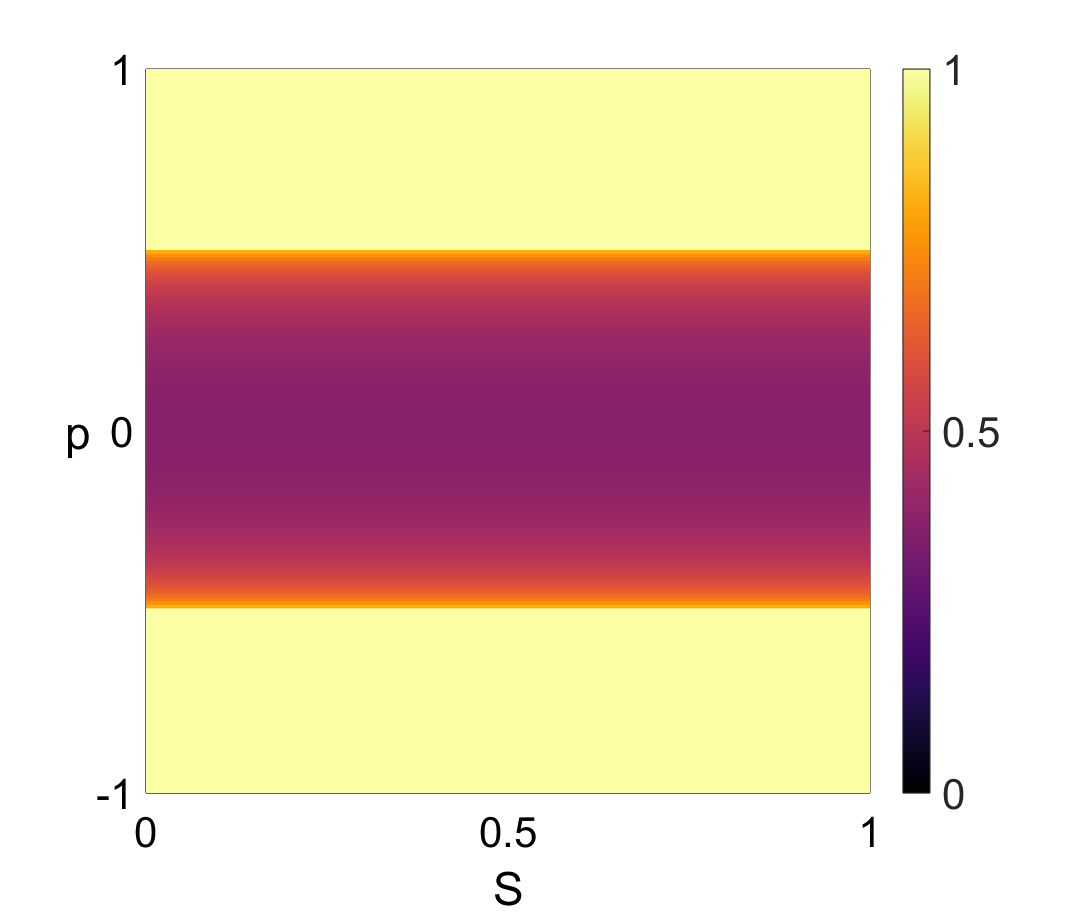}
\end{center}
\caption{Poincar\'e section (left) and intensity landscape after the first iteration (right) of the circle billiard with  $a=1$, $R=0.5$, and absorption strength $\gamma=0.4$. }
\label{diagram:circle_loss_landscape}
\end{figure}

\section{Circle}\label{sec-circle}

We begin with the trivial example of a circular  billiard with circular absorbing region in the centre. The boundary of the billiard in polar coordinates is simply given by
\begin{equation}
     r_{\rm circle}(\phi)=a,
\end{equation}
where $a\in\mathbb{R}^{+}$. The normalised arc-length $S$ as a function of the angle $\phi$ is expressed as 
\begin{equation}
S(\phi)=\frac{\phi}{2\pi}, 
\end{equation}
and the billiard map takes the form
\begin{equation}\label{map:circle}
p_n=p_0\quad {\rm and}\quad S_{n} = S_0+n\frac{{\rm acos}(p_0)}{\pi},
\end{equation}
a regular twist map.

The corresponding Poincar\'e section (that is not influenced by the absorbing region), consists of points aligned on horizontal lines, as shown in the left plot of Figure \ref{diagram:circle_loss_landscape}. The map (\ref{map:circle}) leaves horizontal lines invariant.

The boundary of the set \({\cal S}_{1}\) is given by the two horizontal lines at \(p = \pm p_{c}\) where \(p_{c} =\frac{R}{a}\). Thus,  the sets ${\cal S}_n$ do not change with the iterations, and are simply given by the region ${\cal S}_n=\left\{(S,p)\left|\,|p|\leq p_c\right.\right\}$, i.e., a horizontal strip in phase space. Note that, this is not the case when the absorbing region is displaced from the centre of the billiard, when the boundaries of ${\cal S}_1$ are not horizontal lines. The resulting sets ${\cal S}_n$ are depicted for an example in appendix \ref{appendix:offset_AR}. 

The intensity landscapes, an example of which is depicted in the right plot of Figure \ref{diagram:circle_loss_landscape}, have a simple structure on top of the loss set that is independent of $S$, and given by 
\begin{equation}
 \label{eqn:norm_circle}
      w_{n}(p) = \left\{\begin{array}{cc} 1& {\rm for}\ |p|>p_c,\\
      {\rm exp}(-2n\gamma a\sqrt{p_c^2-p^2})& {\rm otherwise.}
      \end{array} \right.
\end{equation}
Thus, the intensity landscape does not change qualitatively with each iteration.  

\begin{figure}[th!]
    \centering
    \includegraphics[width=0.95\linewidth]{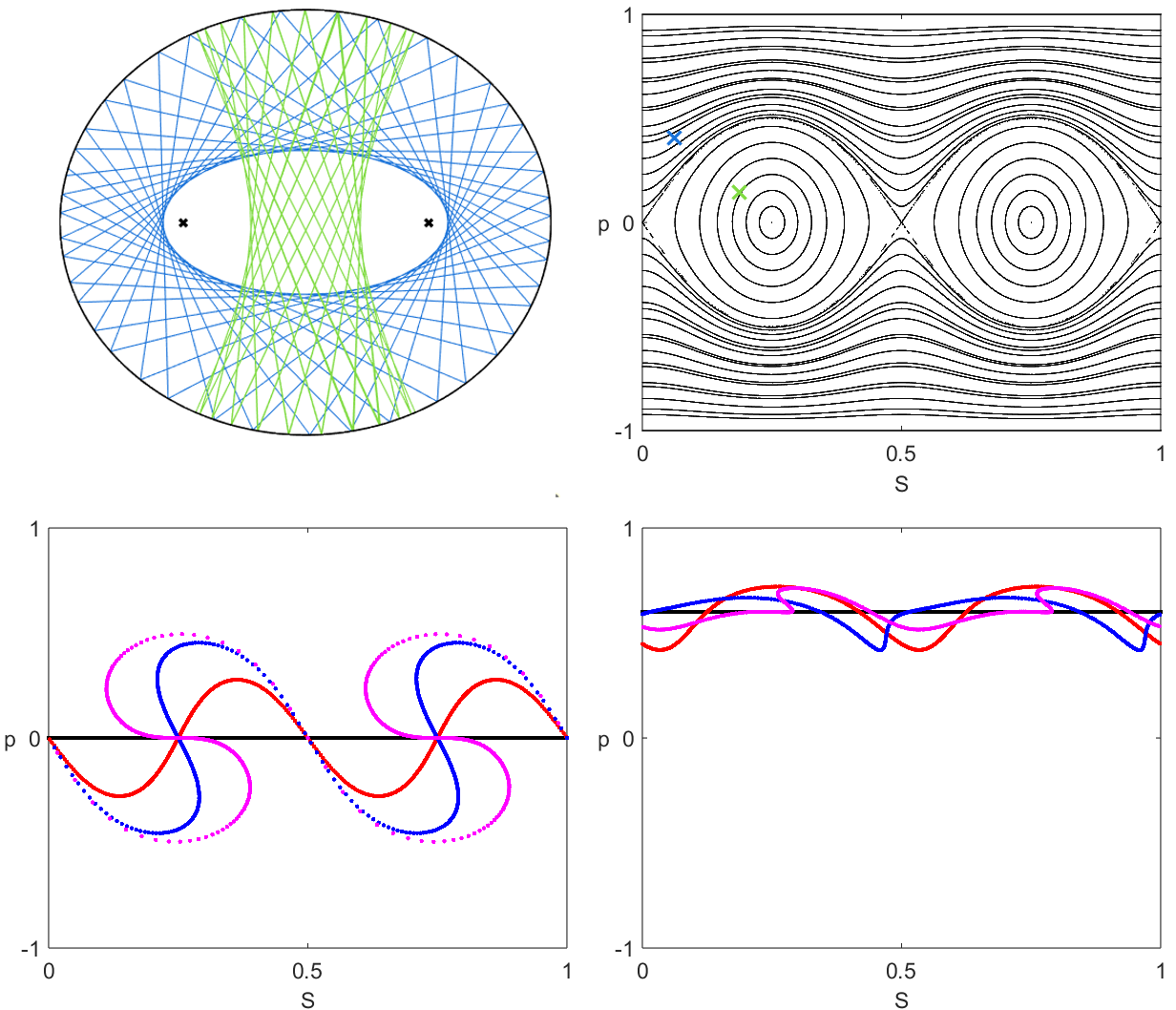}
    \caption{Top left: Billiard boundary (black line) in coordinate space with focal points (black crosses) and two example trajectories (loop - blue, and box - green). Top right: Corresponding Poincar\'e diagram with initial conditions for trajectories (green and blue crosses). Bottom: Forwards propagations of a set of $200$ equidistant points on a horizontal line (black) at \(p=0\) (left) and \(p=0.6\) (right) under the elliptic billiard map \(T\) after \(n= 1\) (red), \(n=2\) (blue) and  \(n=3\) (pink) iterations.}
    \label{Figure_propagated_lines}
\end{figure}

\begin{figure}[t]
\begin{center}
\hspace{-0.2cm}\includegraphics[width=0.95\columnwidth]{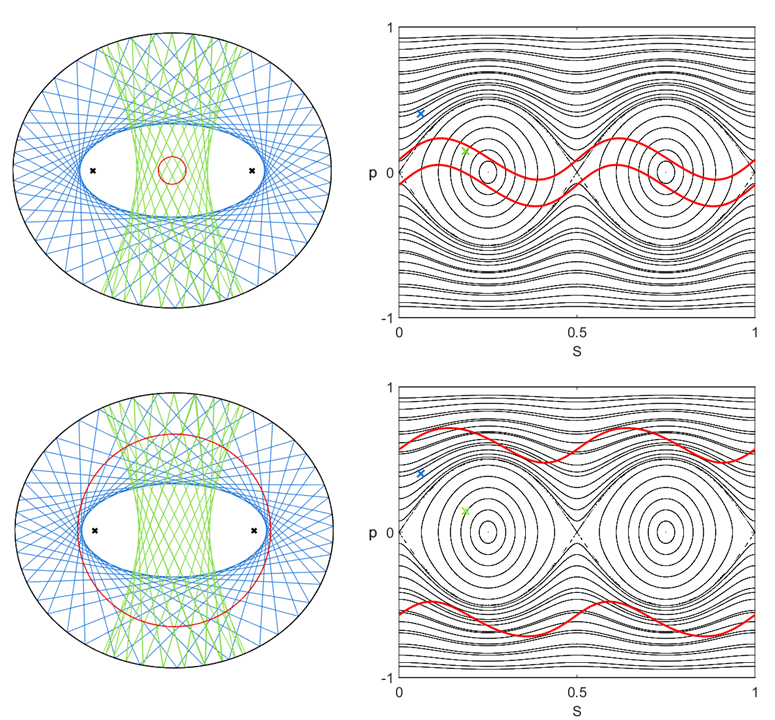}
\end{center}
\caption{Left: Elliptic billiard with $b=1$ and  $\epsilon = 0.5$ and absorbing regions of radii $R=0.1$ (top) and $R=0.7$ (bottom). Right: Corresponding Poincar\'e sections, with initial conditions for trajectories (green and blue crosses) and boundaries of the set ${\cal S}_1$ (red lines).}
\label{diagram:ellipse_patches_and_poincare_IC}
\end{figure}

\begin{figure*}[t]
\includegraphics[width=0.38\columnwidth]{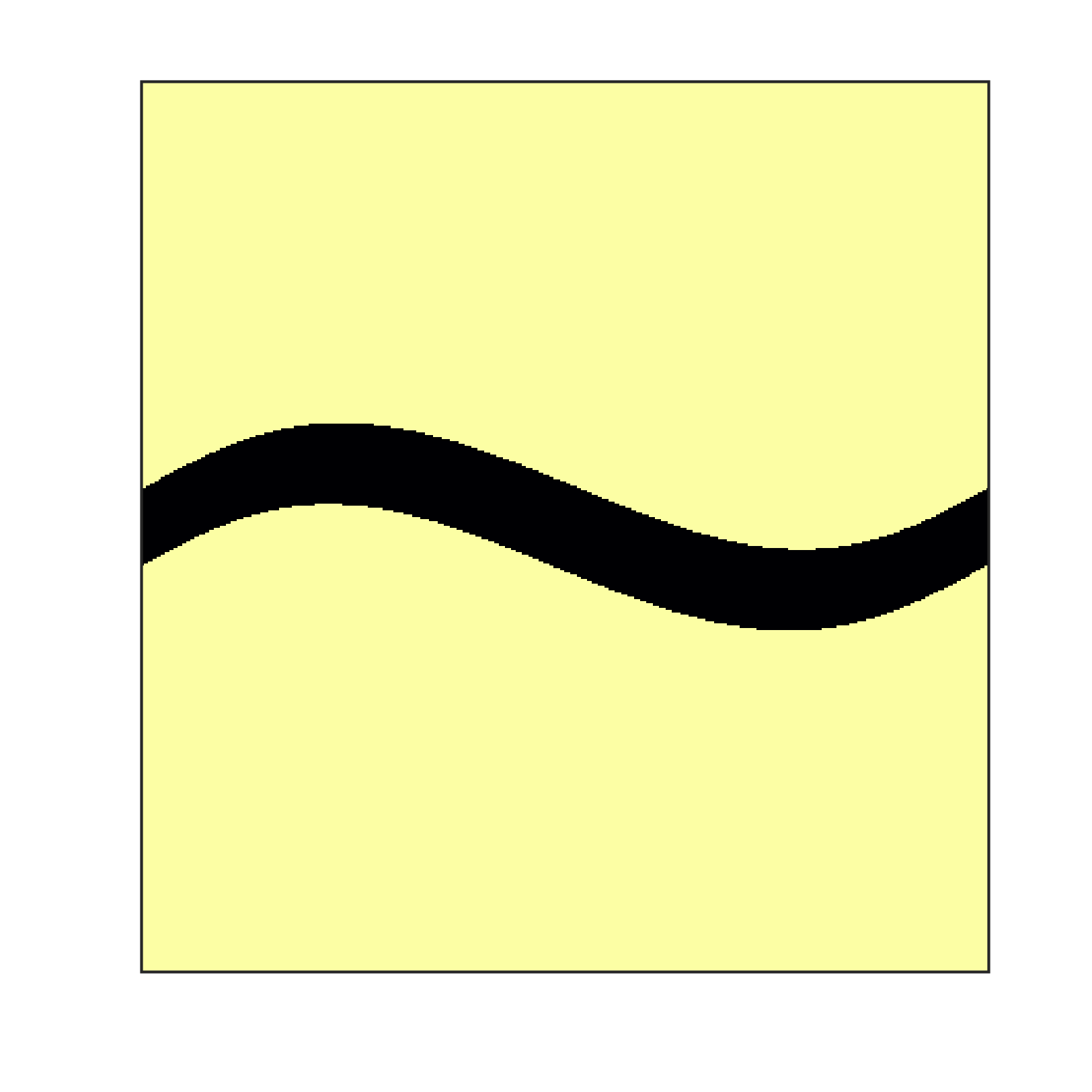}~
\includegraphics[width=0.38\columnwidth]{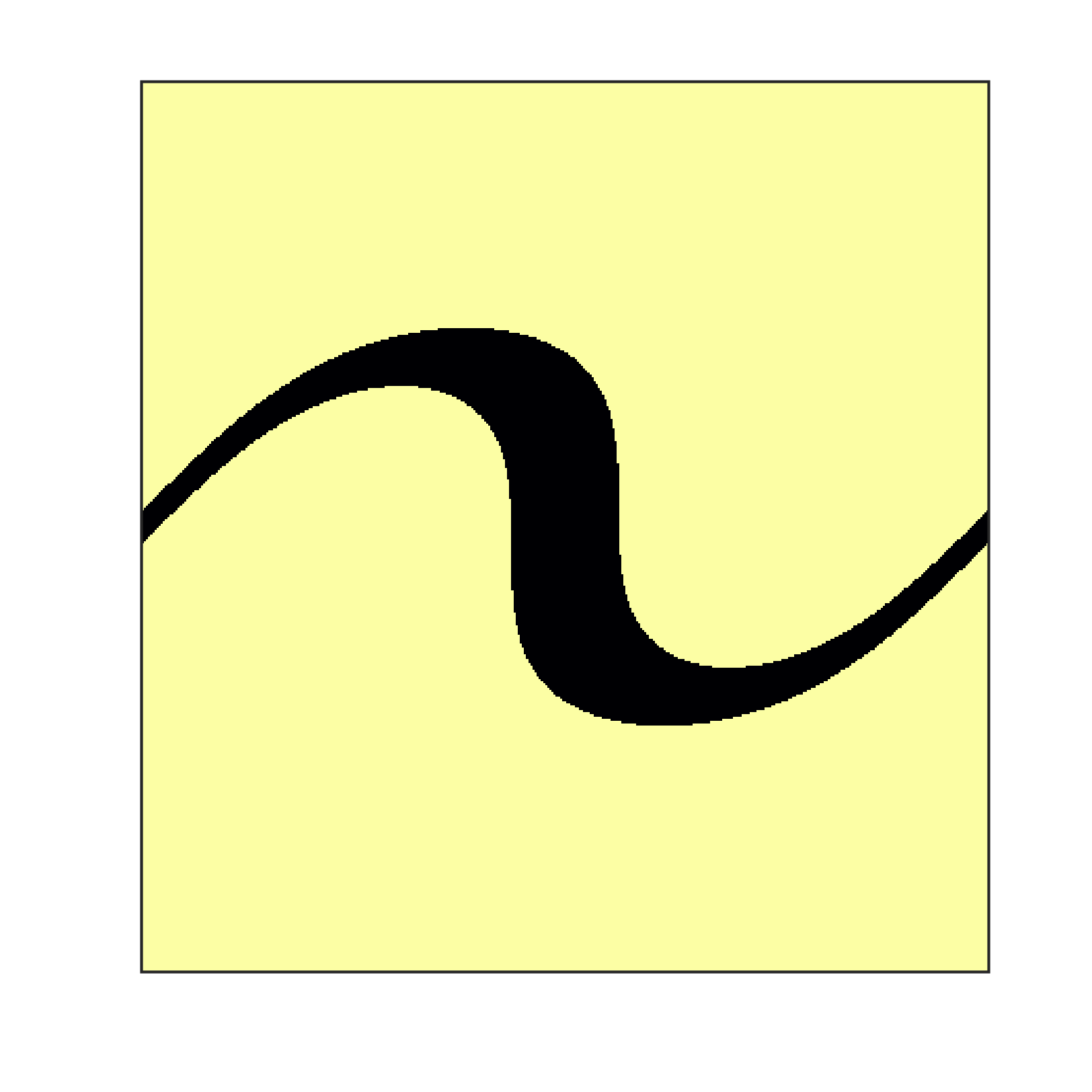}~
\includegraphics[width=0.38\columnwidth]{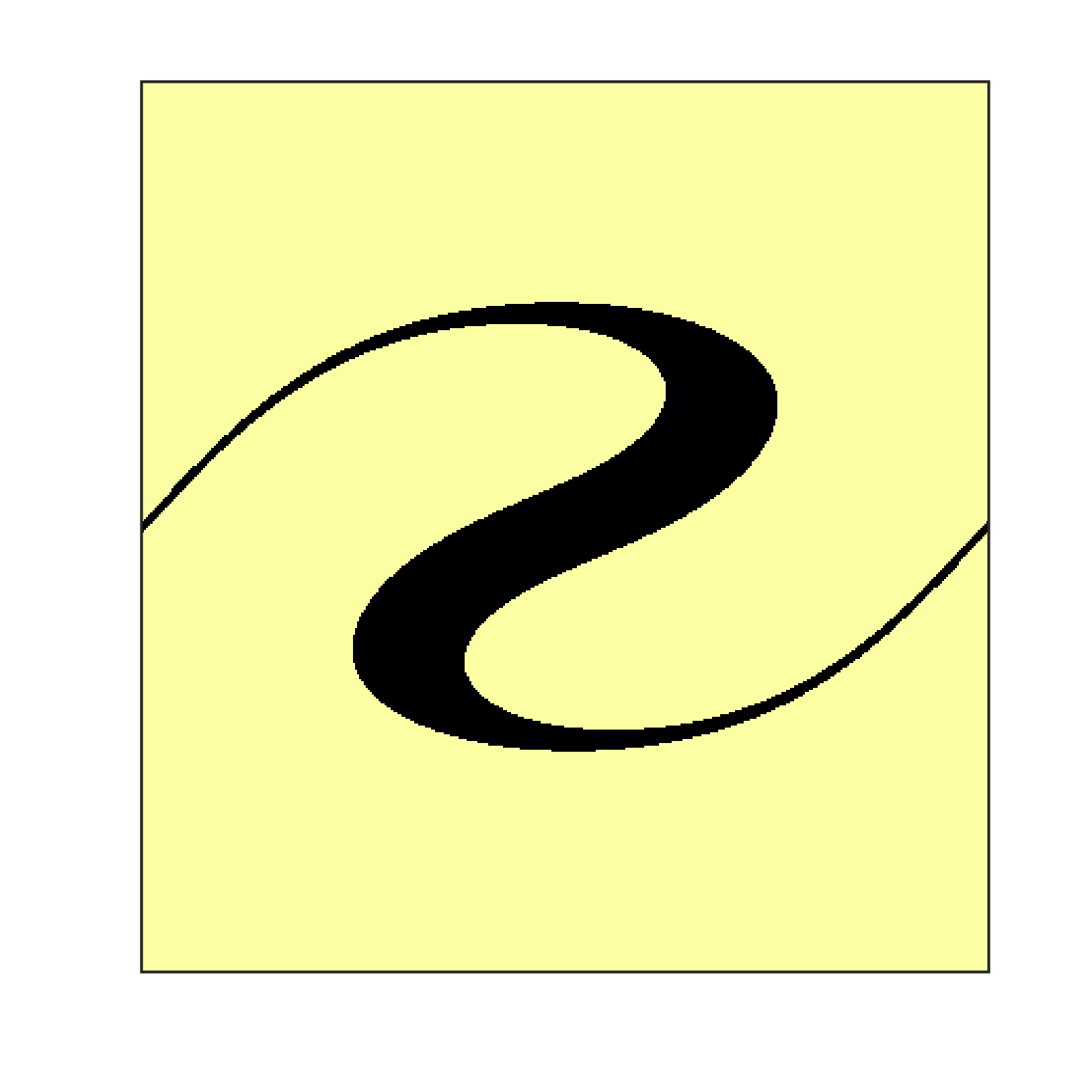}~
\includegraphics[width=0.38\columnwidth]{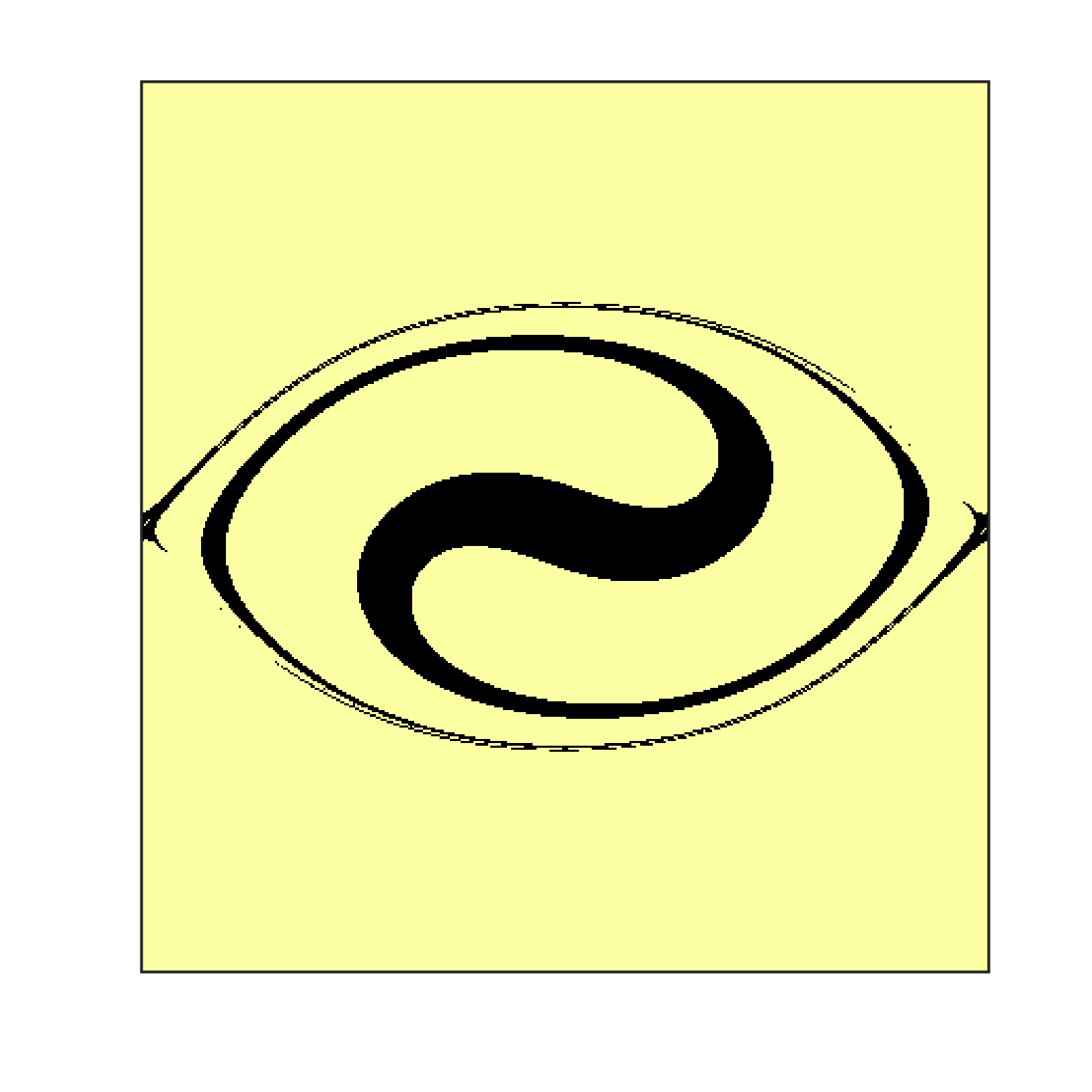}~
\includegraphics[width=0.38\columnwidth]{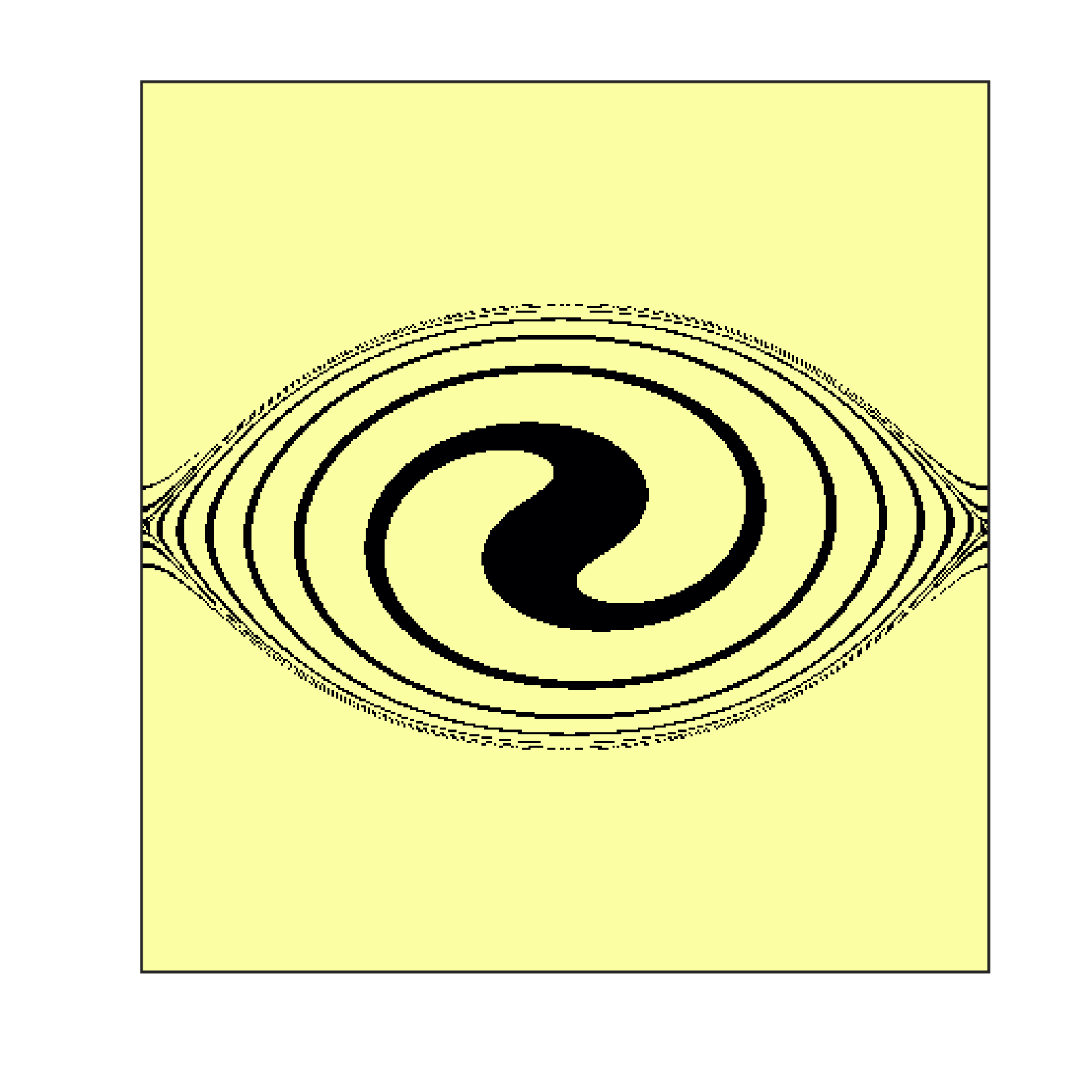}\\
\includegraphics[width=0.38\columnwidth]{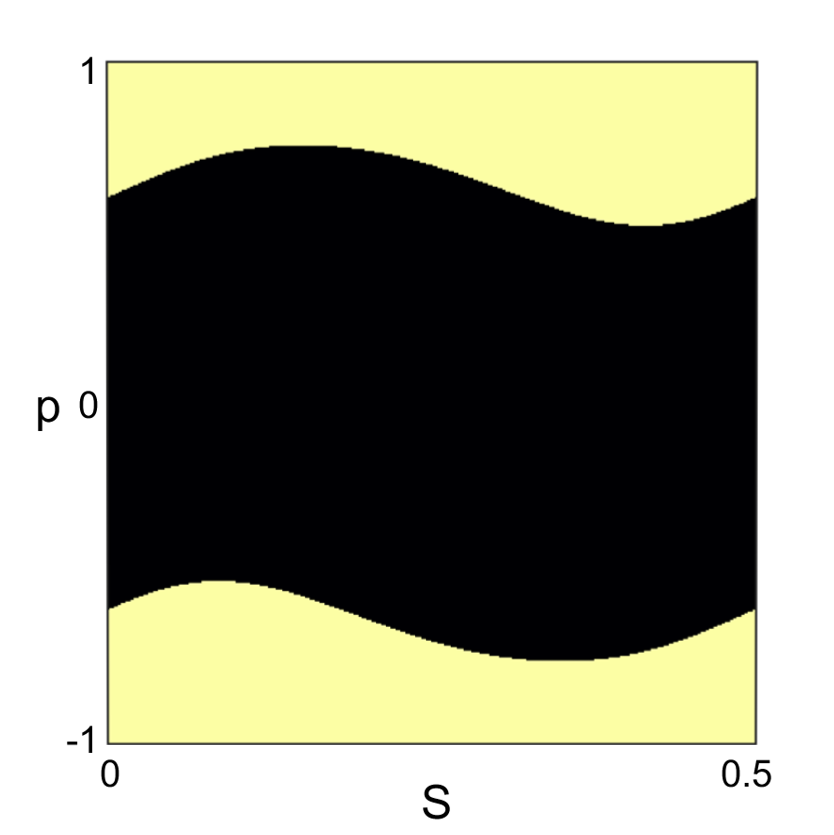}~
\includegraphics[width=0.38\columnwidth]{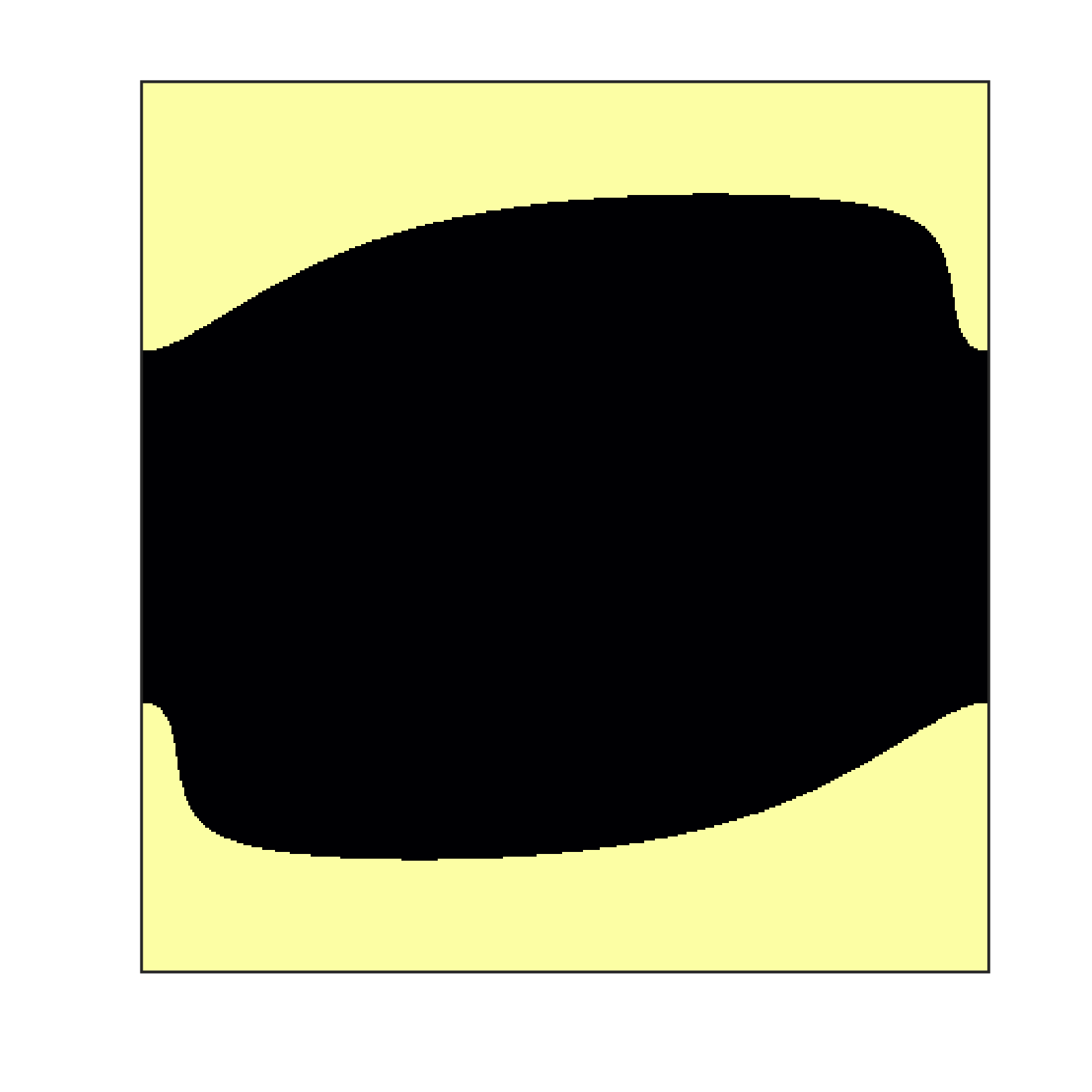}~
\includegraphics[width=0.38\columnwidth]{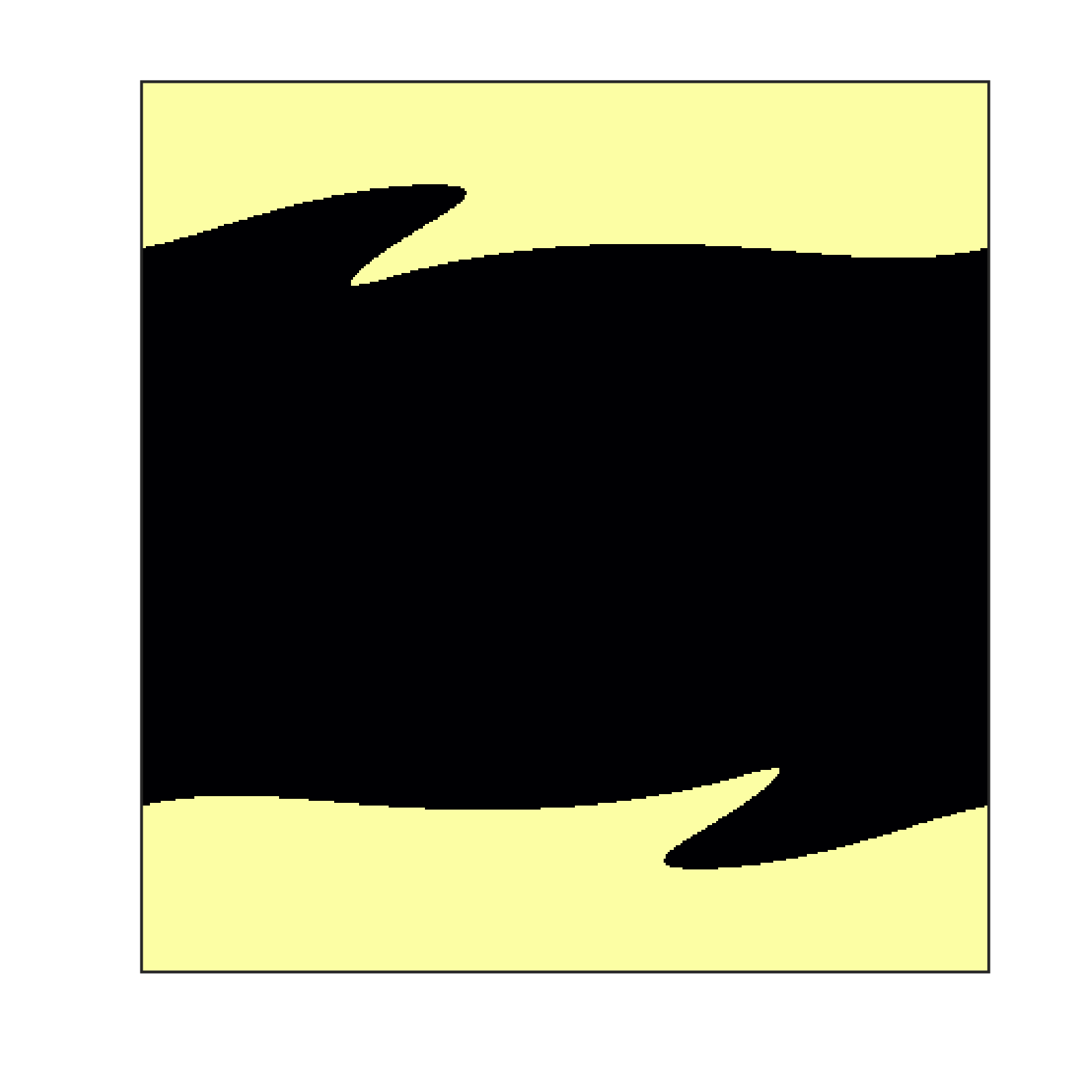}~
\includegraphics[width=0.38\columnwidth]{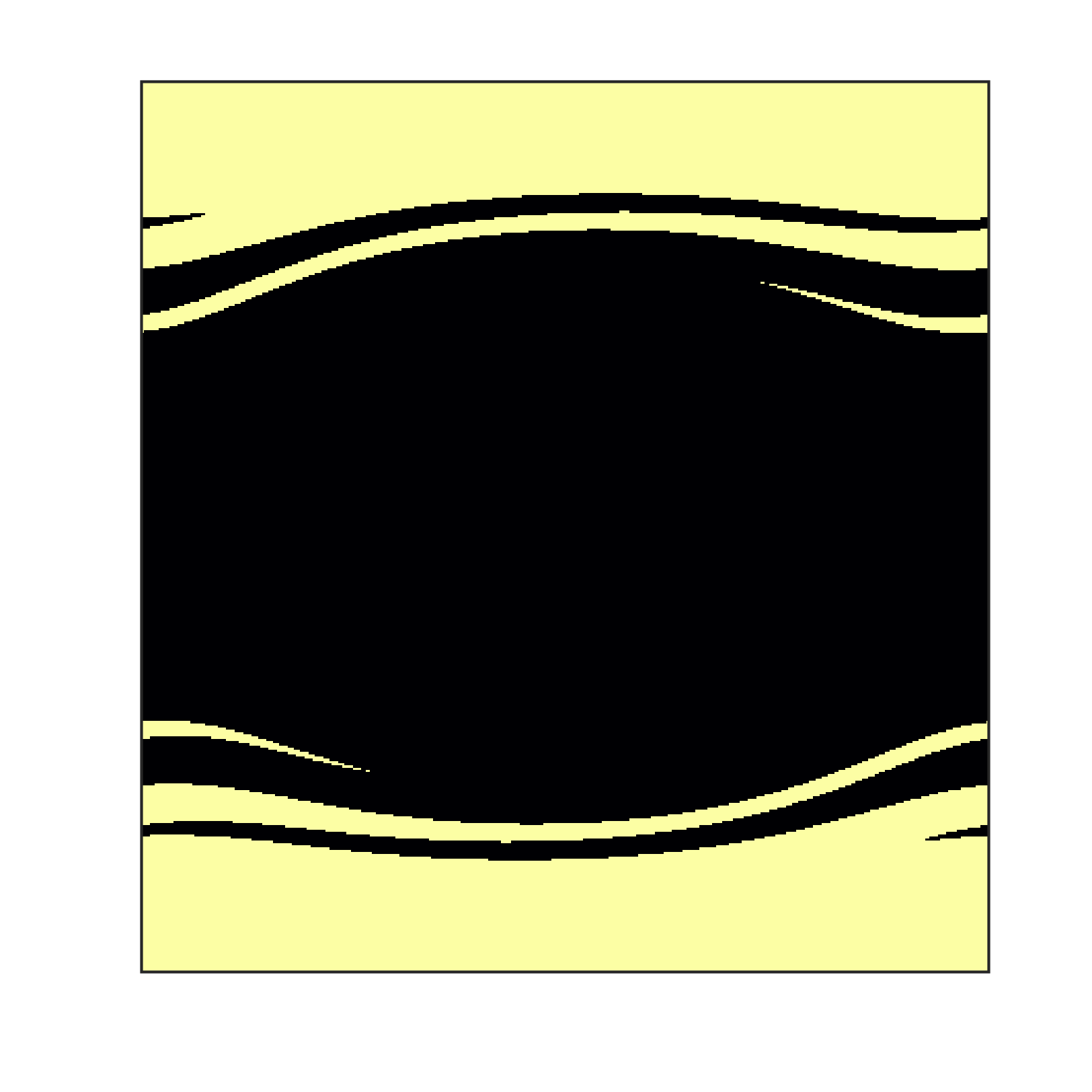}~
\includegraphics[width=0.38\columnwidth]{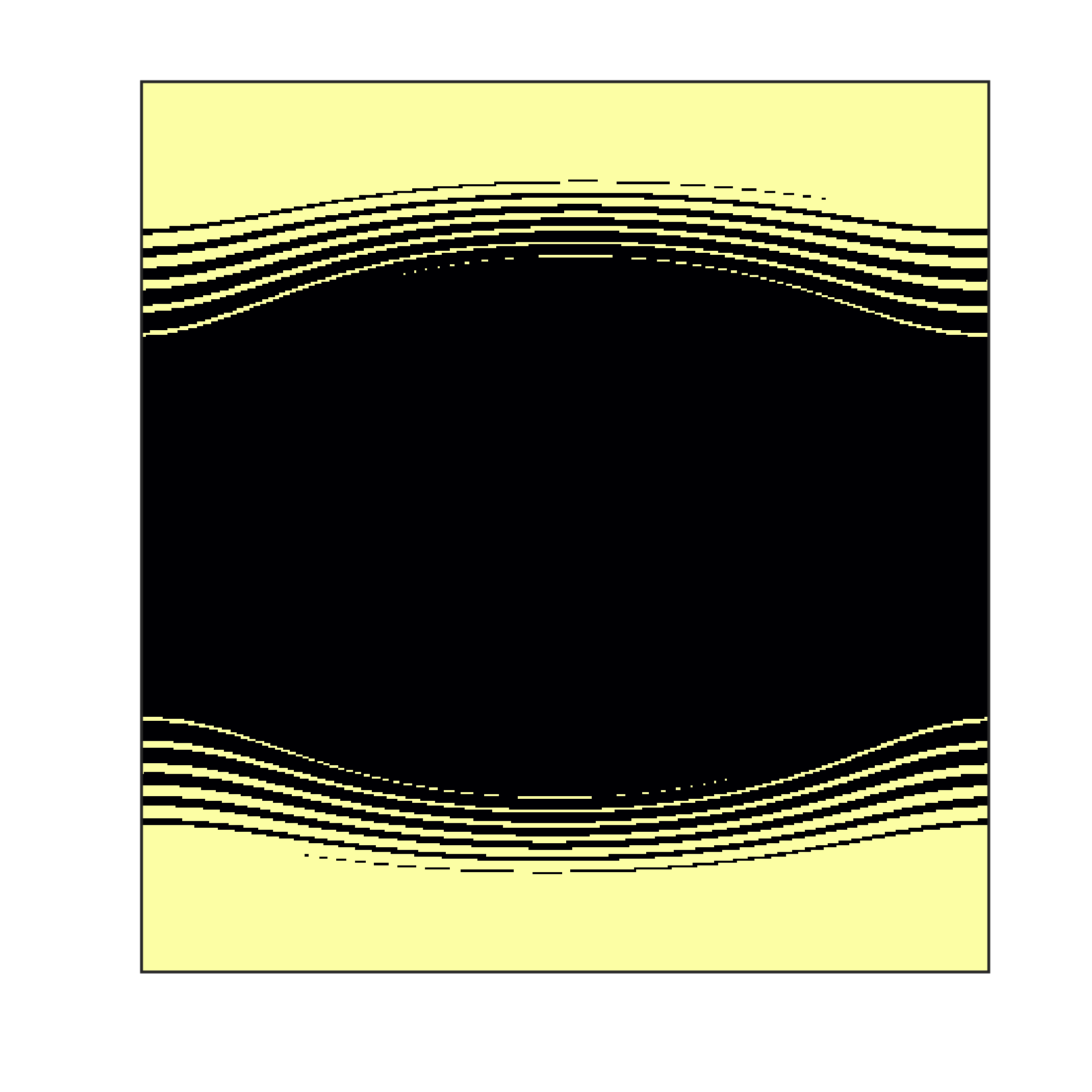}
\caption{The set of initial conditions for the ellipse that intersect the absorbing region at each iteration for $R=0.1$ (top) and $R=0.7$ (bottom) depicted over half of the phase space with \(S\in[0,0.5]\)). From left to right the iterations are $n=1,2,3,10,\rm{and}\ 30$.}
\label{diagram:ellipse_S_n}
\end{figure*}

\section{Ellipse}\label{sec-ellipse}
Let us now consider a billiard with an elliptic boundary, which, although still integrable, leads to more intricate features in the presence of the absorbing region. The boundary is described by 
\begin{equation}
     r_{\rm ellipse}(\phi)=\frac{b}{\sqrt{1-\epsilon^{2}\cos{\phi}^{2}}},
\end{equation}
where $\epsilon\in\mathbb{R}$ with  $0\leq\epsilon<1$, and $b\in\mathbb{R^{+}}$. %$0\leq c\leq a$. $a = \frac{2}{\sqrt{3}}$. 
The arc length of the ellipse is expressed in terms of an elliptic integral as
\begin{equation}
    S(\phi) =\frac{b}{\sqrt{(1-\epsilon^{2})}}\, E(\phi,\epsilon),
\end{equation}
where \(E(\phi,\epsilon)\) denotes the incomplete elliptic integral of the second kind with eccentricity \(\epsilon\).
Throughout this section we will use parameters $b=1$ and  $\epsilon = 0.5$.% and consider absorbing regions of size $R=0.1$ and $R=0.7$.

In contrast to the circle, the dynamics in a billiard with elliptic boundary gives rise to different types of trajectories, known as box and loop trajectories, respectively \cite{berry1981regularity,bandres2004classical,tabachnikov2005geometry}. The top left panel of Figure \ref{Figure_propagated_lines} shows the billiard in coordinate space with two example trajectories; a box trajectory is depicted as a green line, and a loop trajectory is shown in blue. The corresponding initial points are highlighted in the Poincar\'e section, depicted in the top right panel. The Poincar\'e section shows a structural resemblance to the phase space of the mathematical pendulum. It is organised around two 2-cycles, corresponding to two \textit{bouncing-ball} trajectories, that reflect off two opposing points indefinitely. The bouncing-ball trajectory in the vertical direction (iterating between $S=\frac{1}{4}$ and $S=\frac{3}{4}$) is stable, leading to elliptic structures in the Poincar\'e section in the neighbourhood of the corresponding 2-cycle. The corresponding trajectories are the box trajectories, which are restricted to only parts of the boundary corresponding to two distinct intervals of $S$ in the Poincar\'e section. The bouncing ball mode in the horizontal direction, on the other hand, is unstable, and corresponds to the separatrix structure in the Poincar\'e section. This separates the phase space region associated to the box trajectories from that associated to the loop trajectories, that traverse around the whole of the billiard boundary. In the case of large initial values of $|p|$ (corresponding to small angles of incidence), one observes \textit{whispering-gallery} trajectories, that stay close to the boundary (confined to a narrow band of large values of $|p|$).

While the Poincar\'e diagram gives an overview of the dynamical behaviour, it is also instructive to consider a different visualisation of the map \(T\) in which one considers the images for a sub-section of points for select iterations. In the bottom row of Figure  \ref{Figure_propagated_lines} we depict two such examples with $200$ initial equidistant points arranged in two horizontal lines located at \(p=0\) (left) and \(p=0.75\) (right) for one, two, and three iterations. While the dynamics are regular for the elliptic billiard, in this visualisation one observes the appearance of non-trivial whorl and tendril structures \cite{berry1979evolution}. For the horizontal line at \(p=0\) all points are transported along box trajectories leading to different displacements for each initial point, but all to points lying within the separatrix, leading to a characteristic whorl structure after several iterations. Further iterations lead to an increasingly fine whorl structure within the separatrix. In contrast to this, a horizontal line at \(p=0.75\) consists of the initial conditions for only loop trajectories, and after many iterations leads to a filament structure confined to a band outside of the separatrix.

Let us now move on to consider the effects of our absorbing region. Figure \ref{diagram:ellipse_patches_and_poincare_IC} again depicts the billiard in coordinate space (left) with two example trajectories and the corresponding Poincar\'e sections (right) with two different sizes of the absorbing region ($R=0.1$ on the top and $R=0.7$ on the bottom), the boundary of which is marked by red lines in the left panels, and the boundaries of the corresponding sets ${\cal S}_1$ in the Poincar\'e section are depicted by red lines in the right panels. The corresponding initial points of the example trajectories are highlighted in the Poincar\'e sections by crosses, which are either contained within or fall outside of the sets ${\cal S}_1$.

We observe that the set ${\cal S}_1$ contains both 2-cycles for both values of $R$ (and in fact, for any non-vanishing $R$). Thus, in both cases box and loop trajectories contribute to the intensity landscapes. For any non-vanishing $R$ all box trajectories intersect the absorbing region at some point. For loop trajectories, whispering gallery-type modes with sufficiently large initial values of $|p|$ avoid the absorbing region, yet for nonvanishing radius $R$, there are always some loop trajectories that do intersect the absorbing region and experience absorption. 

\begin{figure*}[t]
\includegraphics[height=0.37\columnwidth]{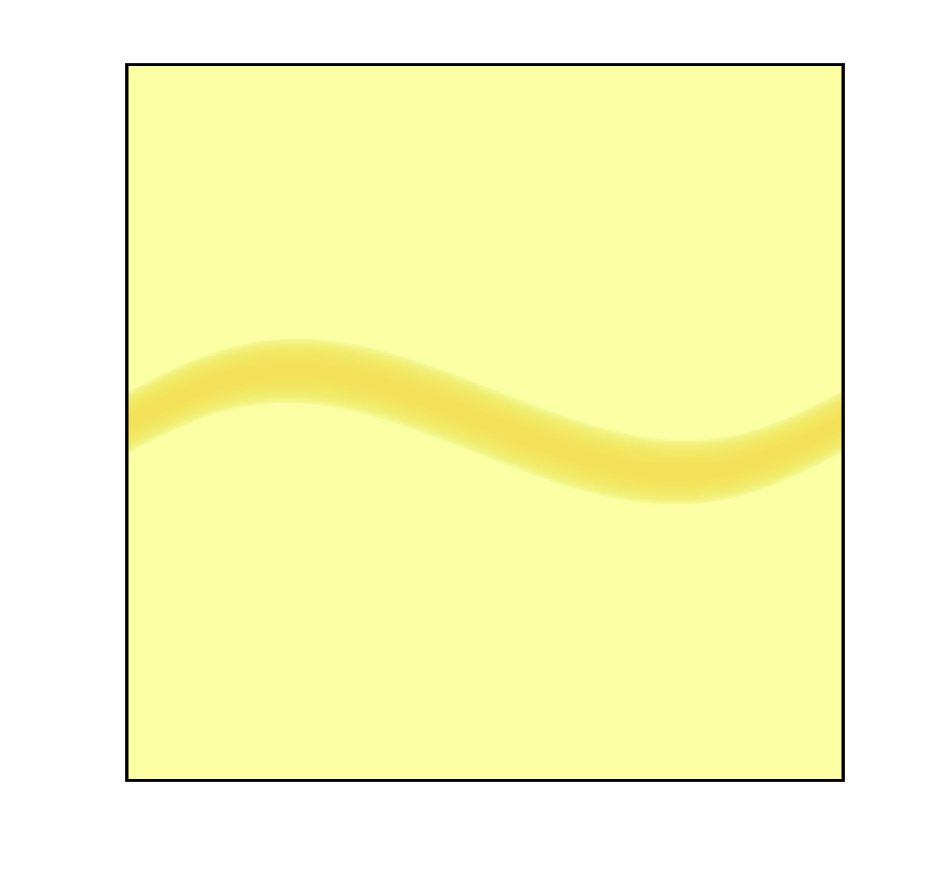}
\includegraphics[height=0.37\columnwidth]{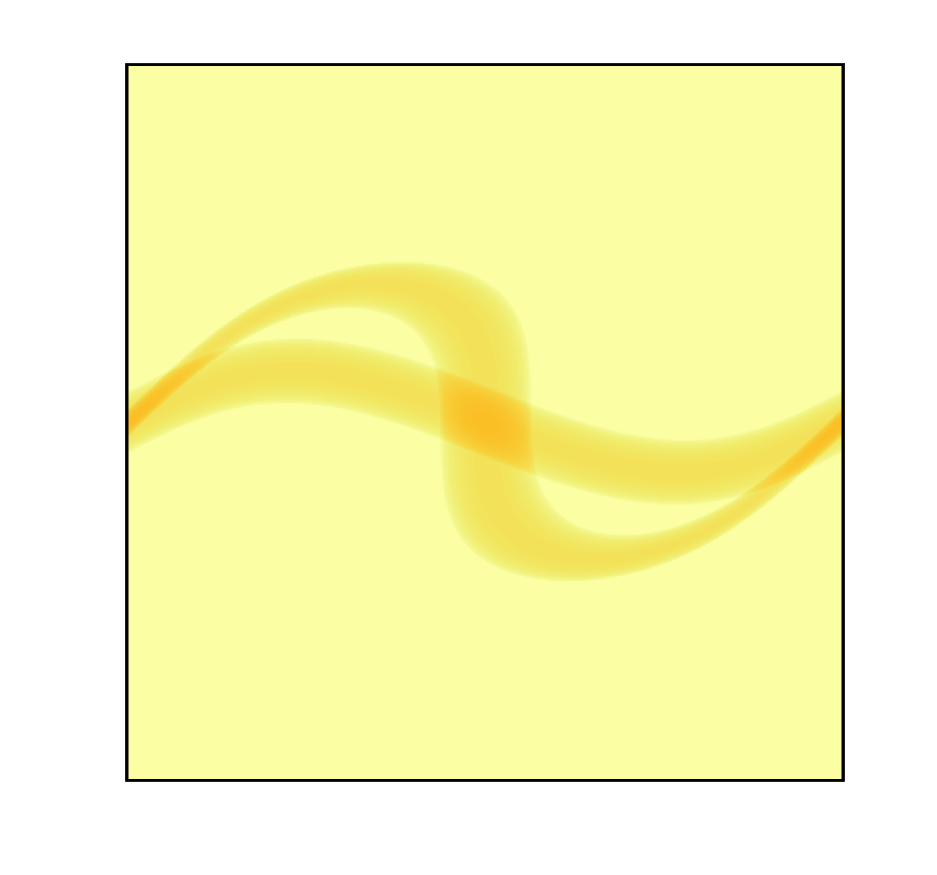}
\includegraphics[height=0.37\columnwidth]{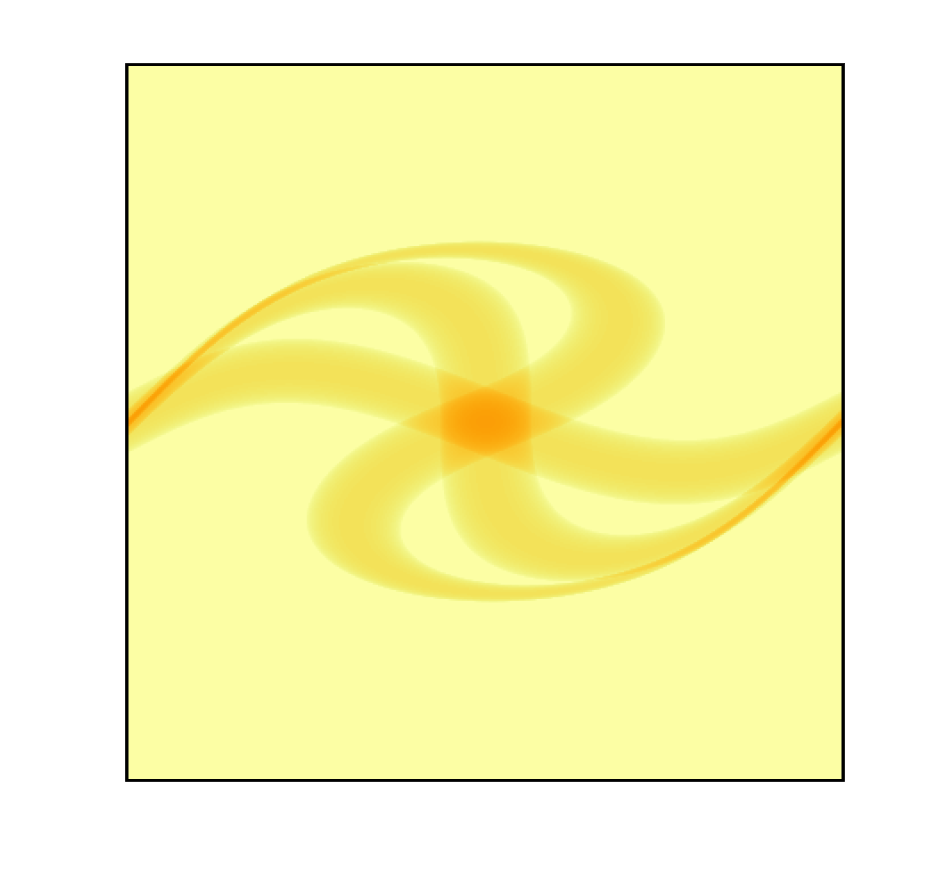}
\includegraphics[height=0.37\columnwidth]{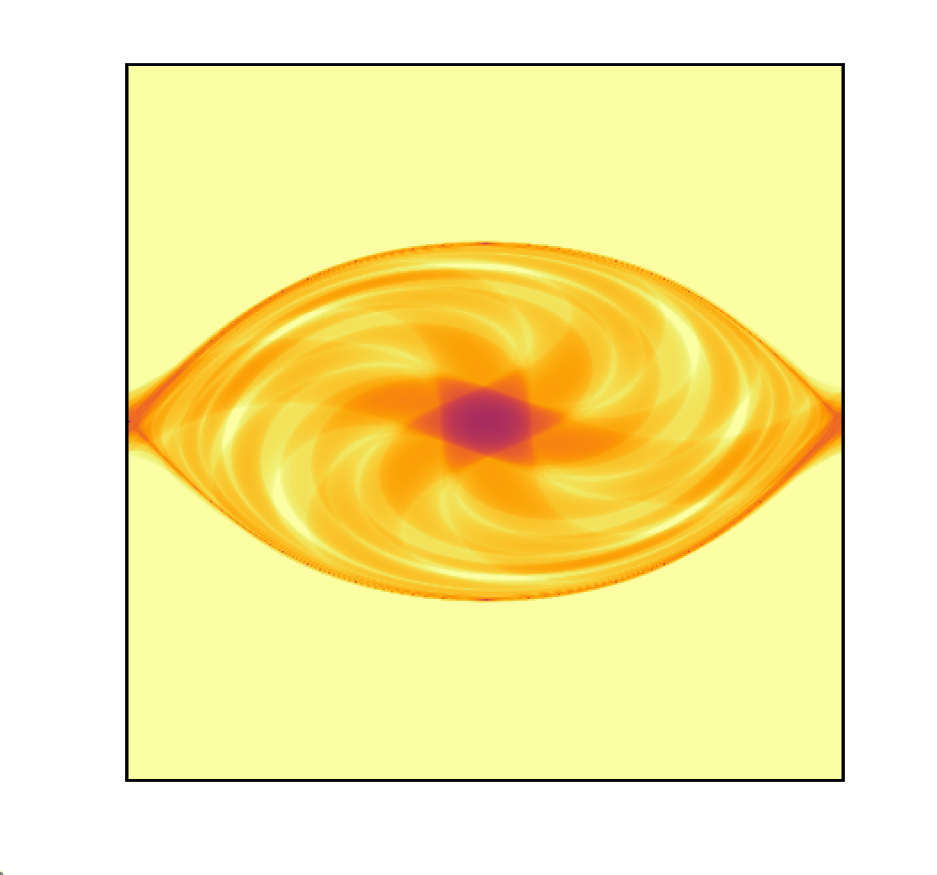}
\includegraphics[height=0.37\columnwidth]{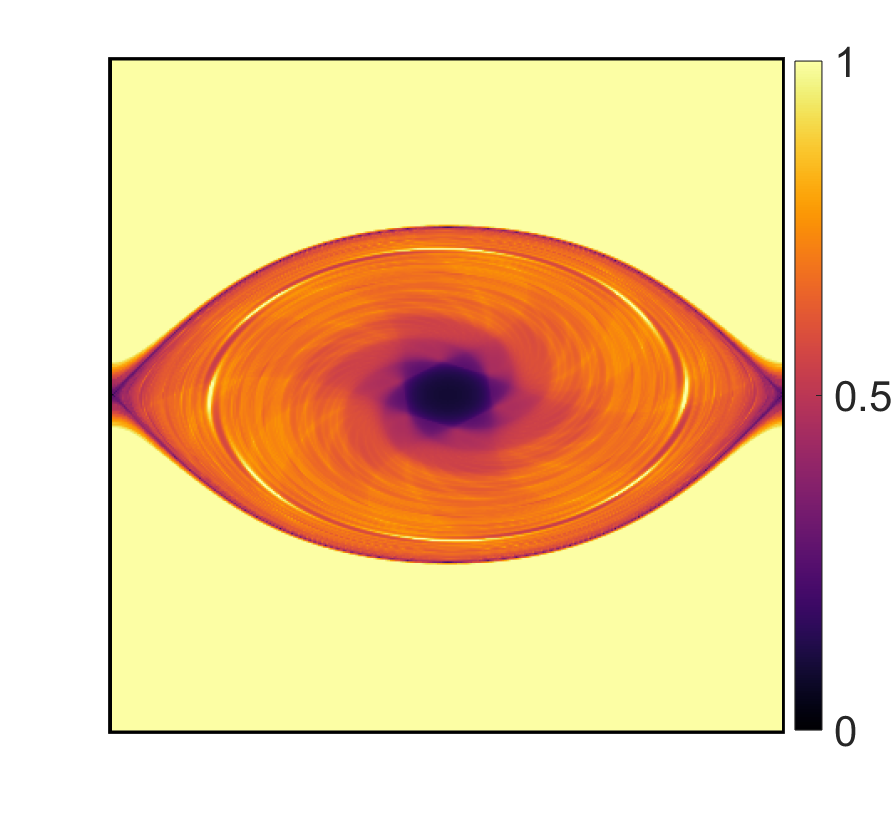}\\
\includegraphics[height=0.37\columnwidth]{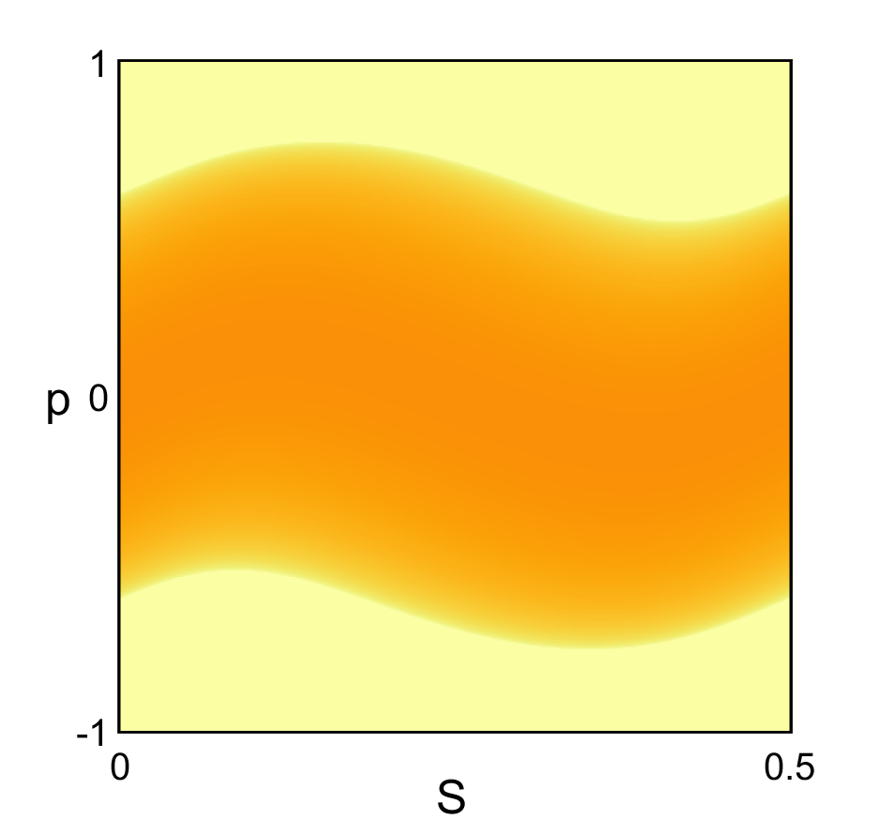}
\includegraphics[height=0.37\columnwidth]{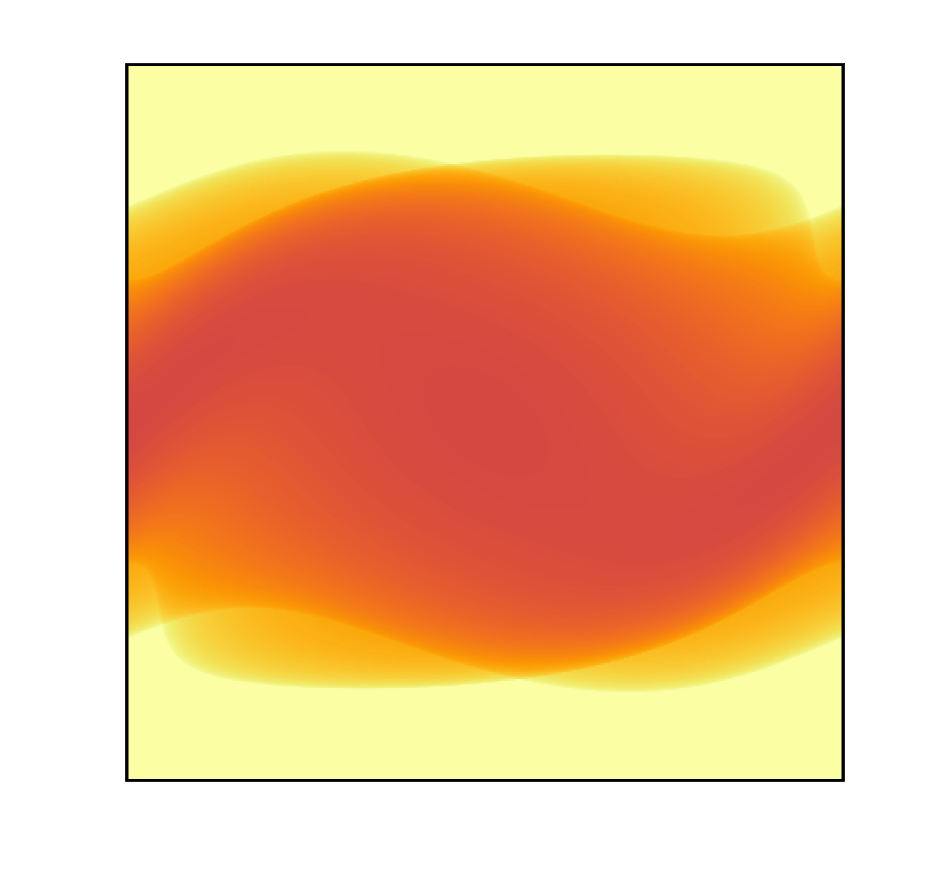}
\includegraphics[height=0.37\columnwidth]{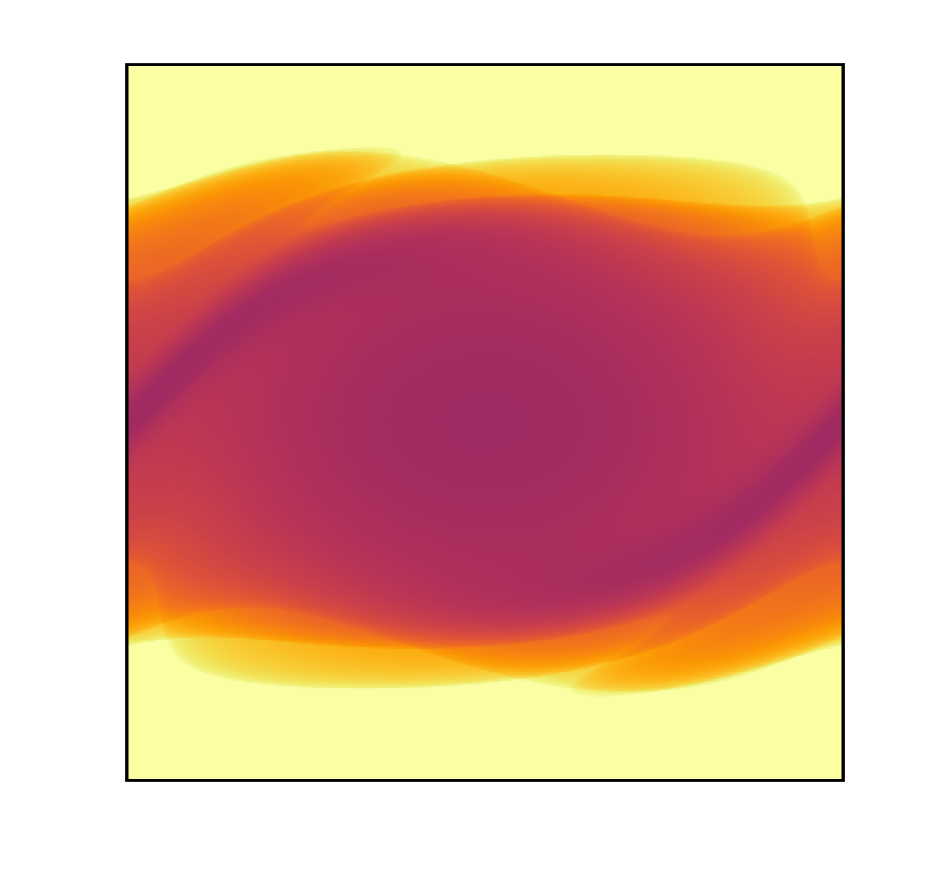}
\includegraphics[height=0.37\columnwidth]{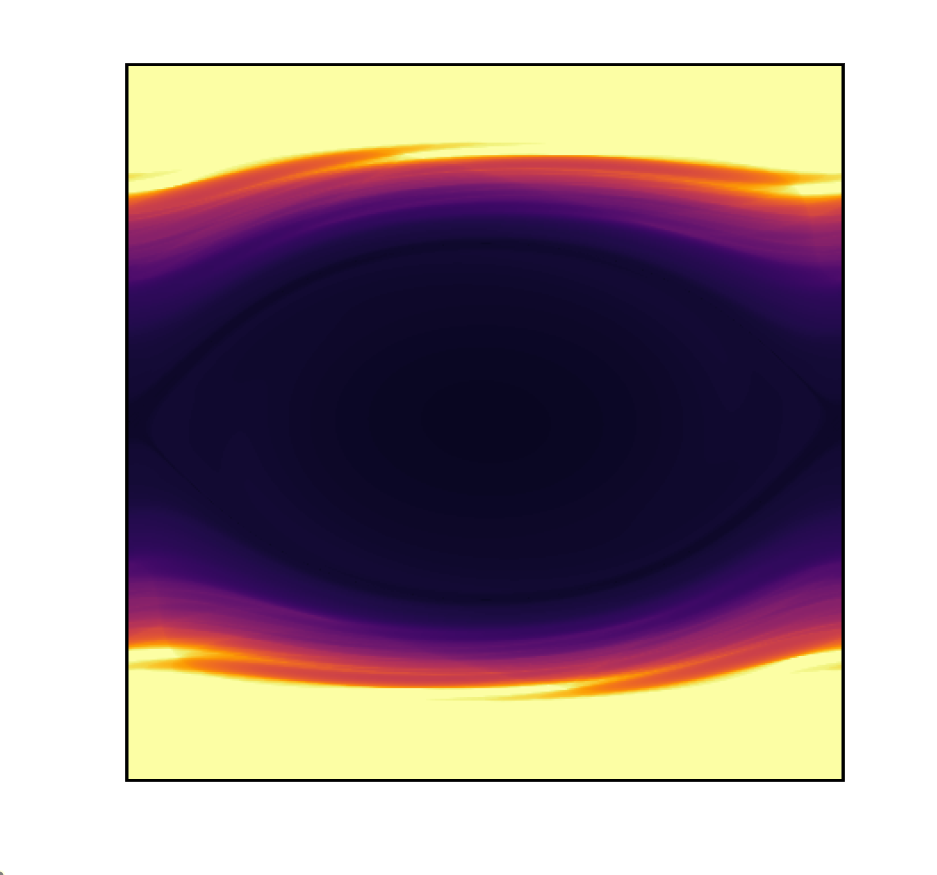}
\includegraphics[height=0.37\columnwidth]{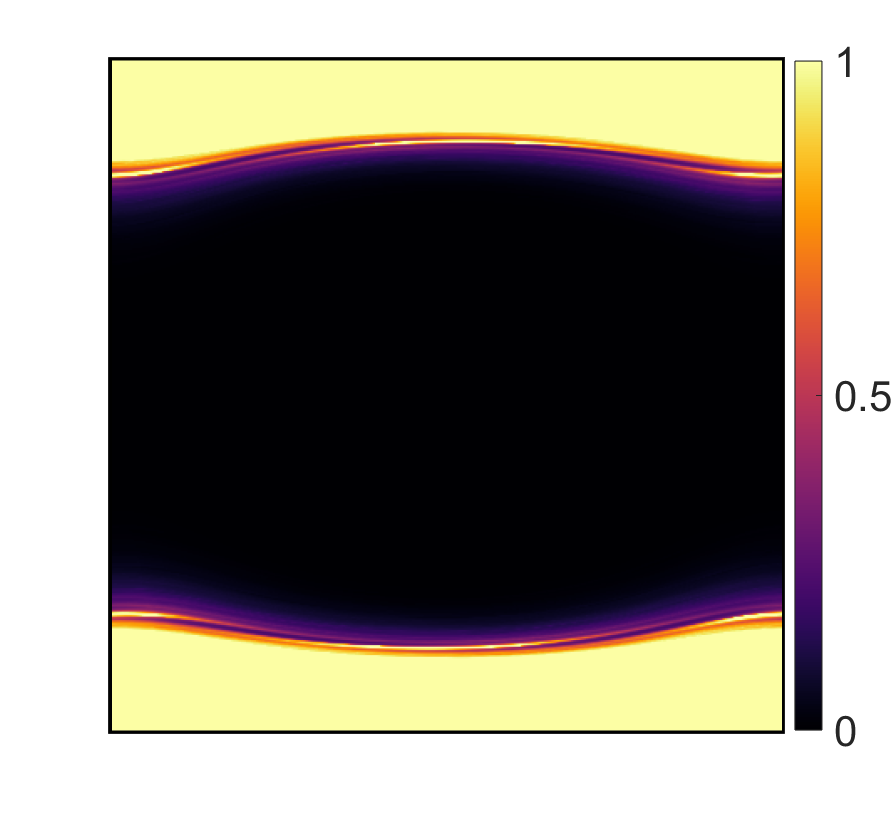}
\caption{Intensity landscape for the ellipse billiard with absorbing region with absorption rate $\gamma=0.4$ and radius $R=0.1$ (top) and absorption rate $\gamma=0.2$ and radius $R=0.7$ (bottom) for different numbers of iterations depicted over half of the phase space \(S\in[0,0.5]\). From left to right: $n_f=1,2,3,10,\rm{and}\ 30$. }
\label{diagram:ellipse_norm_evolution}
\end{figure*}

Trajectories that intersect the absorbing region in one iteration may not do so at the next. An example of a trajectory that intersects the absorbing region in the first iteration, but not in the second, has been plotted in Figure \ref{diagram:trajectory_with_patch}.  As a result, the intensity landscapes change from iteration to iteration. Figure \ref{diagram:ellipse_S_n} shows the sets ${\cal S}_n$ for five different numbers of iteration ($n=1,\, 2,\, 3,\, 10$, and $30$) for the two different radii of the absorbing region considered here ($R=0.1$ on the top and $R=0.7$ on the bottom). Due to space constraints we only depict half of the phase space here for the region \(S\in[0,0.5]\). The behaviour on the remaining half can be deduced from the symmetry of the system.  
Let us first focus on the top row, depicting the ${\cal S}_n$ for the smaller radius $R=0.1$. We observe that at each iteration there is a set of initial conditions near $(S,p) = (0.25,0)$ (which corresponds to the bouncing ball trajectories) that intersects the absorbing region in every iteration. Outside of this set, we see the development of filament structures around the central region. Recall that the ${\cal S}_n$ are related to ${\cal S}_1$ by successive applications of the inverse flow $T^{-1}$, where the forwards map \(T\) is depicted in Figure \ref{Figure_propagated_lines}. The filaments seen are the transported set $\mathcal{S}_1$ after different numbers of backwards iterations. In the case of $R=0.7$, depicted in the bottom row, the set ${\cal S}_1$ contains the separatrix, i.e. it contains the entire set of phase-space points belonging to box trajectories, that is,  every box trajectory intersects the absorbing region in every iteration. This area is thus invariant. The evolution of the ${\cal S}_{n>1}$ is therefore confined to those initial conditions in the area of the Poincar\'e section outside the separatrix related to loop trajectories with initial conditions in the set ${\cal S}_1$. In higher iterations, we observe the development of thin filament structures on the edges of the bulk region of ${\cal S}_1$. A displacement of the absorbing region away from the centre, breaks the symmetry of the system, leading different patterns in the sets ${\cal S}_n$. An example is provided in Appendix \ref{appendix:offset_AR}.

The intensity landscapes, depicted in Figure \ref{diagram:ellipse_norm_evolution}, for $R=0.1$ (top) and $R=0.7$ (bottom), are structured around the union of the sets ${\cal S}_n$. 
Recall that the intensity landscapes for different (finite) values of $\gamma$ are trivially related to each other. Here we chose the values $\gamma=0.4$ for the smaller radius and the value $\gamma=0.2$ for the larger radius (to compare the intensities on the same false-colour scale), and depict the intensity on the Poincar\'e section after 1, 2, 3, 10, and 30 iterations, respectively. The intensity landscape after the first iteration is equal to one on the complement of the set ${\cal S}_1$, and shows modulations in value on the set ${\cal S}_1$, due to the different length of trajectory segments passing through the absorbing region for different initial conditions. The logarithm of the intensity landscape for successive iterations is then simply the sum of that for the first iteration and its backwards images under the billiard map. Initial conditions that intersect the absorbing region in more iterations, i.e. the intersections of the sets ${\cal S}_n$ lead to a stronger loss of intensity, as is clearly visible for example in the star-shaped structure in the centre of the Poincar\'e section for $R=0.1$. As a result of this, for larger numbers of iterations, we observe a pronounced valley structure along the separatrix and the two-cycles of the Poincar\'e dynamics, related to trajectories that have long segments in the absorbing region on all or most iterations. 

The intensity landscape for an absorbing region of radius \(R=0.7\) shown in Figure \ref{diagram:ellipse_norm_evolution} displays a similar structure and evolution, however, since all box trajectories and some of the loop trajectories intersect the absorbing region in every iteration  and in long segments, there is a more extended region of very low intensity values at the centre of the phase space, and the separatrix does not stand out.   

\section{Oval Billiard}\label{sec-oval}

\begin{figure}[t]
\hspace{-0.2cm}\includegraphics[width=0.95\columnwidth]{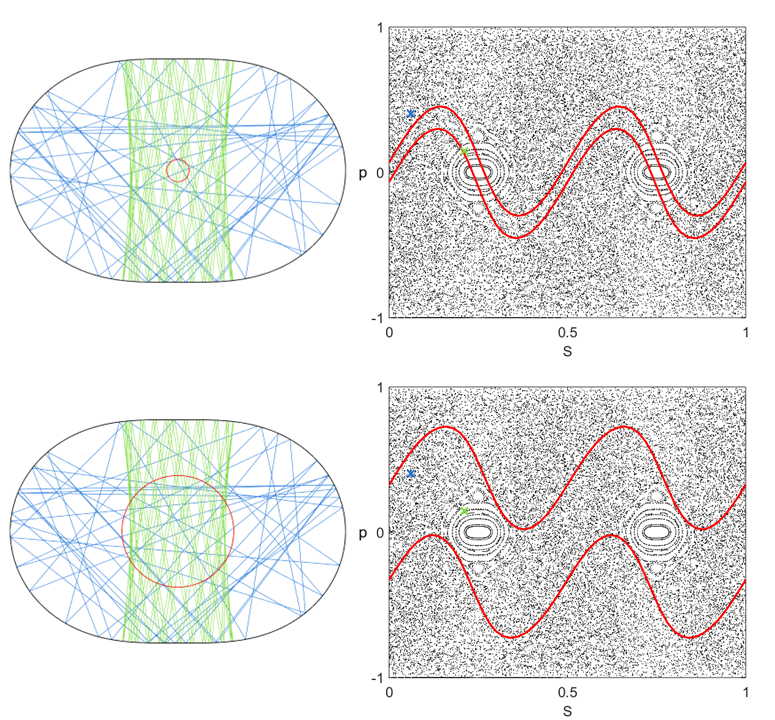}
\caption{Oval billiard with $c=0.25$ and absorbing regions of radius $R=0.1$ (top) and $R=0.5$ (bottom). Left: Billiard boundary (black line) in coordinate space with boundary of the absorbing region (red line), and two example trajectories: one periodic trajectory (blue) and a chaotic (green). Right: Corresponding Poincar\'e sections with initial conditions for trajectories on top (green and blue crosses) and boundaries of the set ${\cal S}_1$ (red lines). }
\label{diagram:oval_patches_and_poincare_IC}
\end{figure}

Finally, we consider an oval billiard with boundary
\begin{equation}
    r_{\text{oval}}(\phi) = c + \cos(2\phi),
\end{equation}
where $c\in\mathbb{R}^{+}$. Throughout this section we will use parameters $c=0.25$ and consider absorbing regions of size $R=0.1$ and $R=0.5$.
The oval billiard has a mixed regular-chaotic dynamics. Figure \ref{diagram:oval_patches_and_poincare_IC} depicts the billiard in coordinate space (top) and the corresponding Poincar\'e sections (bottom) with two different sizes of the absorbing region ($R=0.1$ on the left and $R=0.7$ on the right), the boundary of which is marked by red lines in the top panels, and the boundaries of the corresponding sets ${\cal S}_1$ in the Poincar\'e section are depicted by red lines in the bottom panels. Two example trajectories are shown in the top panels; a trajectory belonging to the two-chain of regular islands, i.e. an elliptic trajectory is depicted as a green line, and a chaotic trajectory is shown in blue. The corresponding initial points are highlighted in the Poincar\'e sections by crosses. 
The Poincar\'e section consists of a prominent regular island embedded in a large chaotic sea. The bouncing ball trajectory still gives rise to two elliptic structures in the Poincar\'e section, belonging to trajectories that bounce back and forth between the top and the bottom boundaries and do not explore the rest of the boundary. In contrast to the circular and elliptic billiard, most of the phase space of the oval considered here is covered by a chaotic sea. Trajectories within the chaotic sea are dynamically connected and ergodically cover the available phase-space area. 

The set $\mathcal{S}_1$ of initial conditions that intersect the absorbing region in the first iteration, contains the elliptic 2-cycle belonging to the bouncing ball trajectory at the centre of the regular island in the Poincar\'e section as well as a portion of the chaotic sea for both radii $R=0.1$ and $R=0.5$, and in fact for arbitrary small non-vanishing $R>0$.  Thus, regardless of the size of the absorbing region $\mathcal{S}_1$ will always contain both regular and chaotic trajectories. If the radius of the absorbing region is large enough, all regular trajectories are contained in $\mathcal{S}_1$, as can be seen in figure \ref{diagram:oval_patches_and_poincare_IC} for $R=0.5$. 

\begin{figure}[htb]
\begin{center}
\hspace{-0.2cm}\includegraphics[width=0.95\columnwidth]{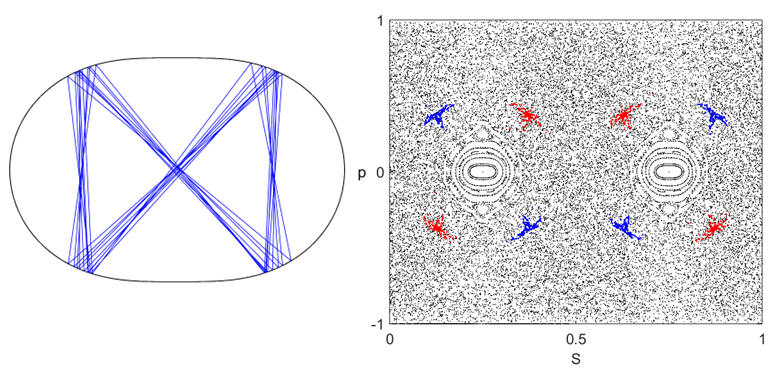}
\end{center}
\caption{Sticky trajectories in the oval billiard. The left hand plot is an example of a sticky trajectory for the first $40$ iterations. The right hand plot shows the Poincar\'e section of the oval billiard with  two sets of phase-space points corresponding to sticky trajectories highlighted in blue and red, respectively. The phase-space points of the example trajectory (left) belong to the blue regions on the Poincar\'e section. }
\label{diagram:Poincare_diagram}
\end{figure}

Within the chaotic sea there are sticky structures, that lead to trajectories spending a large proportion of iterations in the same region of phase space \cite{bunimovich2012many}. This is the case for the trajectories belonging to the transport chains outside of the main island and chaotic trajectories that are transiently caught in the neighbourhood of these islands - displayed in red and blue within the right hand plot of Figure \ref{diagram:Poincare_diagram}. The first $40$ iterations of a trajectory corresponding to the blue transport chain is depicted in the left plot of the same figure. For short iteration numbers the trajectory stays close to a bow tie shape, that, if the absorbing region is below a critical radius, intersects the absorbing region in every other iteration. Note that the trajectories belonging to the island chain marked in red in the Poincar\'e section have a similar structure but traverse the billiard in the opposite direction. 

\begin{figure*}[hbt!]
\includegraphics[height=0.38\columnwidth]{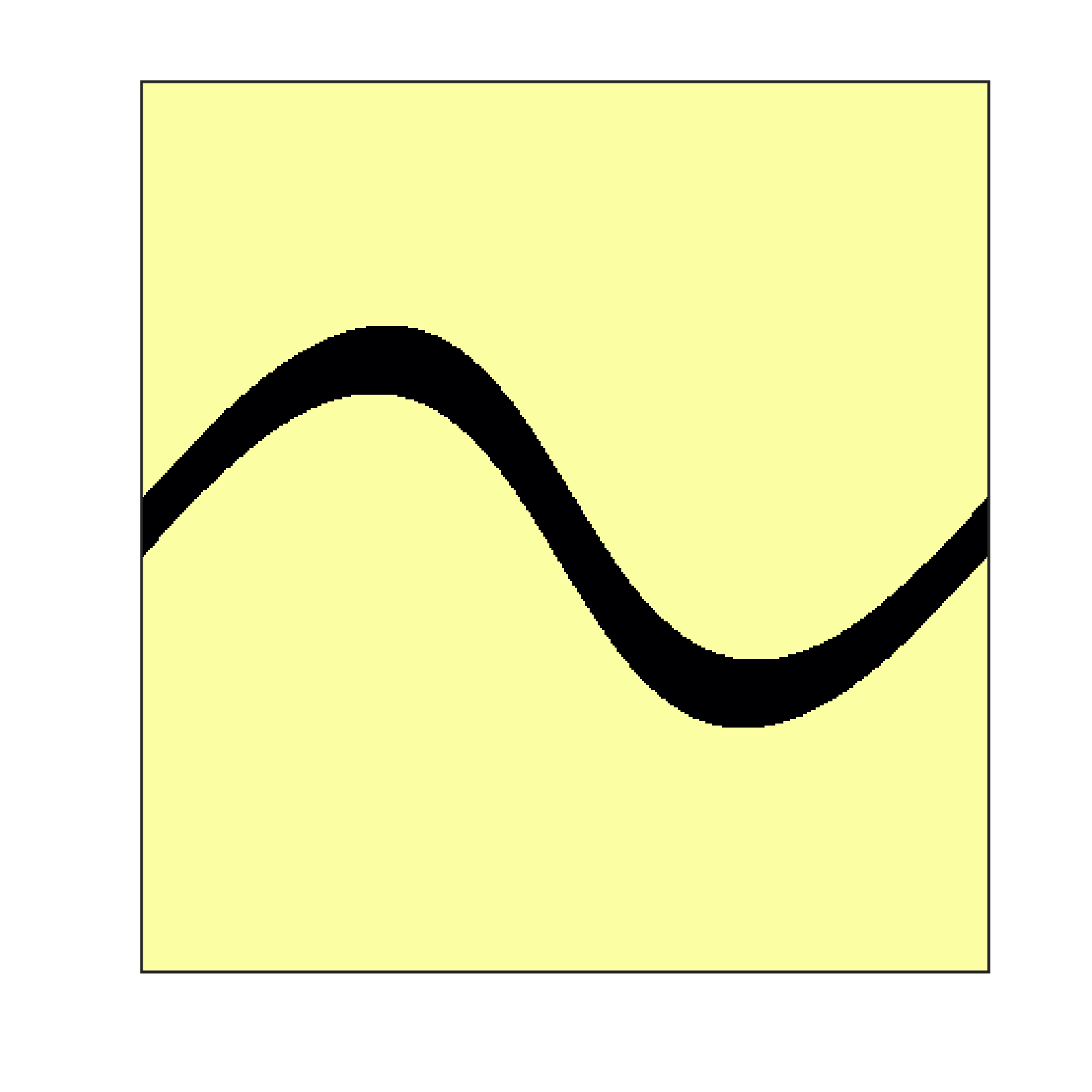}~
\includegraphics[height=0.38\columnwidth]{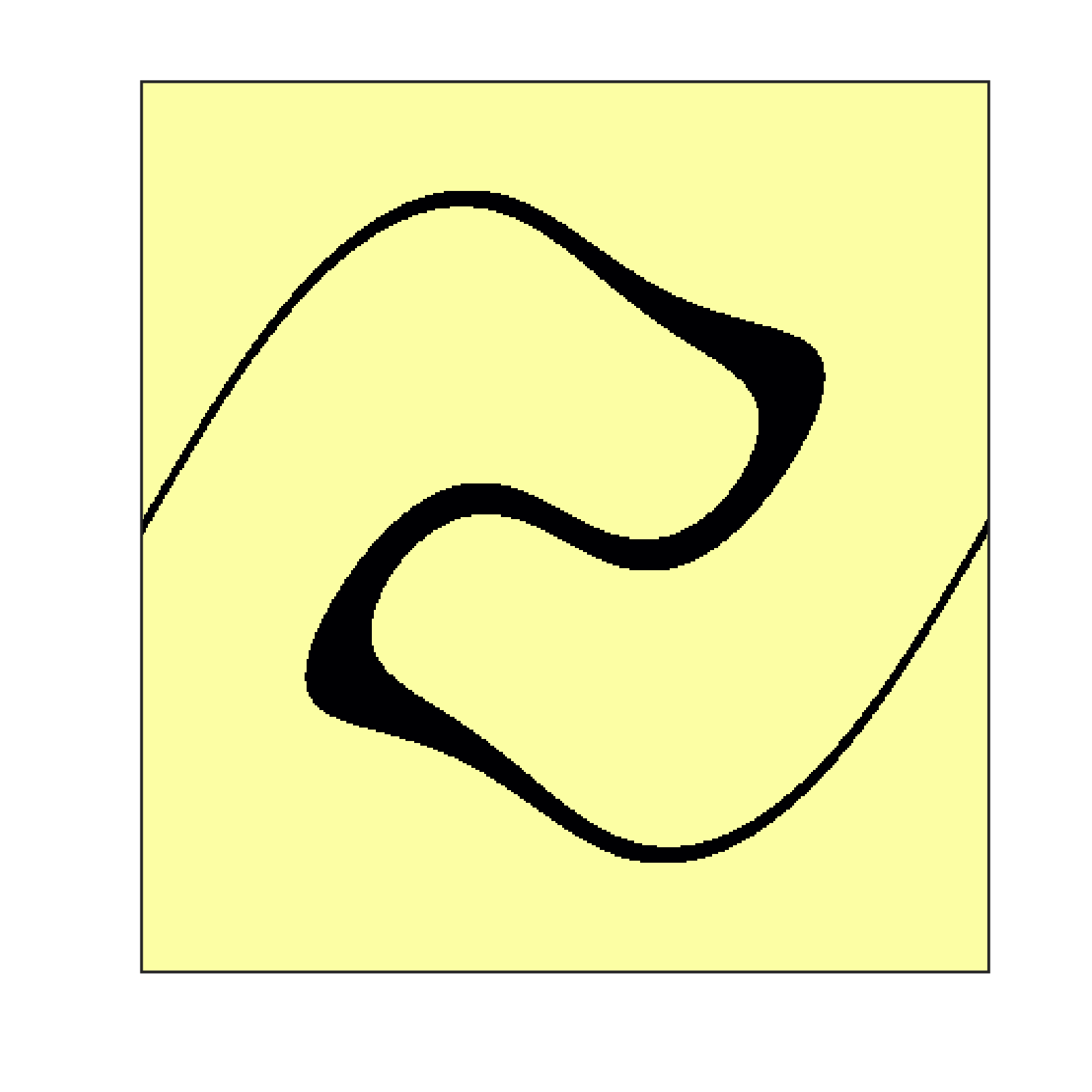}~
\includegraphics[height=0.38\columnwidth]{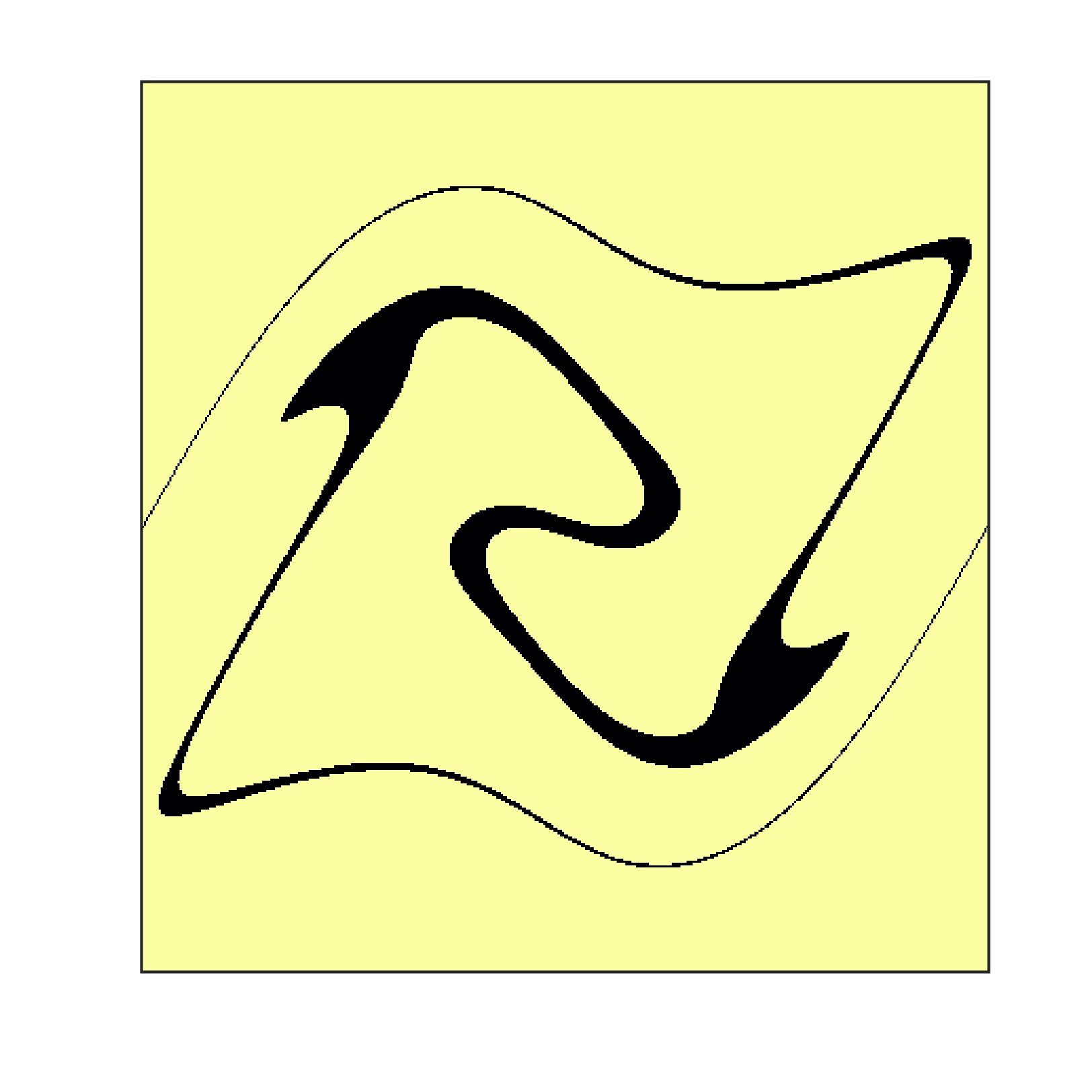}~
\includegraphics[height=0.38\columnwidth]{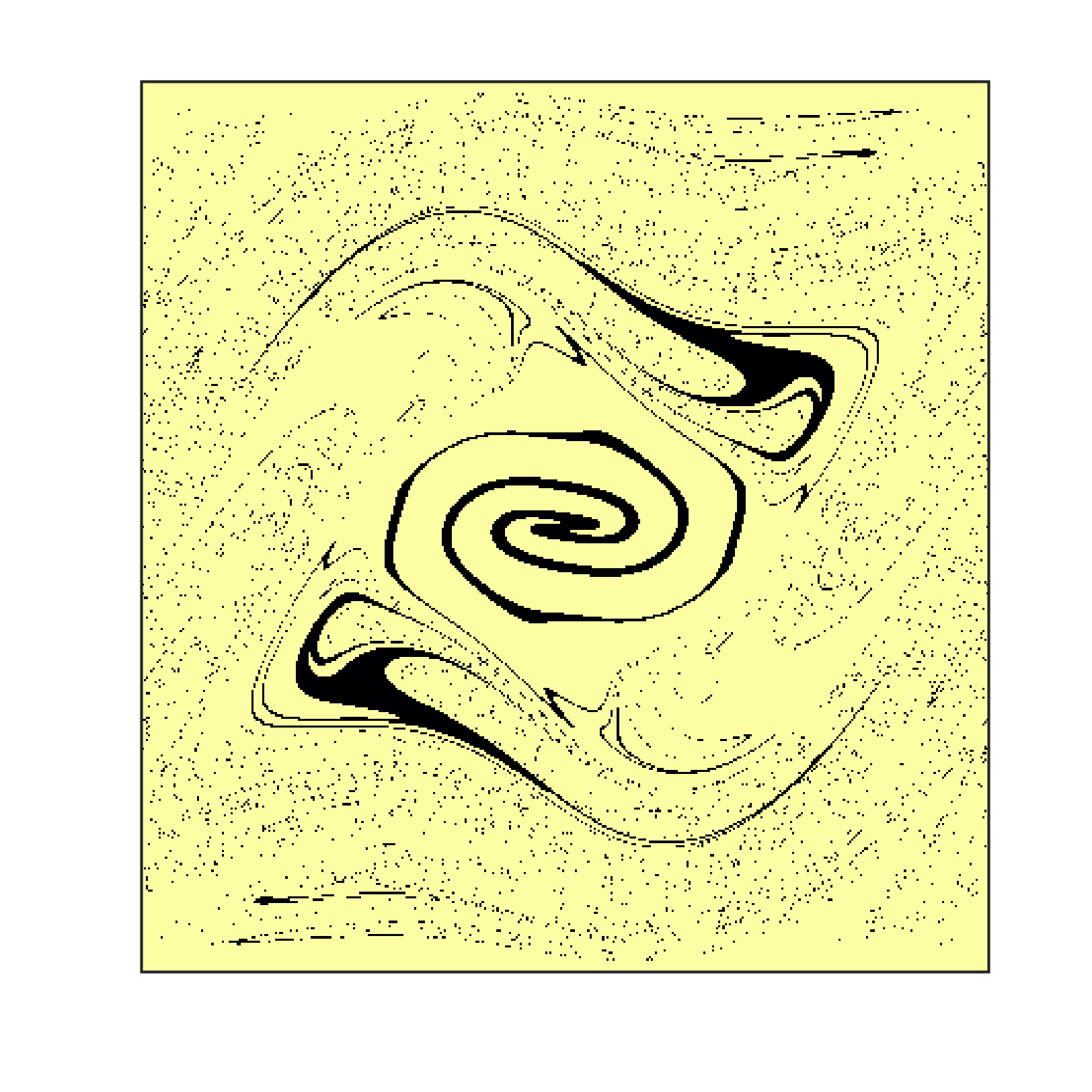}~
\includegraphics[height=0.38\columnwidth]{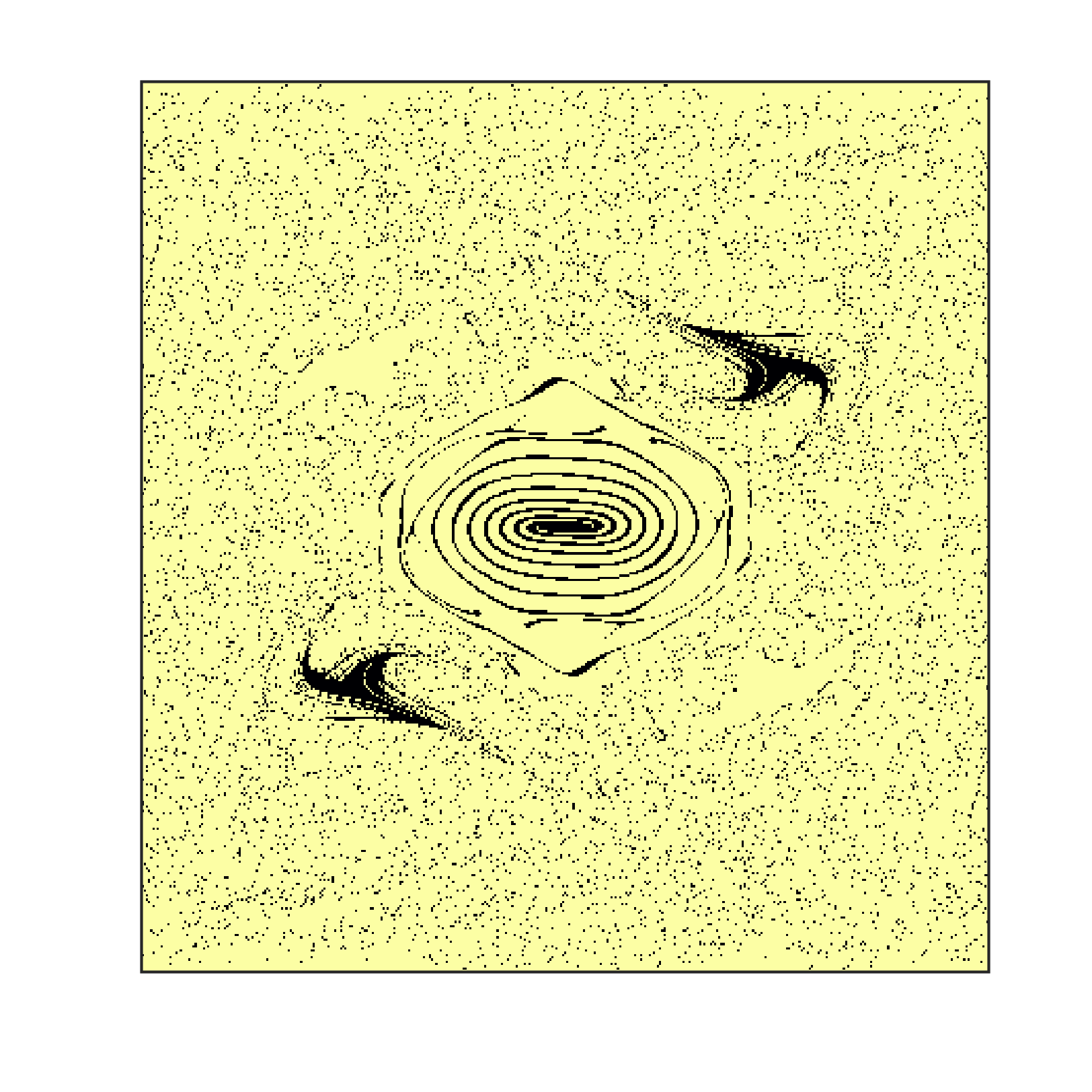}\\
\includegraphics[width=0.38\columnwidth]{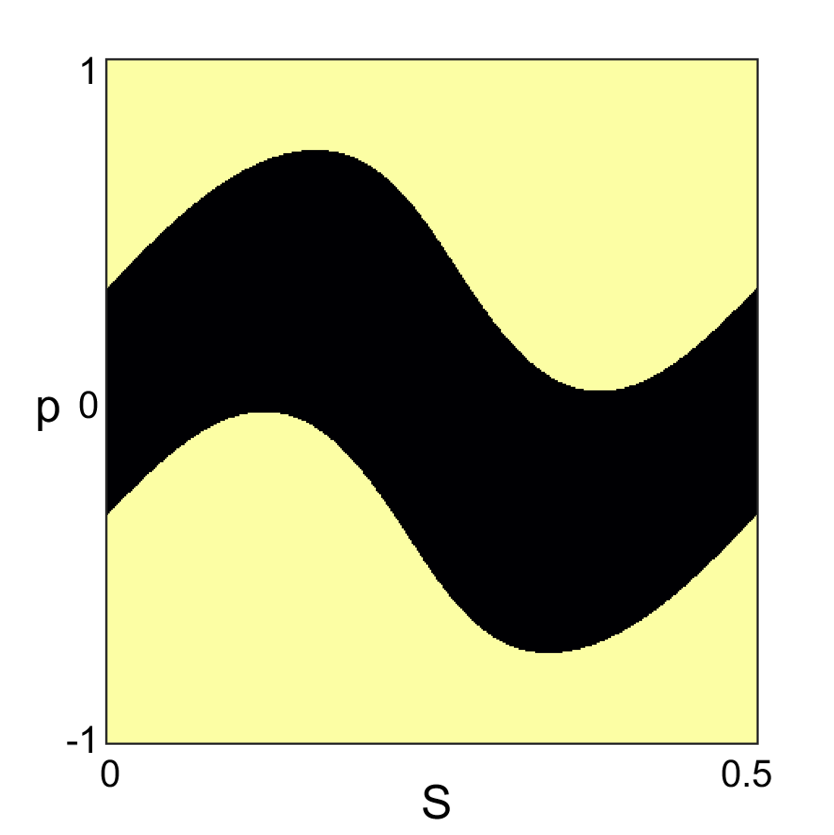}~
\includegraphics[width=0.38\columnwidth]{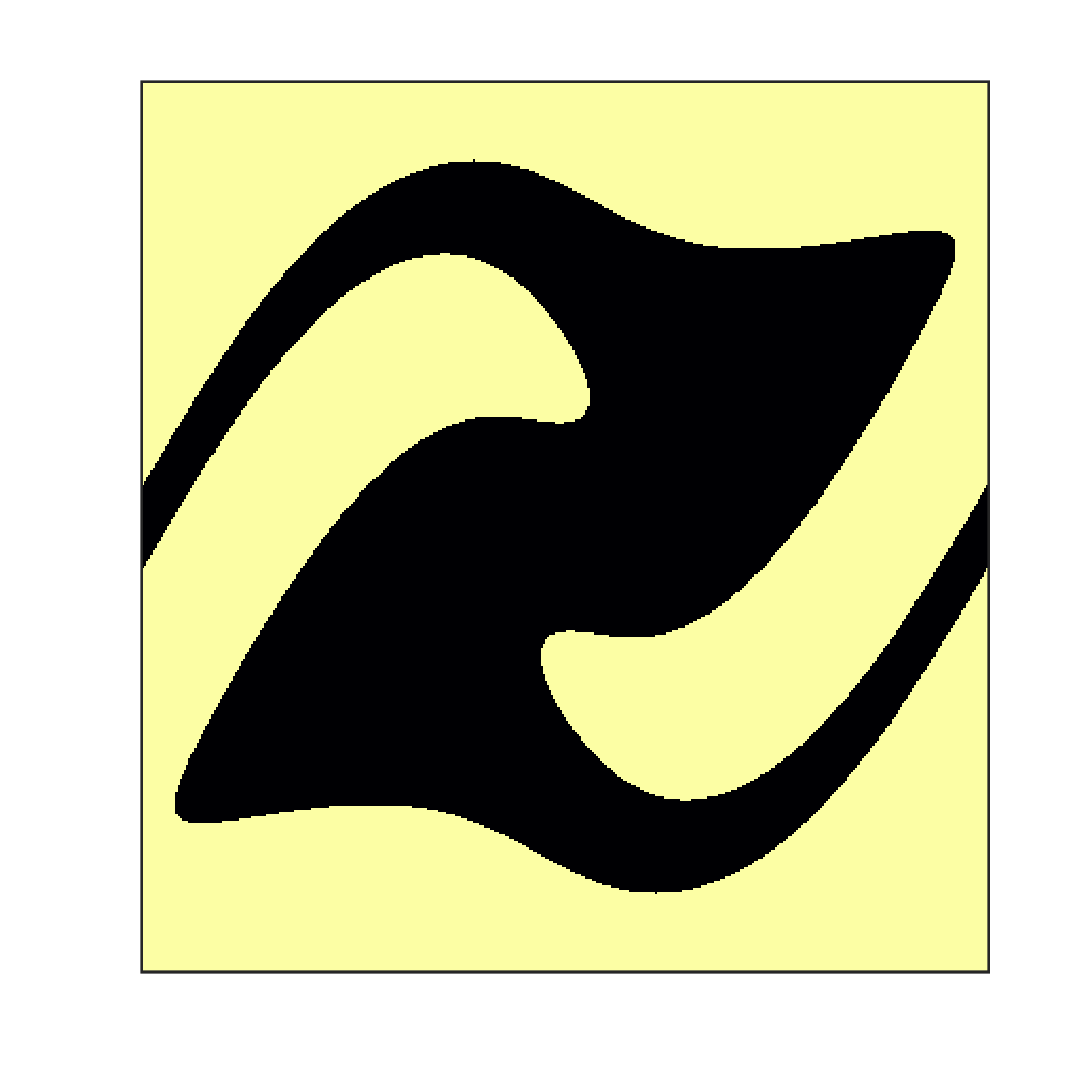}~
\includegraphics[width=0.38\columnwidth]{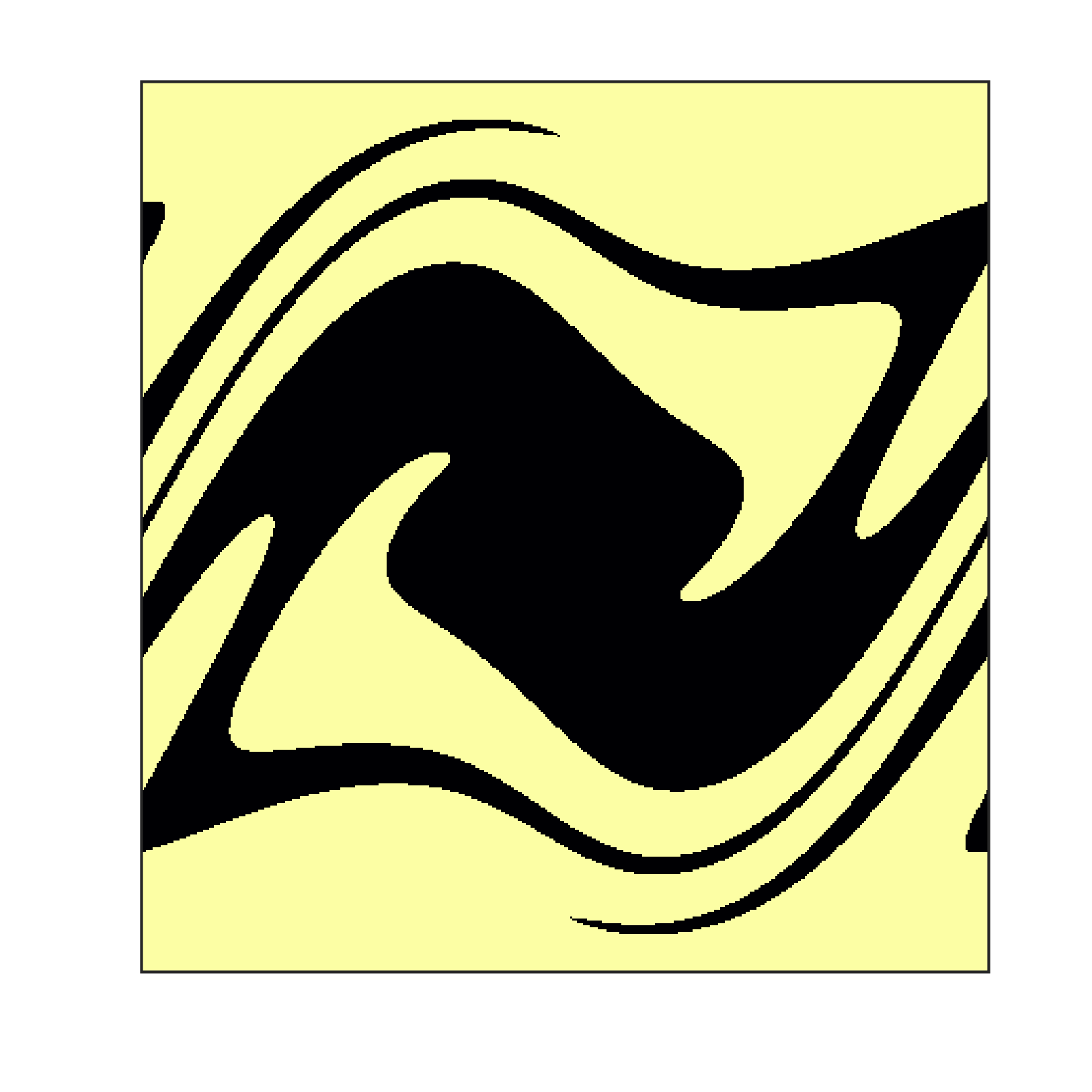}~
\includegraphics[width=0.38\columnwidth]{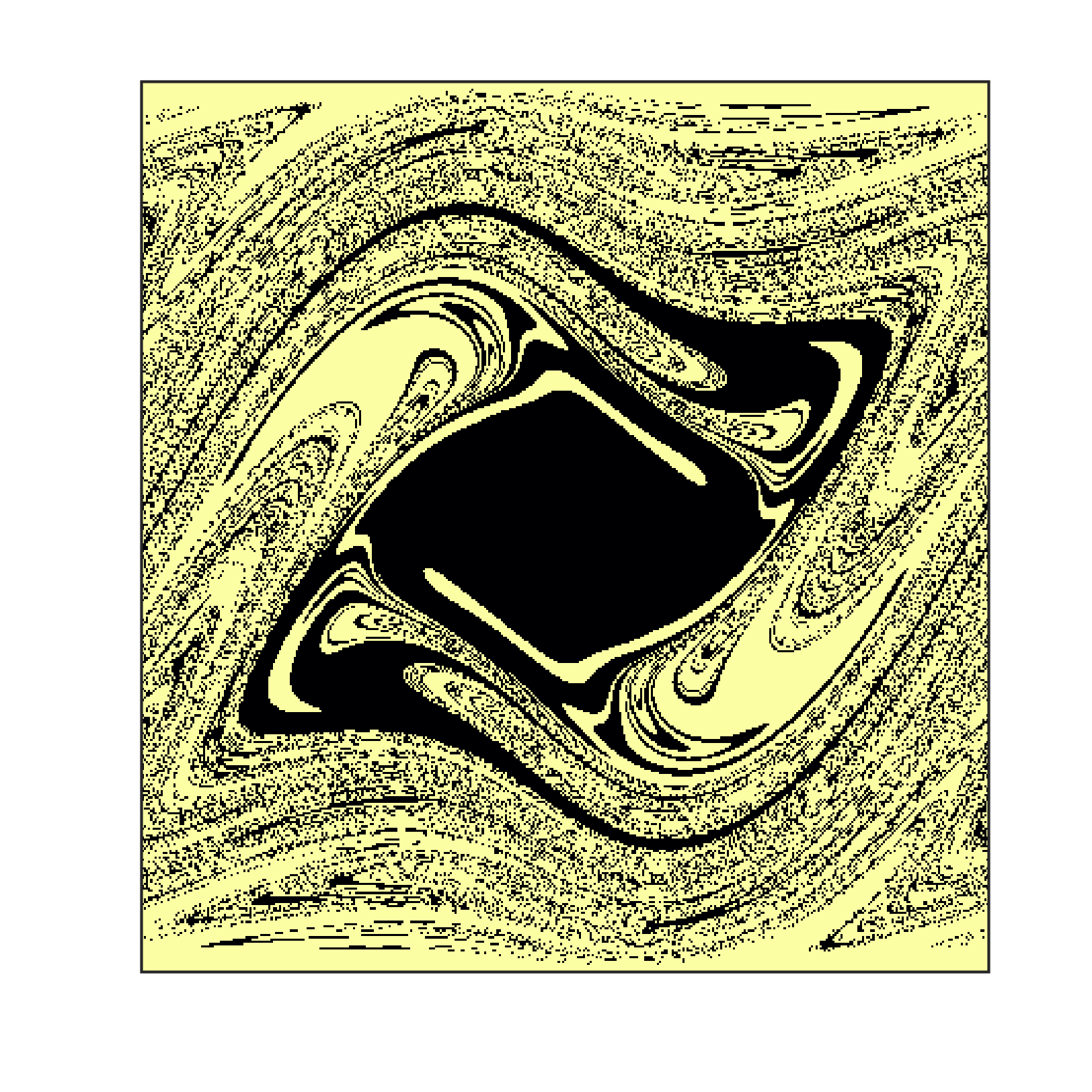}~
\includegraphics[width=0.38\columnwidth]{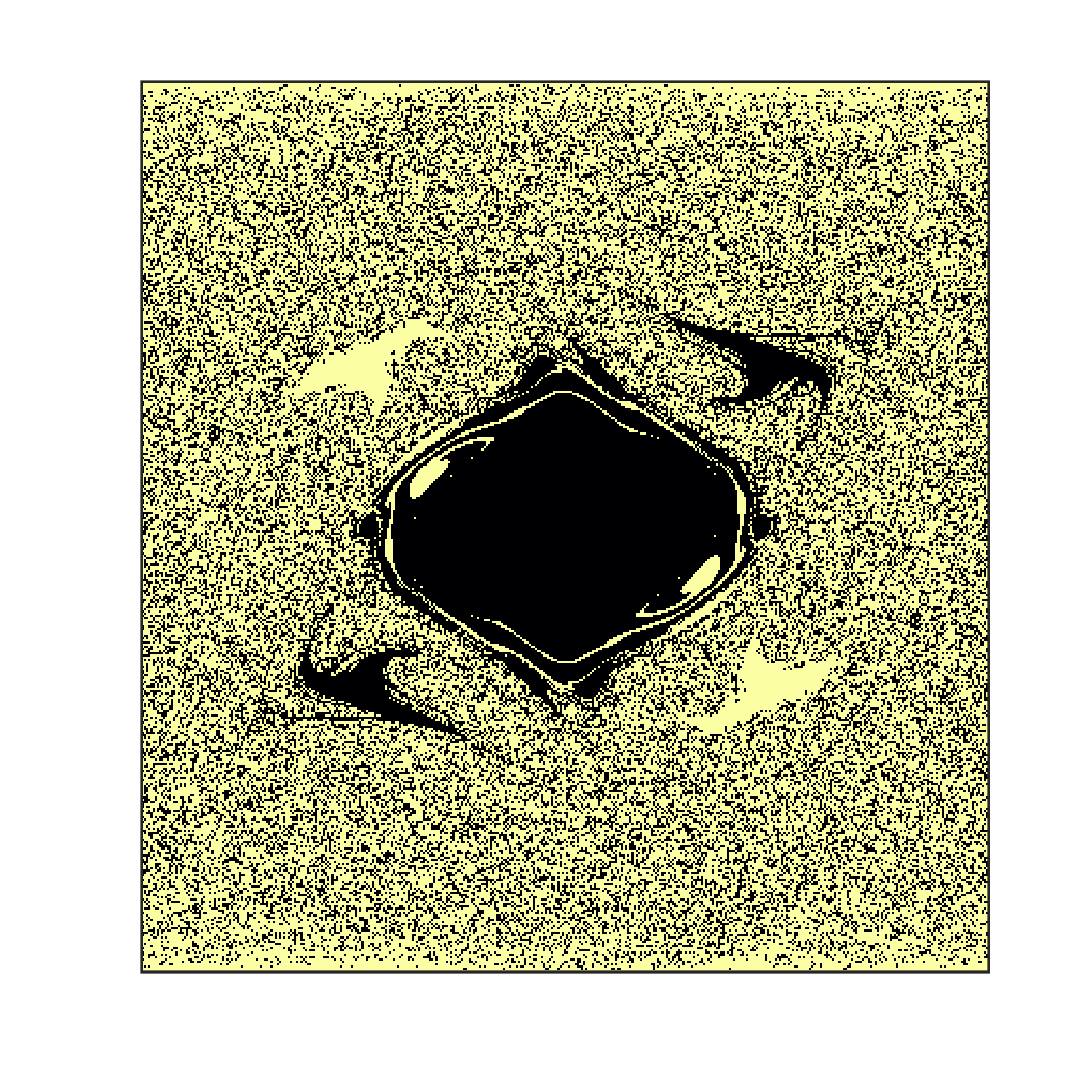}
\caption{The set of initial conditions for the oval that intersect the absorbing region at each iteration for $R=0.1$ (top) and $R=0.5$ (bottom) depicted over half of the phase space \(S\in[0,0.5]\). From left to right the iterations are $n=1,2,3,10,\rm{and}\ 30$.}
\label{diagram:oval_patch_rep_rp01}
\end{figure*}   

\begin{figure*}[hbt!]
\includegraphics[height=0.37\columnwidth]{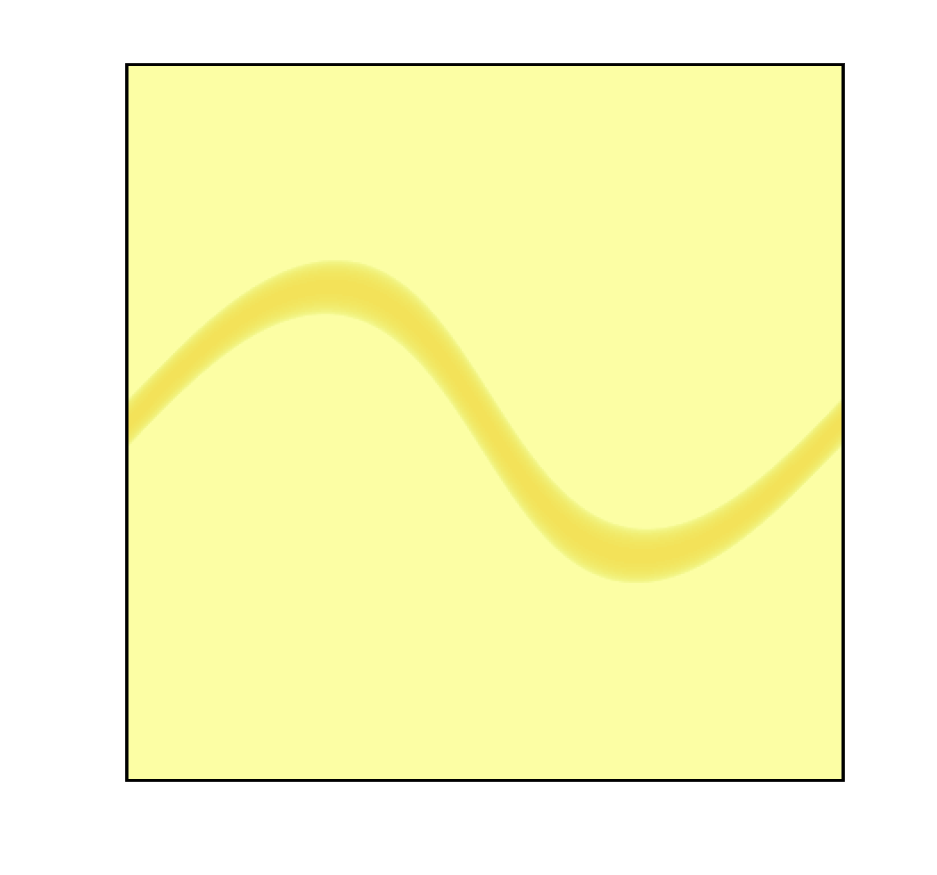}
\includegraphics[height=0.37\columnwidth]{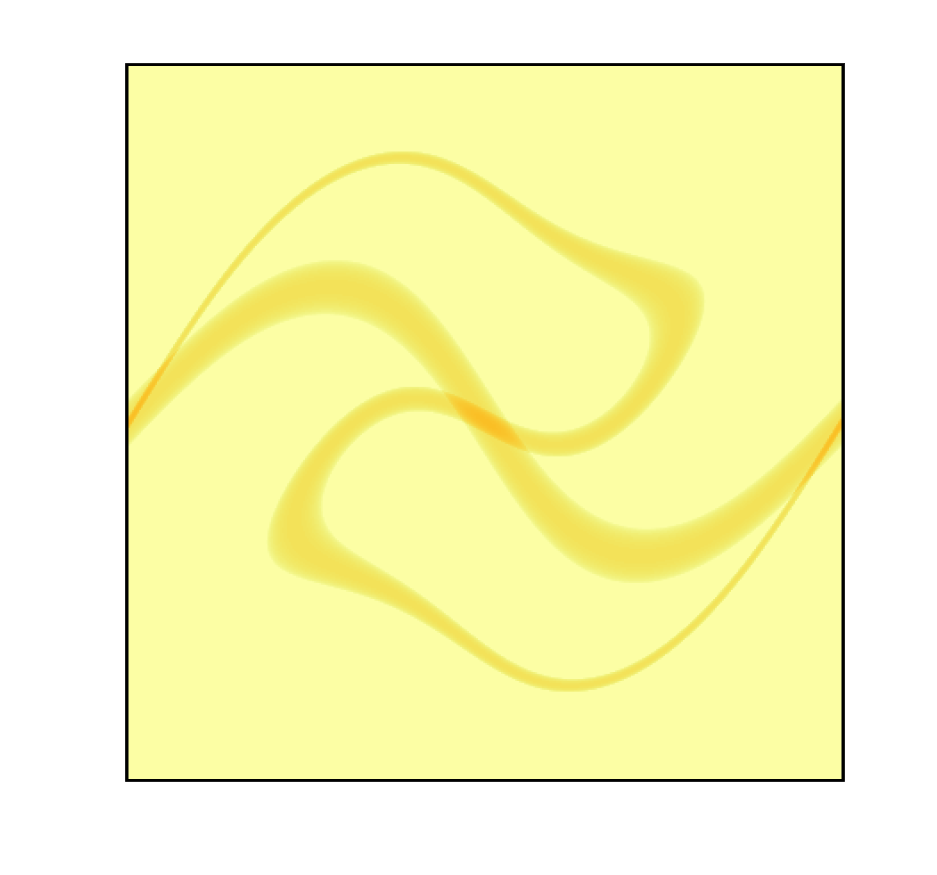}
\includegraphics[height=0.37\columnwidth]{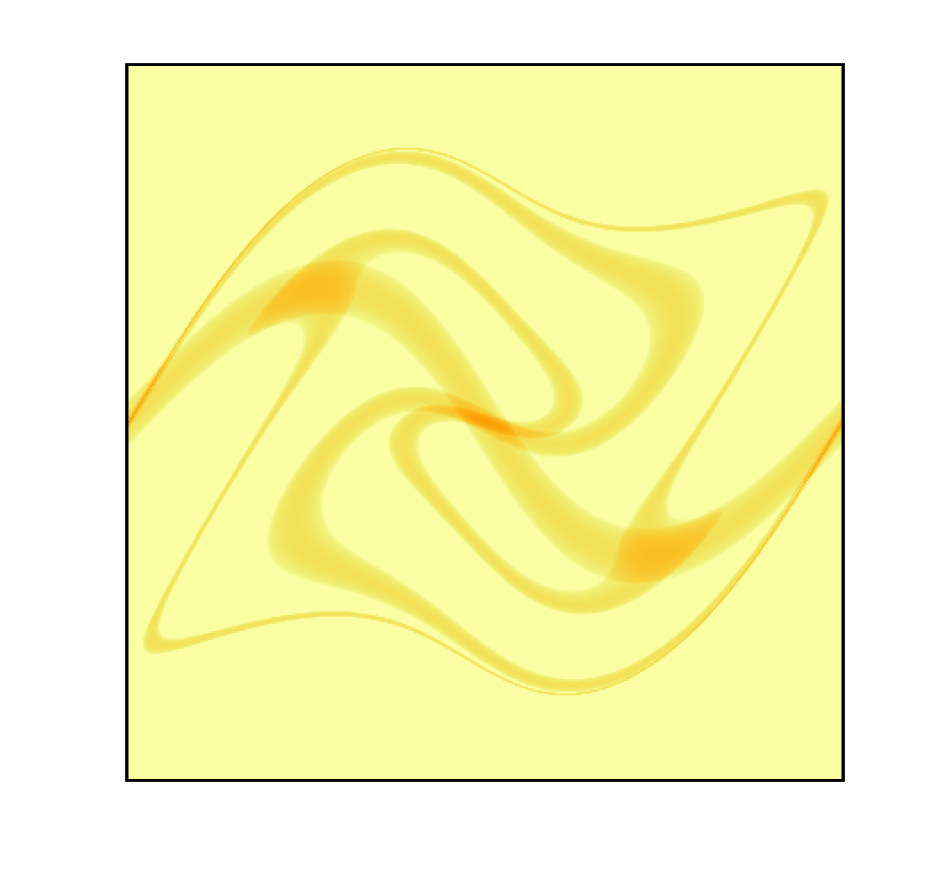}
\includegraphics[height=0.37\columnwidth]{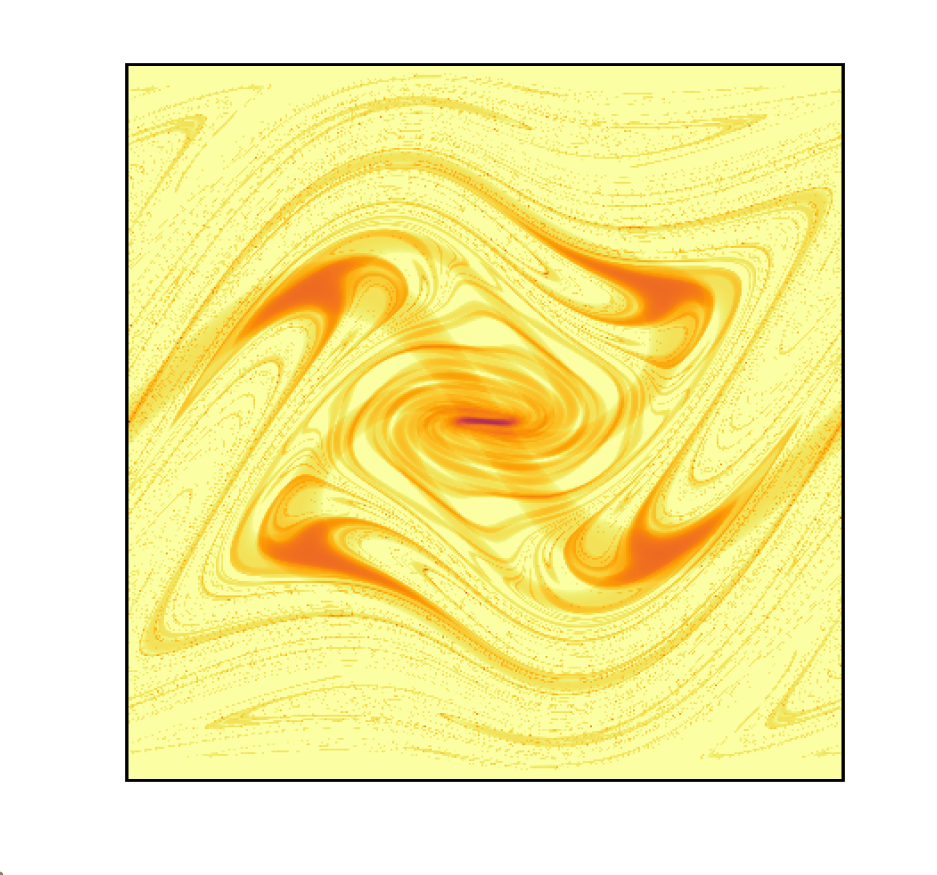}
\includegraphics[height=0.37\columnwidth]{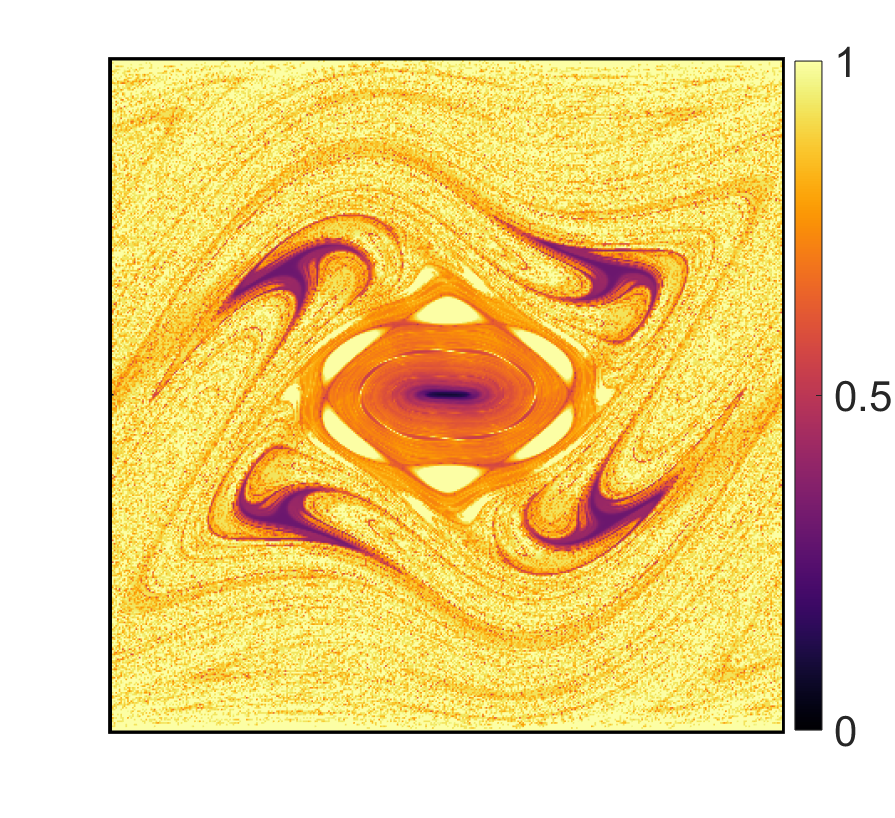}\\
\includegraphics[height=0.37\columnwidth]{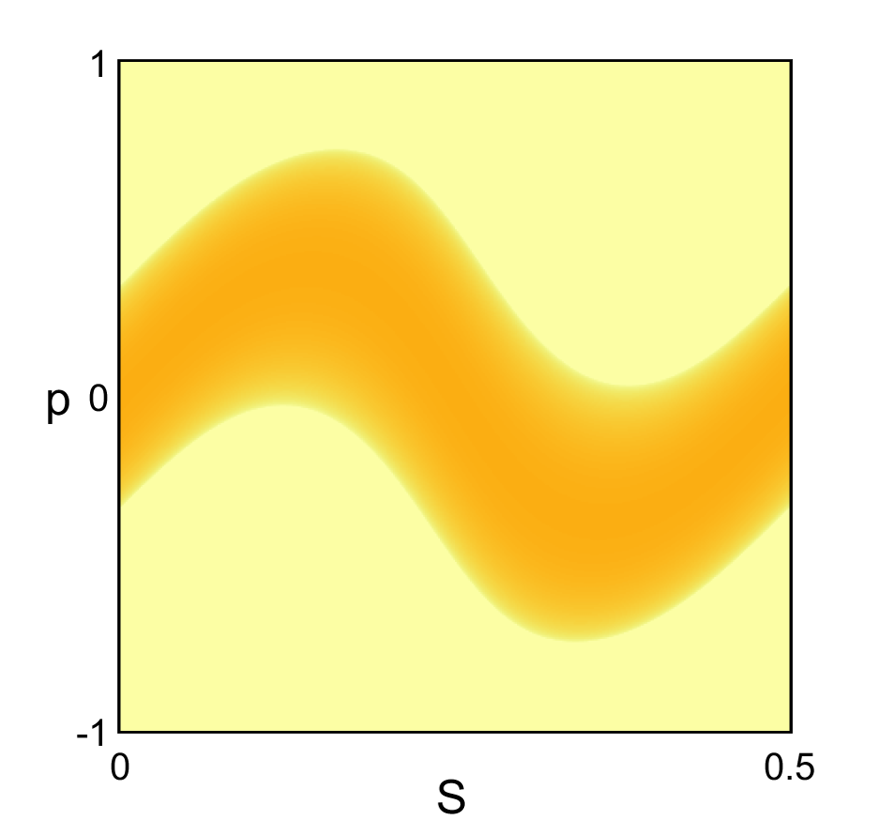}
\includegraphics[height=0.37\columnwidth]{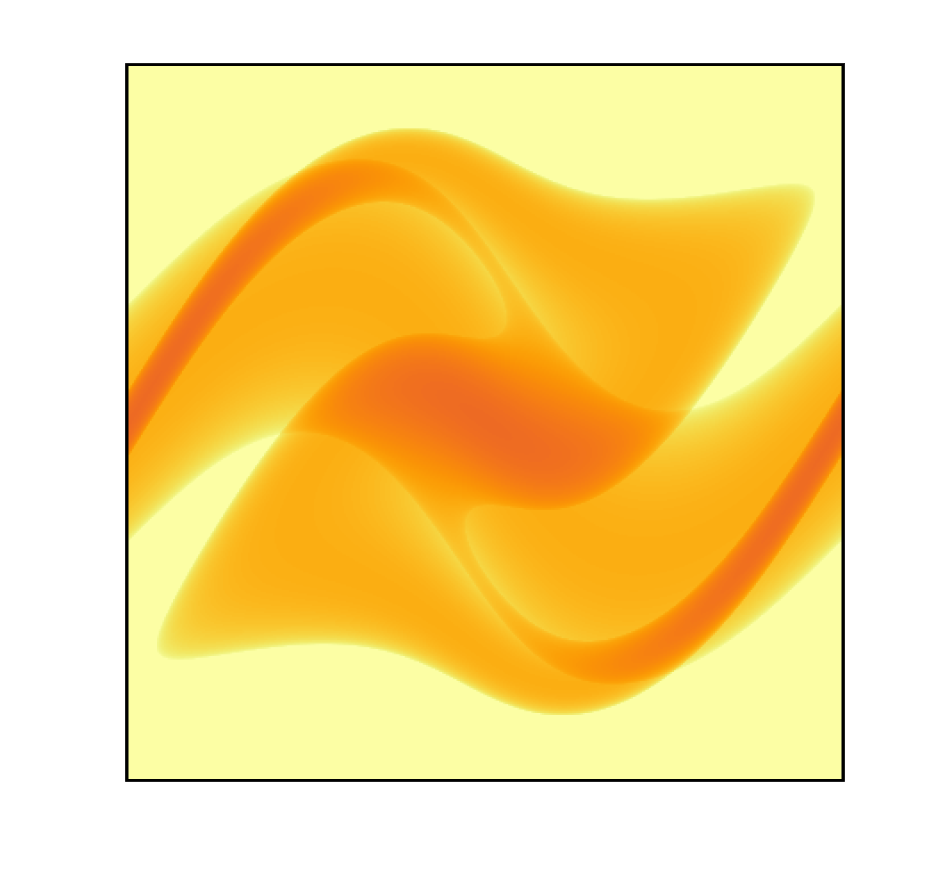}
\includegraphics[height=0.37\columnwidth]{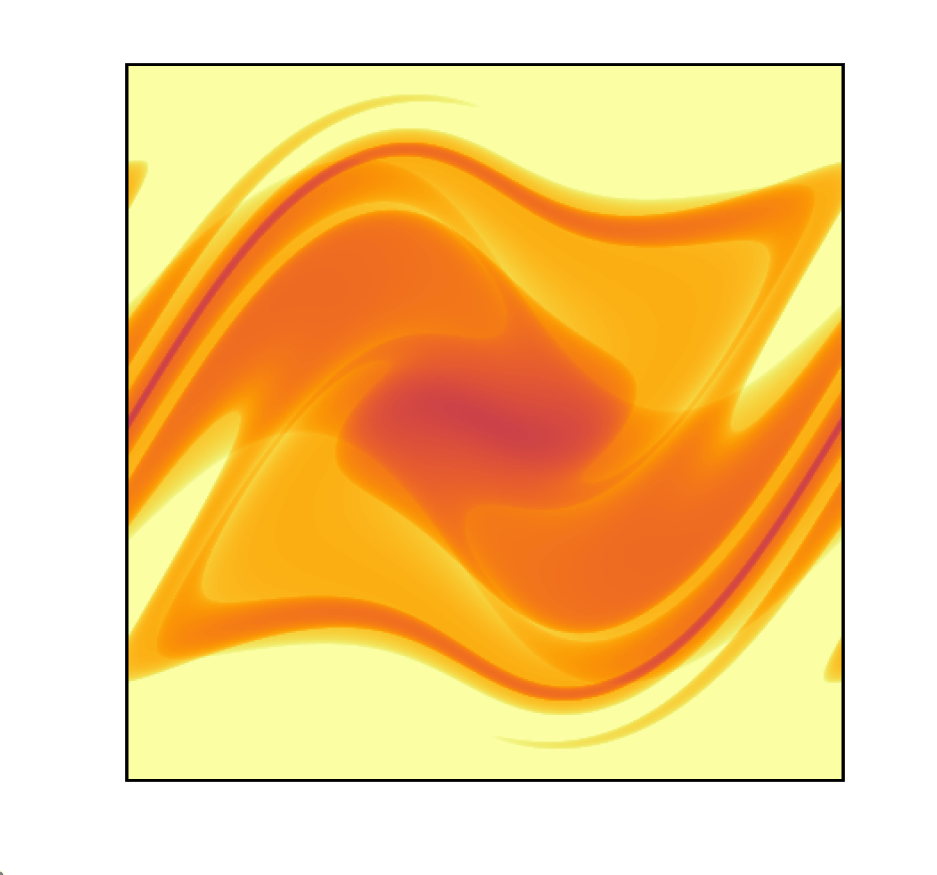}
\includegraphics[height=0.37\columnwidth]{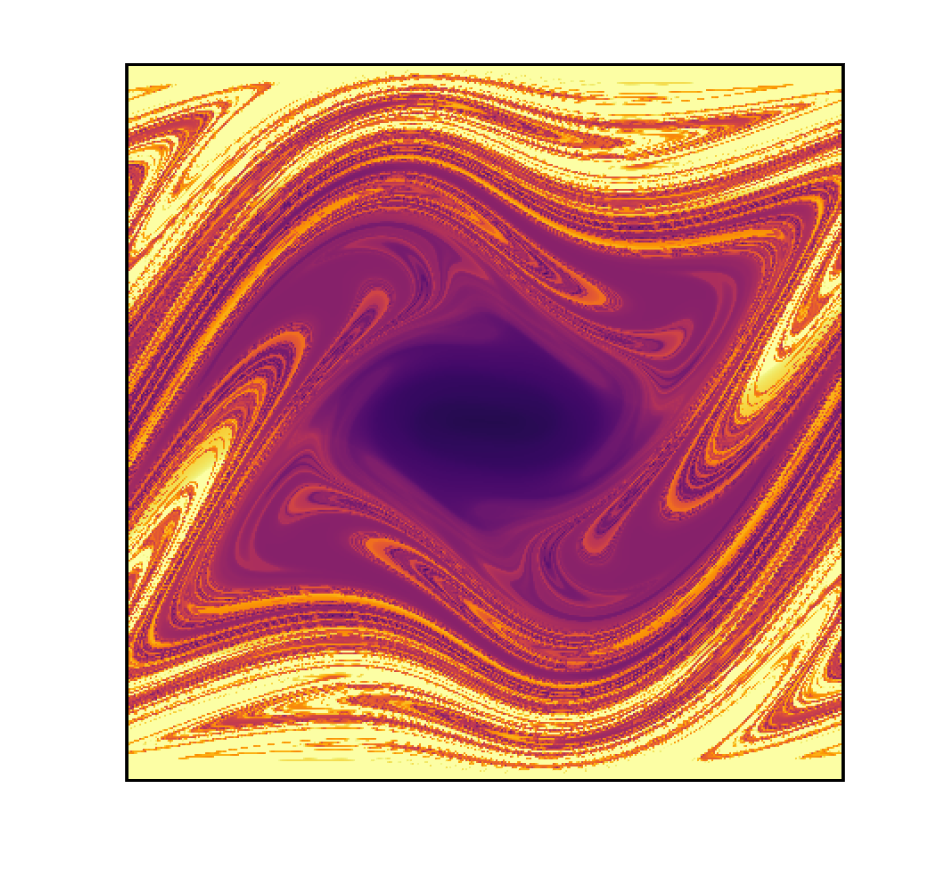}
\includegraphics[height=0.37\columnwidth]{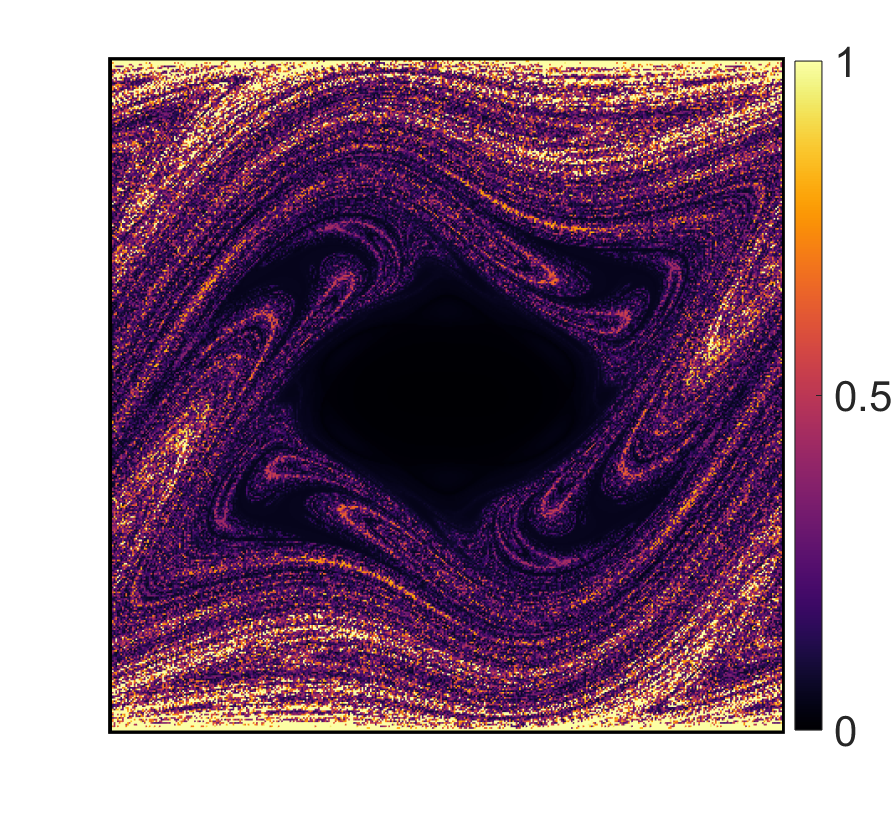}
\caption{Intensity landscape for the oval billiard with absorbing region with absorption rate $\gamma=0.4$ and radius $R=0.1$ (top), and absorption rate $\gamma=0.2$ and radius $R=0.5$ (bottom) for different numbers of iterations depicted over half of the phase space \(S\in[0,0.5]\). From left to right: $n_f=1,2,3,10,\rm{and}\ 30$.}
\label{diagram:oval_norm_evolution_gp1}
\end{figure*} 

Dynamical invariants e.g. stable and unstable manifolds and asymptotic curves of unstable periodic orbits, form the skeleton of the chaotic motion. These invariants play a prominent role in the local dynamics and are responsible for transient phenomenon, namely, stickyness in chaotic trajectories. While these sets are obscured within the long time Poincar\'e dynamics, they leave a clear mark on the intensity landscapes, via the sets $\mathcal{S}_n$, which we depict in figure \ref{diagram:oval_patch_rep_rp01} for $R=0.1$ (top) and $R=0.5$ (bottom) for various values of $n$. We remind ourselves that these sets are the backwards propagation of the set $\mathcal{S}_1$. For short and intermediate iteration numbers ($n\lesssim 5$), we clearly observe the filament structures arising from the repeated backwards iteration of the set $\mathcal{S}_1$, organised around the dynamical invariants embedded in the chaotic sea of the absorption-free dynamics. For longer times the sets $\mathcal{S}_n$ show a clear division between the regular island and the chaotic sea, where the filament structure outside the regular region slowly makes way for the natural measure of the chaotic sea, which eventually becomes uniform. The structures within the region of the regular island still display finer and finer regular whorl structures for the smaller radius $R=0.1$, while in the case of $R=0.5$ the whole region of the regular island is contained within all of the sets. Appendix \ref{appendix:offset_AR} contains additional examples for the oval billiard with a displaced absorbing region.

The union of the sets $\mathcal{S}_n$ provides the skeleton of the intensity landscapes, which become highly non-trivial for finite values of the loss parameter $\gamma$. In Figure \ref{diagram:oval_norm_evolution_gp1} we depict examples for $\gamma=0.4$ for $R=0.1$ in the top row and $\gamma=0.2$ for $R=0.5$ in the bottom row, for the same iteration numbers as in Figure \ref{diagram:oval_patch_rep_rp01}. We observe that over the course of its evolution the intensity landscapes develops intricate structures reflecting the fractal character of the corresponding absorption-free dynamics. The loss of intensity is largest in phase-space points corresponding to trajectories that intersect the absorbing region in every or most iterations, i.e., on regions in phase space that lie on the intersection of many of the $\mathcal{S}_n$, such as trajectories within close vicinity of the bouncing ball trajectory in the centre of the regular island, for both radii. Depending on the radius of the absorbing region there may or may not be a two-dimensional subset of trajectories that never encounter the absorbing region. For the smaller radius $R=0.1$ we observe that a prominent chain of islands around the elliptic islands avoids the absorbing region, and forms a structure of stable islands in the intensity landscape. In the case of the larger radius, on the other hand, the trajectories associated to the same chain of islands intersect the absorbing region in every iteration, as do all other surrounding trajectories, and the corresponding phase-space region appears as an extended area of minimal intensity. Amongst the chaotic trajectories the vast majority will eventually intersect the absorbing region through their natural ergodic coverage of the chaotic sea. The set of initial points that belong to trajectories that do not intersect the absorbing region up to a given number of iterations develops a fractal structure with large number of iterations, as can be seen for example for $n_f=30$ in Figure \ref{diagram:oval_norm_evolution_gp1}. We also recognise the prominent structure of low intensities associated to the sticky region around the bow-tie trajectories, which intersect the absorbing region in every other iteration.

\section{Summary and outlook}\label{sec-sumout}

We have introduced billiard dynamics with absorption in a central circular region, using an intensity variable that we monitor throughout the time evolution. Within the absorbing region, the loss of intensity is homogeneous and thus the billiard dynamics are unchanged from those of the absorption-free system. Due to the finite size of the absorbing region, the intensity profiles of the trajectories are augmented with a true-time component derived from the time each trajectory spends in the absorbing region. We have studied the resulting intensity landscapes, where the intensity resulting from the iteration of an initial point in phase space after a fixed number of iterations is depicted as false colour on phase space, in regular and chaotic billiard geometries. We have demonstrated how many of their features can be understood through the dynamics of the absorption-free system. The examples considered here are by no means exhaustive and the investigation of different geometries and placements of the absorbing region (that we briefly touch upon in Appendix \ref{appendix:offset_AR}), as well as absorbing regions that change with time, is a task for future work. We expect that the intensity landscapes hold crucial information for the behaviour of corresponding quantum systems, which will be the topic of a separate forthcoming study. 

\begin{figure*}[htb!]
\includegraphics[width=0.38\columnwidth]{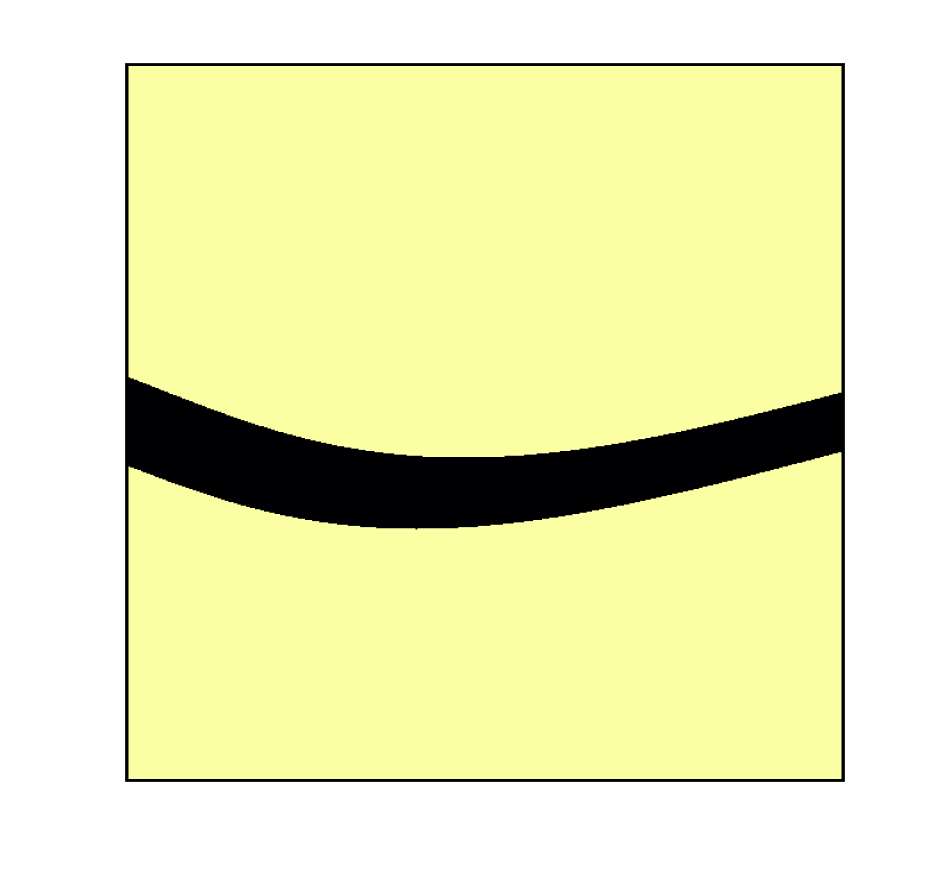}
\includegraphics[width=0.38\columnwidth]{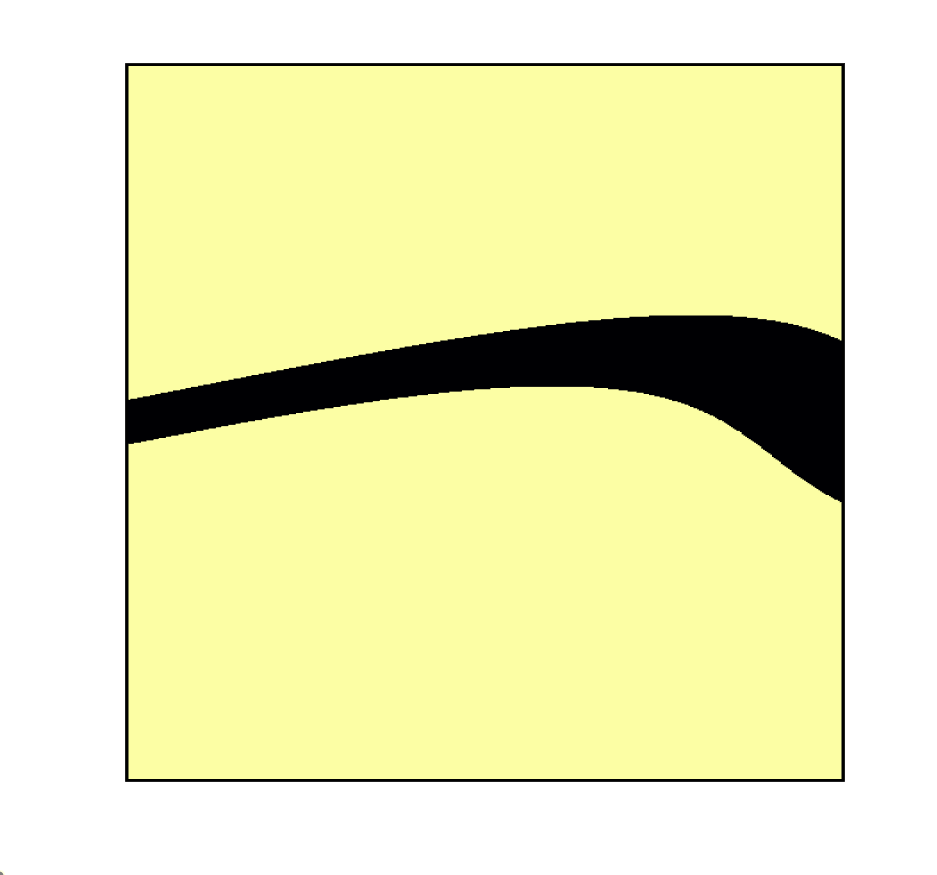}
\includegraphics[width=0.38\columnwidth]{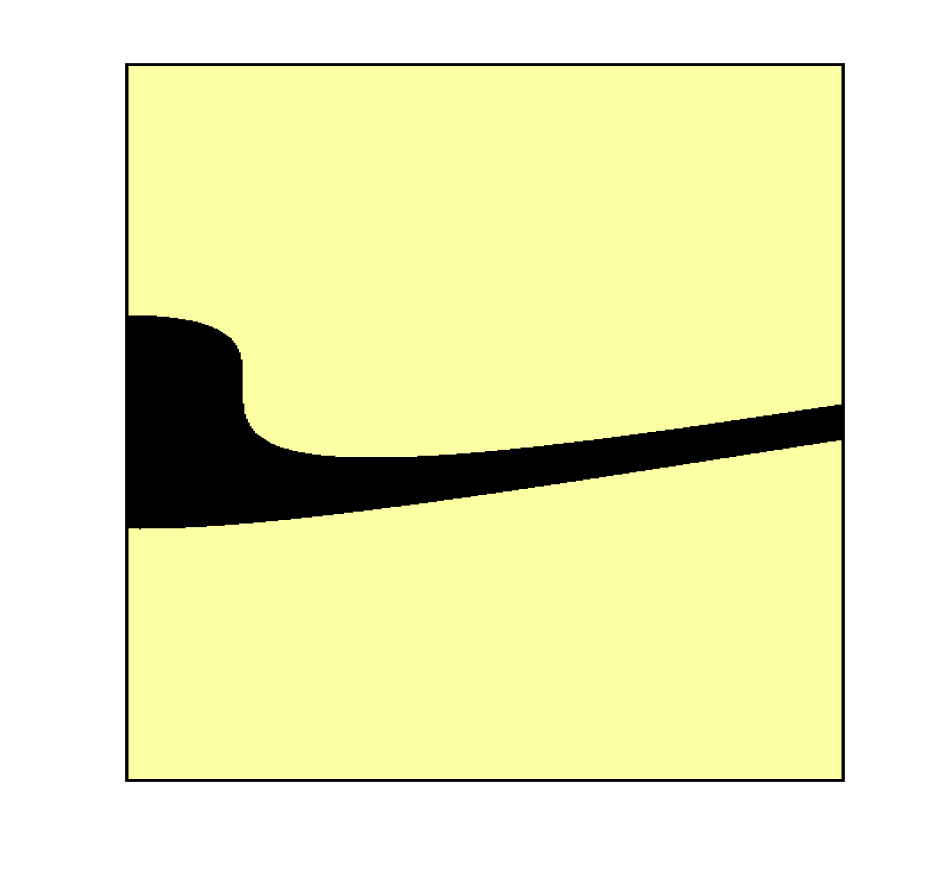}
\includegraphics[width=0.38\columnwidth]{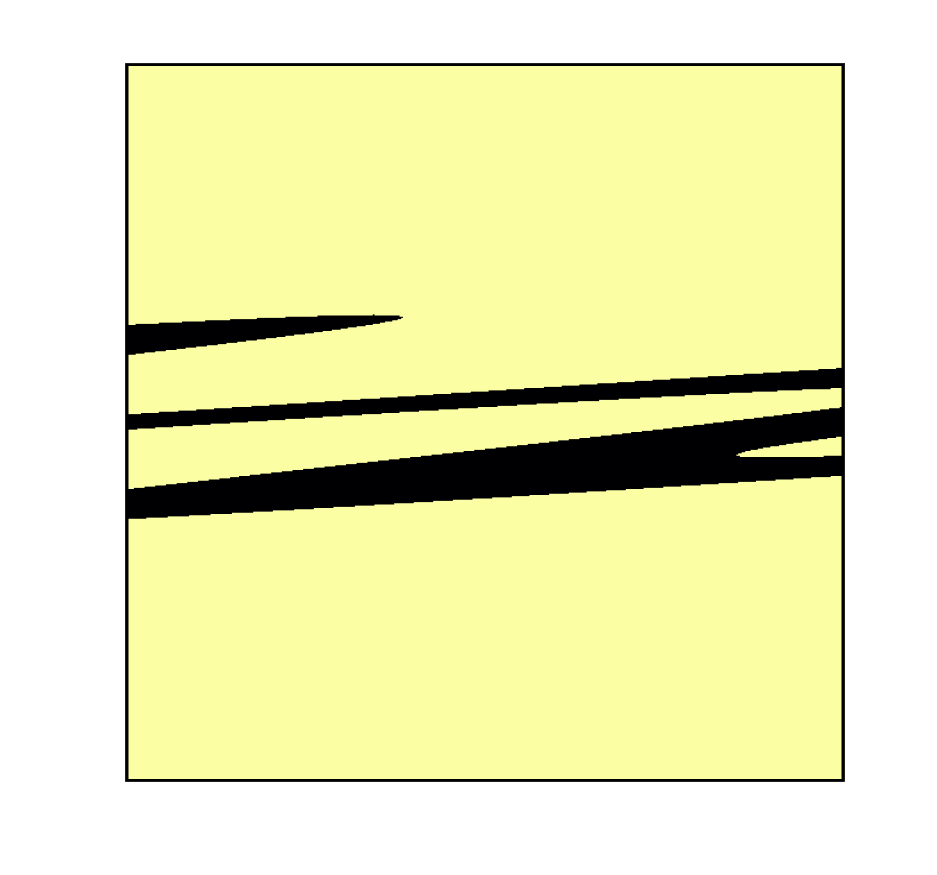}
\includegraphics[width=0.38\columnwidth]{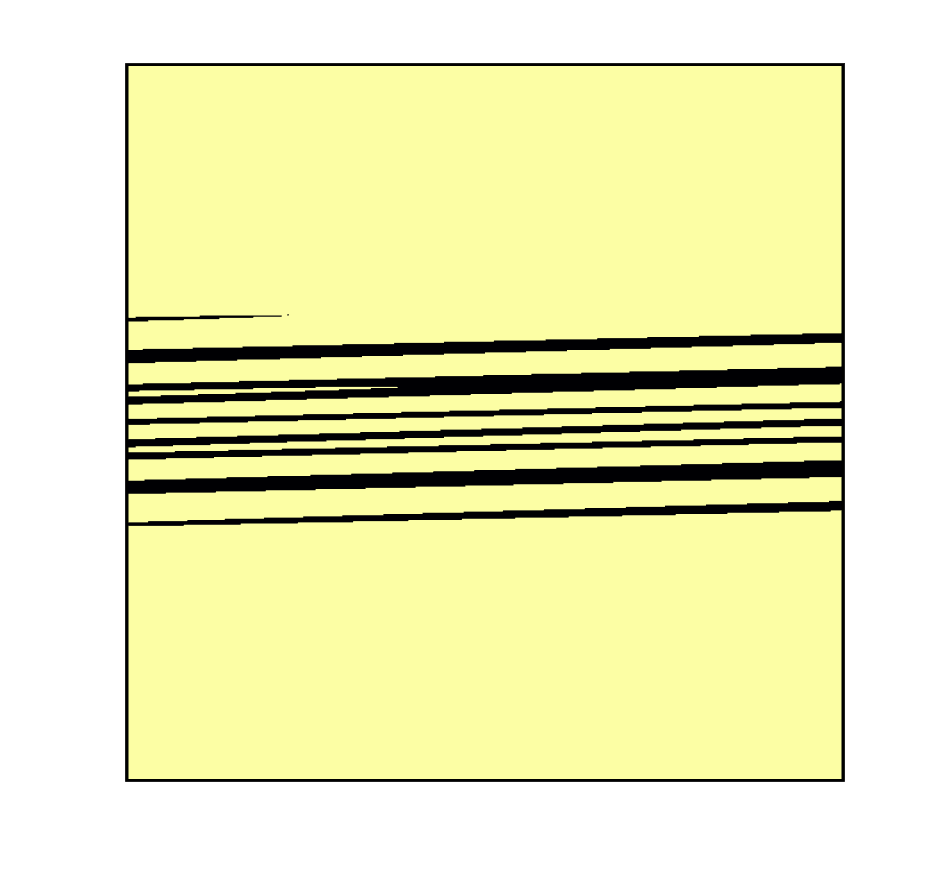}\\
\includegraphics[width=0.38\columnwidth]{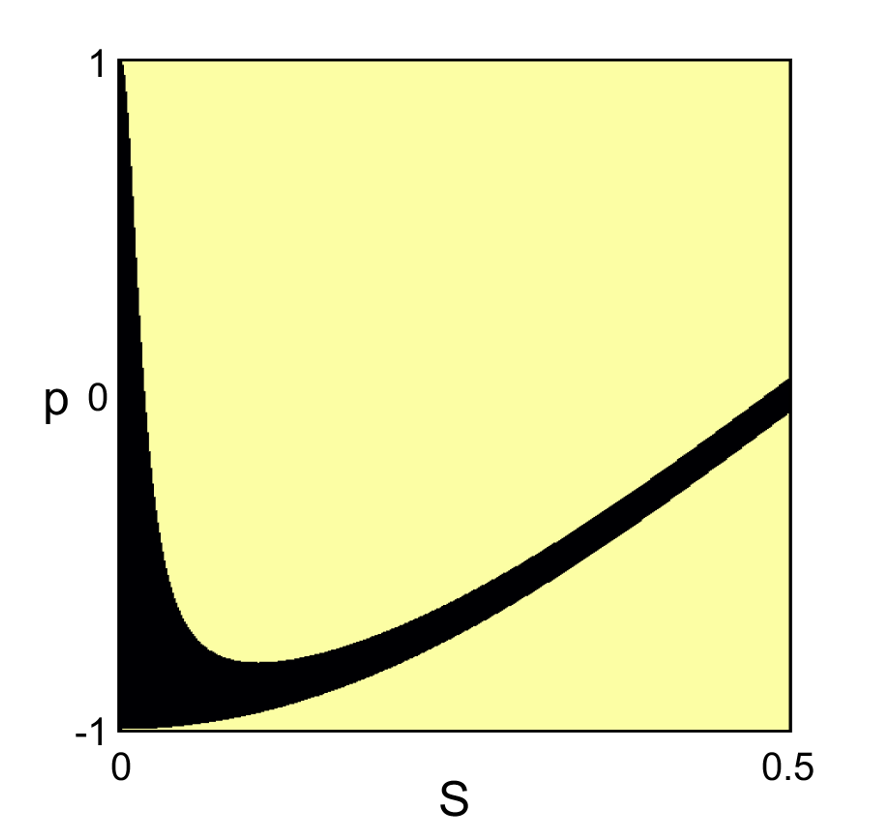}
\includegraphics[width=0.38\columnwidth]{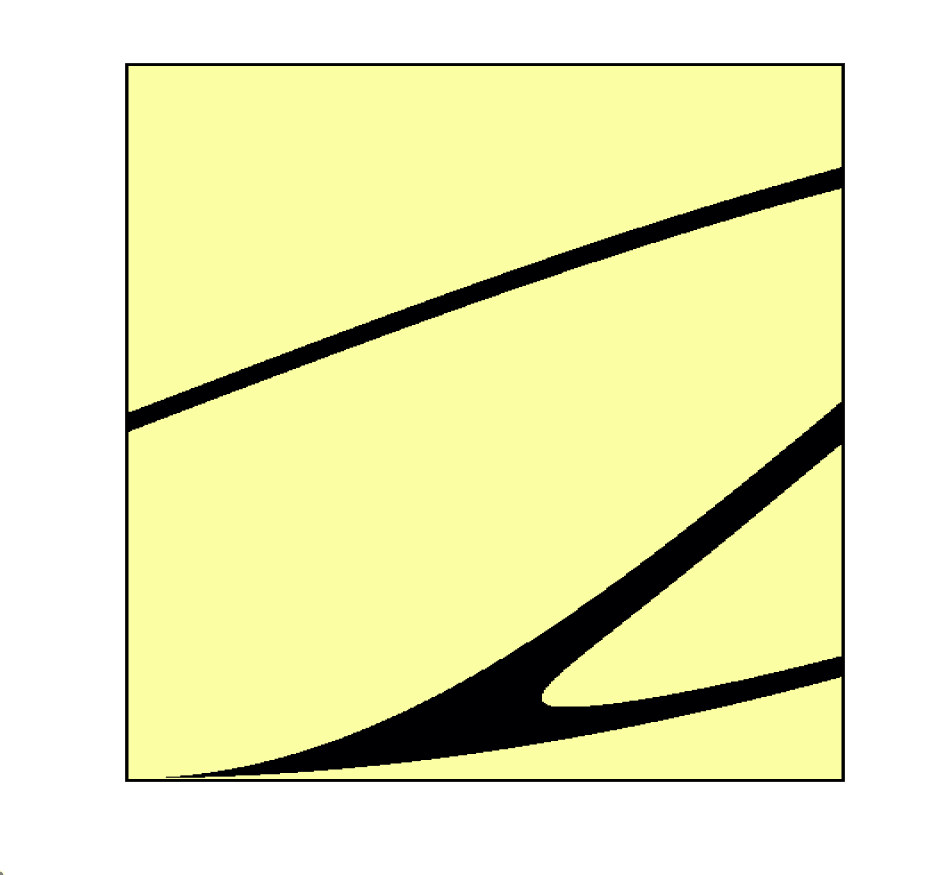}
\includegraphics[width=0.38\columnwidth]{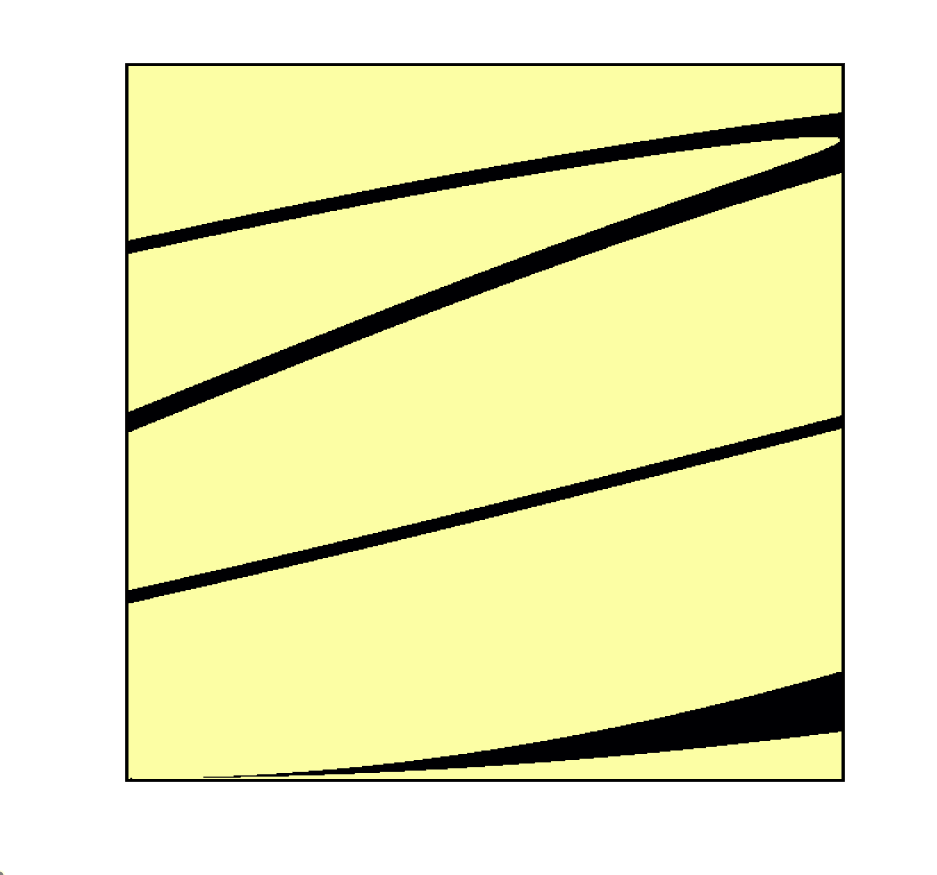}
\includegraphics[width=0.38\columnwidth]{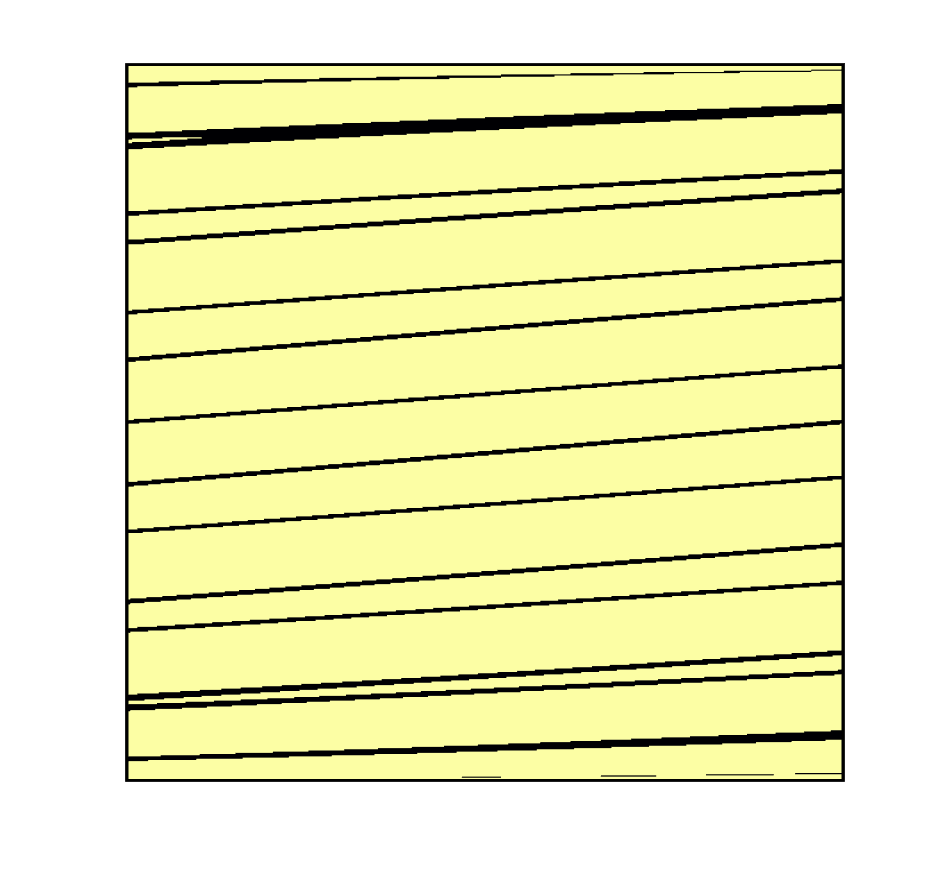}
\includegraphics[width=0.38\columnwidth]{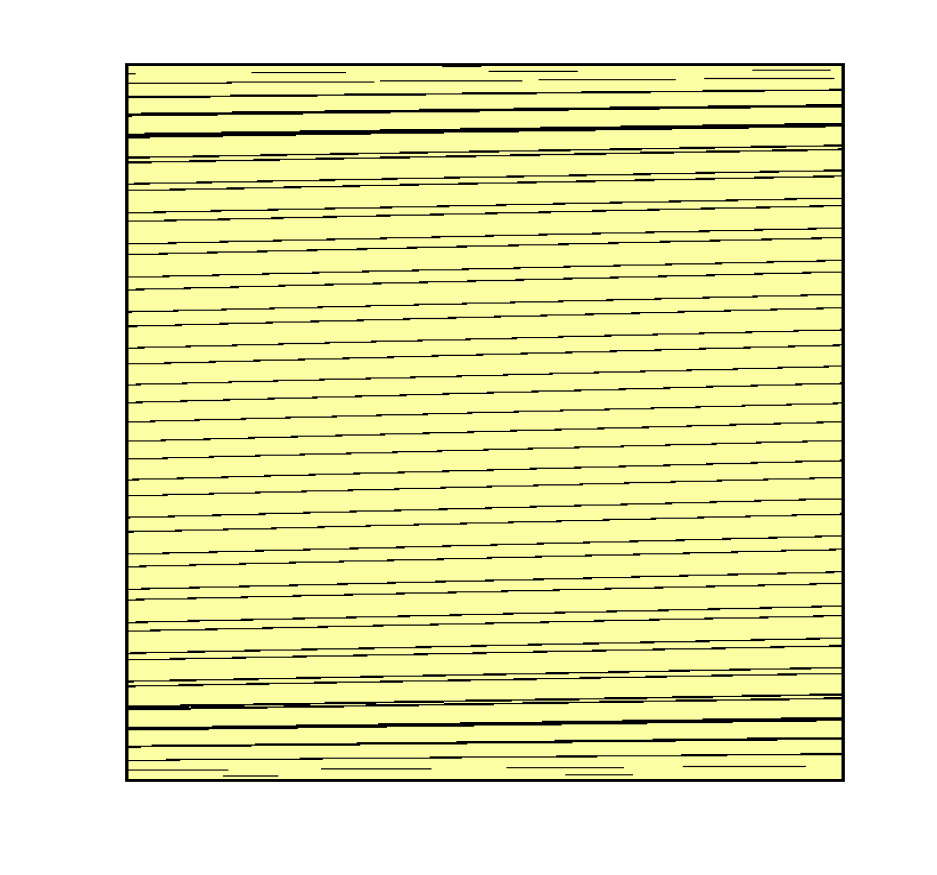}
\caption{The set of initial conditions for the circle that intersects the absorbing region at each iteration for radius $R=0.1$ centred at values $d=0.2$ (top) and $d=0.9$ (bottom) to the right of the billiard centre, depicted over half of the phase space \(S\in[0,0.5]\). From left to right the iterations are $n=1,2,3,10$ and $30$.}
\label{diagram:offset_circle_Sn}
\end{figure*}

\begin{figure*}[htb!]
\includegraphics[width=0.38\columnwidth]{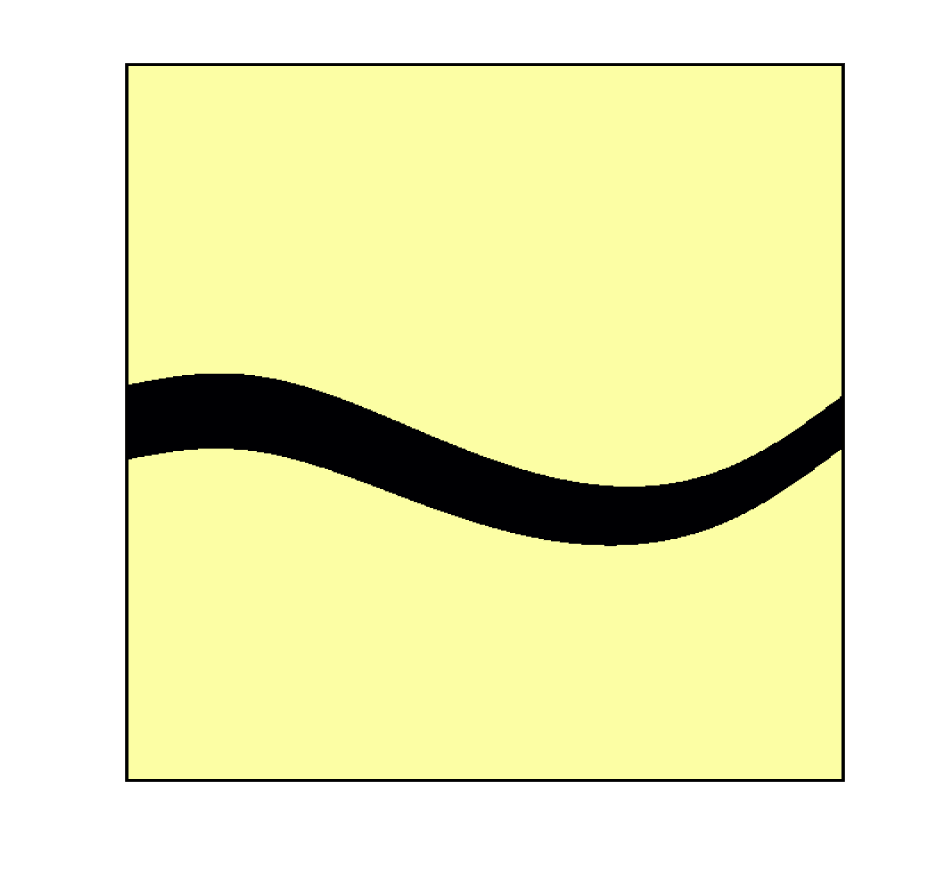}
\includegraphics[width=0.38\columnwidth]{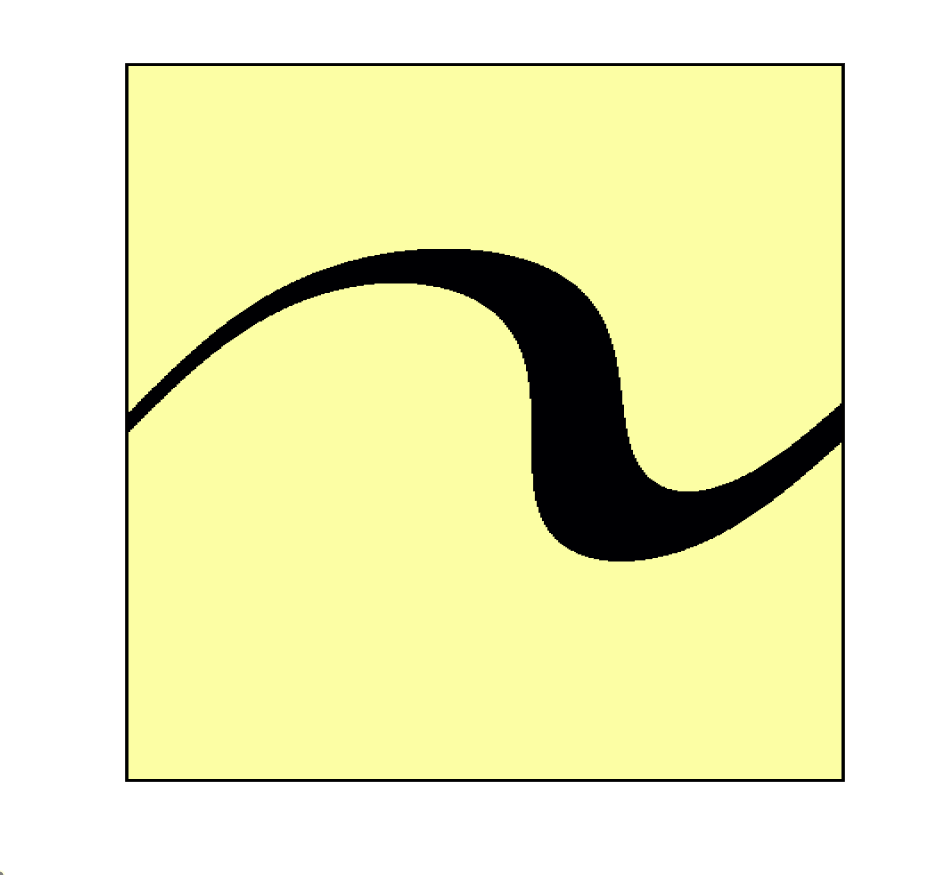}
\includegraphics[width=0.38\columnwidth]{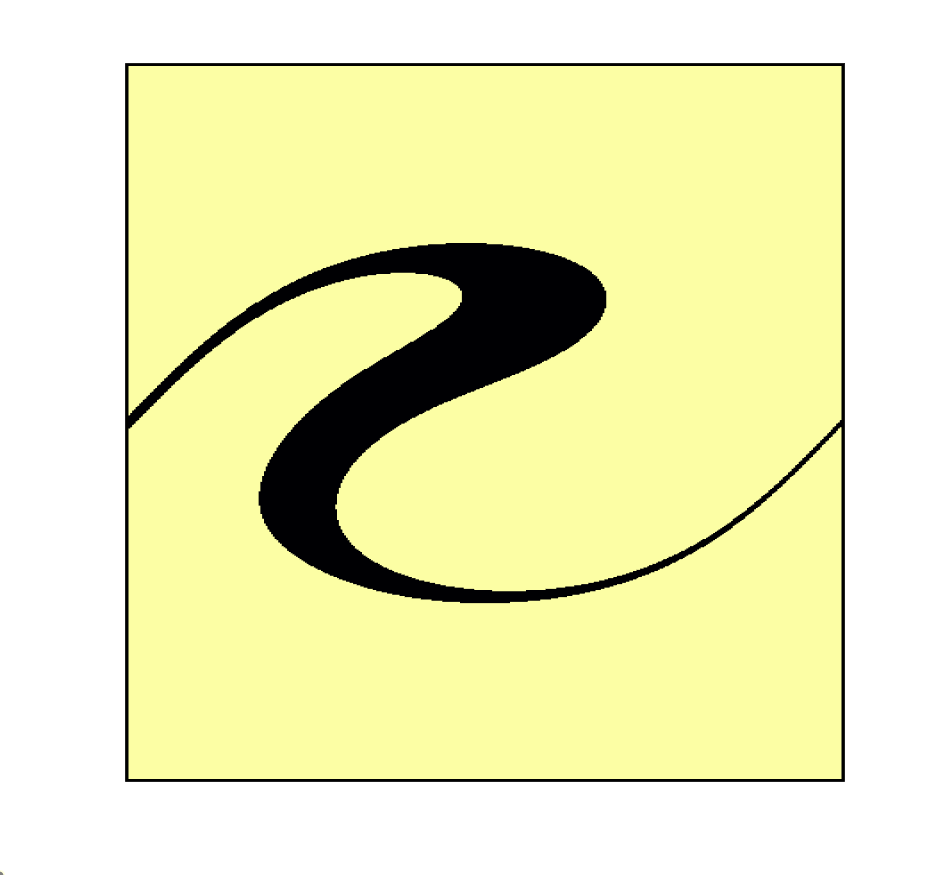}
\includegraphics[width=0.38\columnwidth]{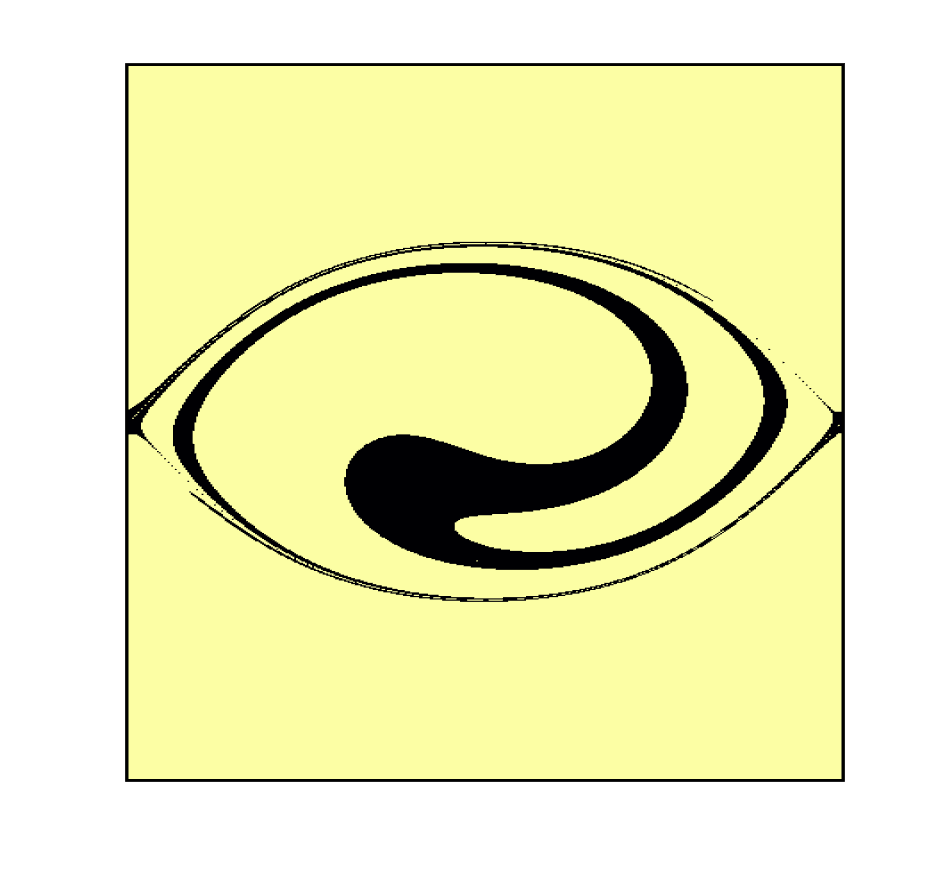}
\includegraphics[width=0.38\columnwidth]{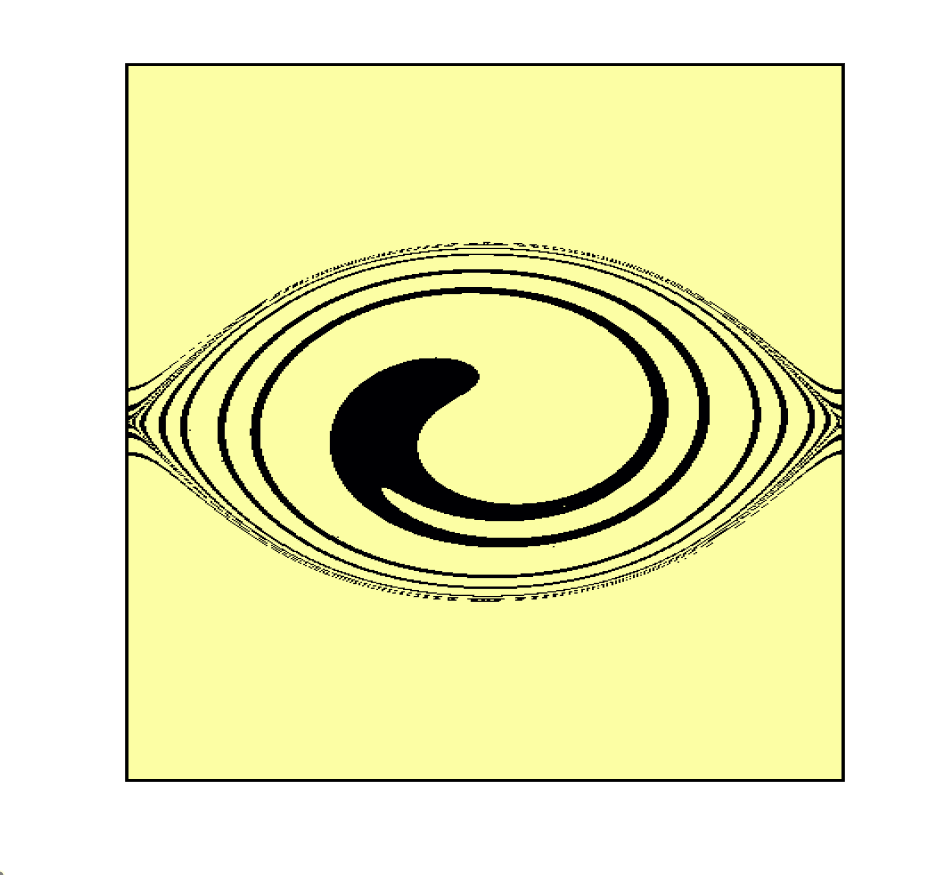}\\
\includegraphics[width=0.38\columnwidth]{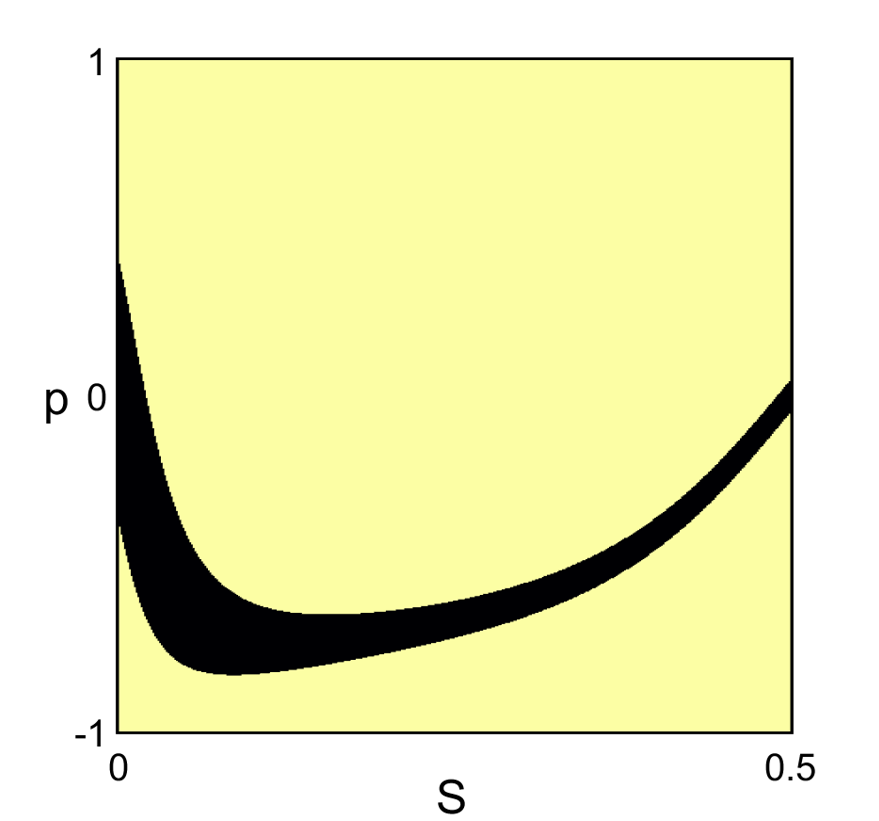}
\includegraphics[width=0.38\columnwidth]{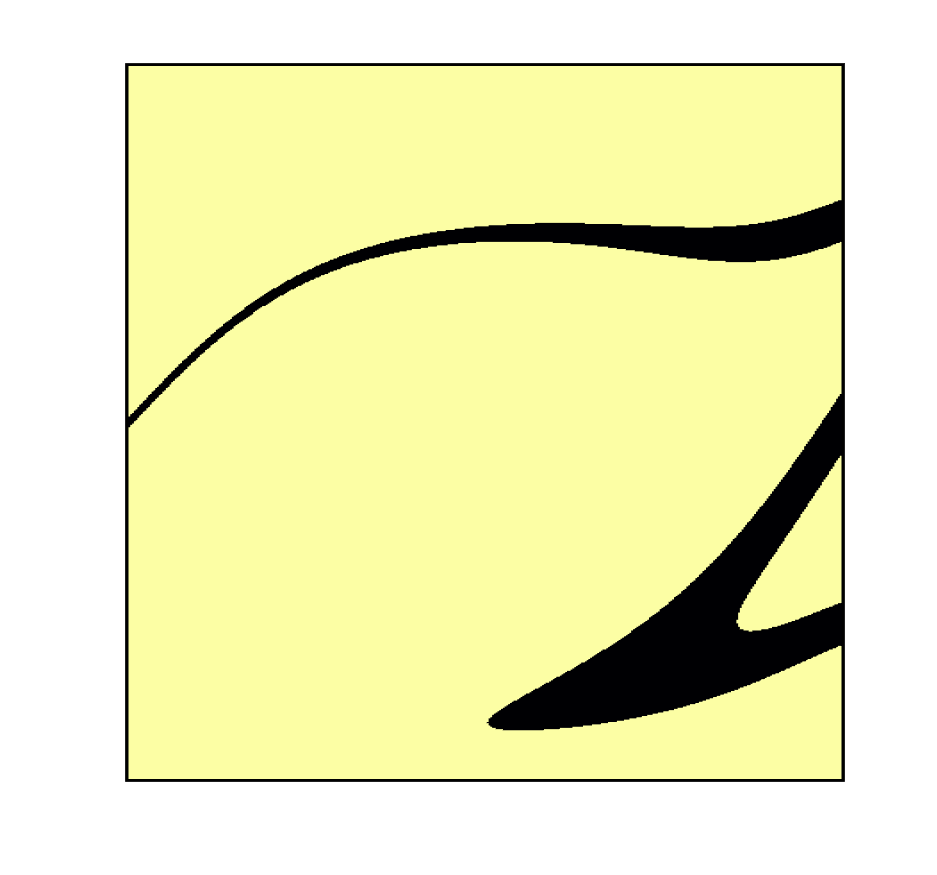}
\includegraphics[width=0.38\columnwidth]{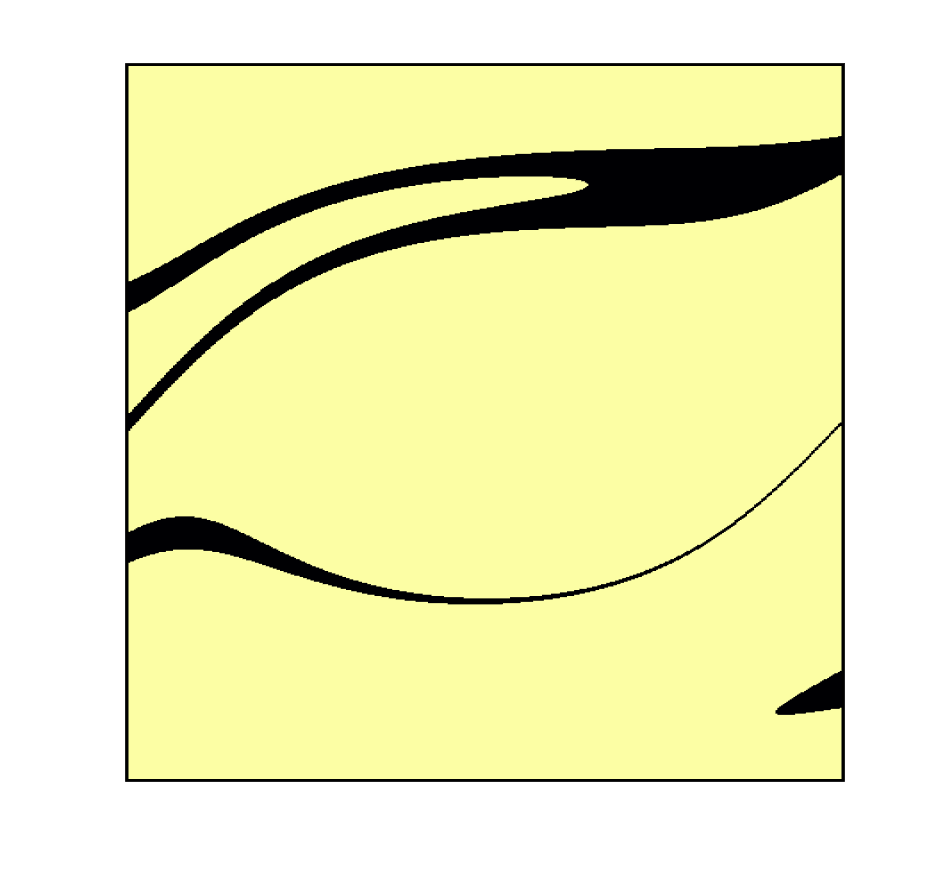}
\includegraphics[width=0.38\columnwidth]{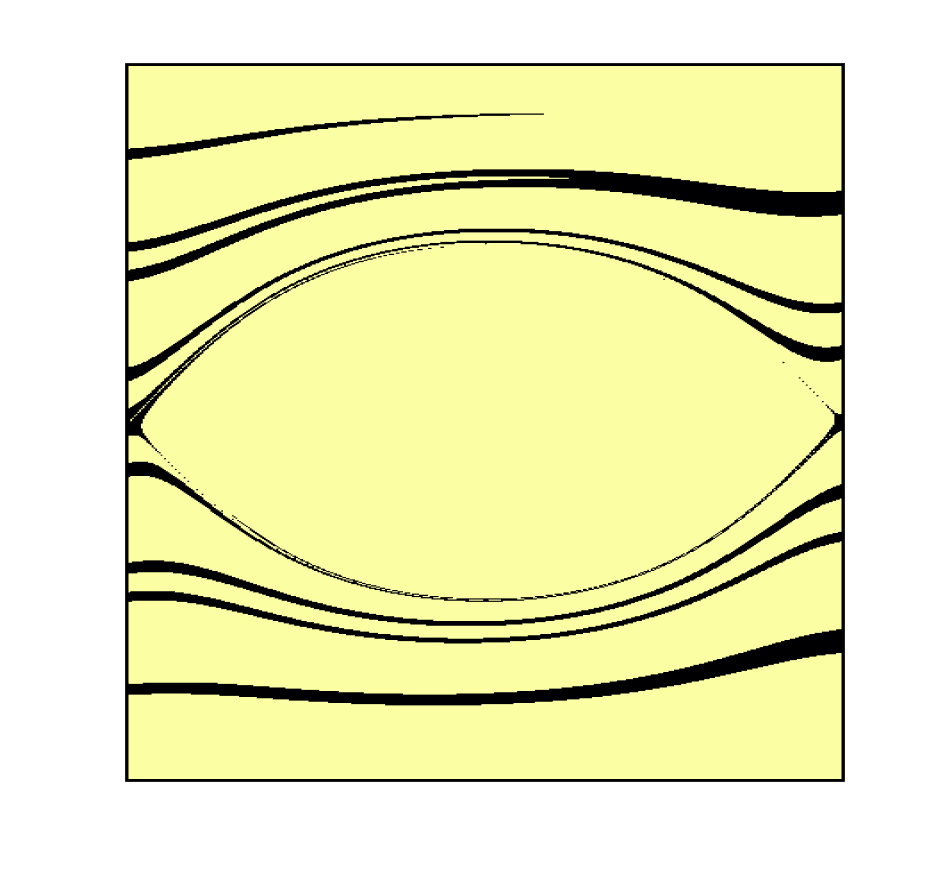}
\includegraphics[width=0.38\columnwidth]{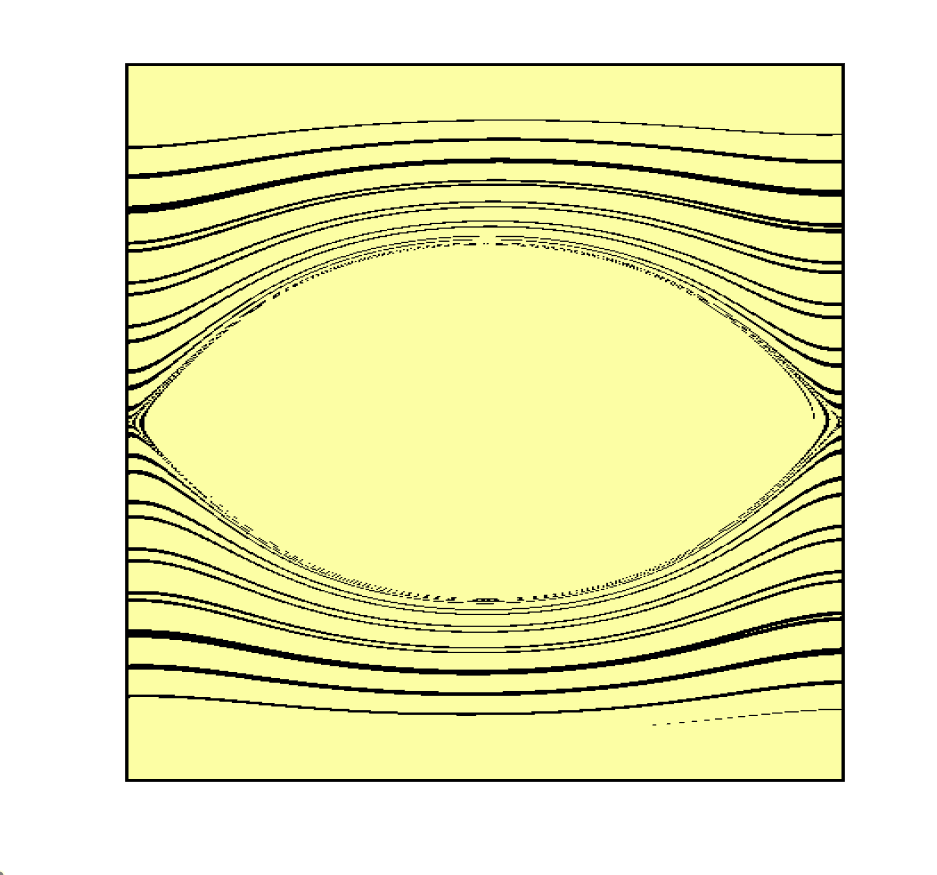}
\caption{The set of initial conditions for the elliptic billiard that intersects the absorbing region at each iteration for radius $R=0.1$ centred at values $d=0.2$ (top) and $d=0.9$ (bottom) to the right of the billiard centre, depicted over half of the phase space \(S\in[0,0.5]\). From left to right the iterations are $n=1,2,3,10$ and $30$.}
\label{diagram:offset_ellipse_Sn}
\end{figure*}

\section*{Acknowledgements}
 E.M.G. and J.H. acknowledge support from the Royal Society (Grant. No. URF\textbackslash R\textbackslash 201034).

\appendix
\section{Billiards with an offset Absorbing Region}\label{appendix:offset_AR}

\begin{figure*}[htb!]
\includegraphics[width=0.38\columnwidth]{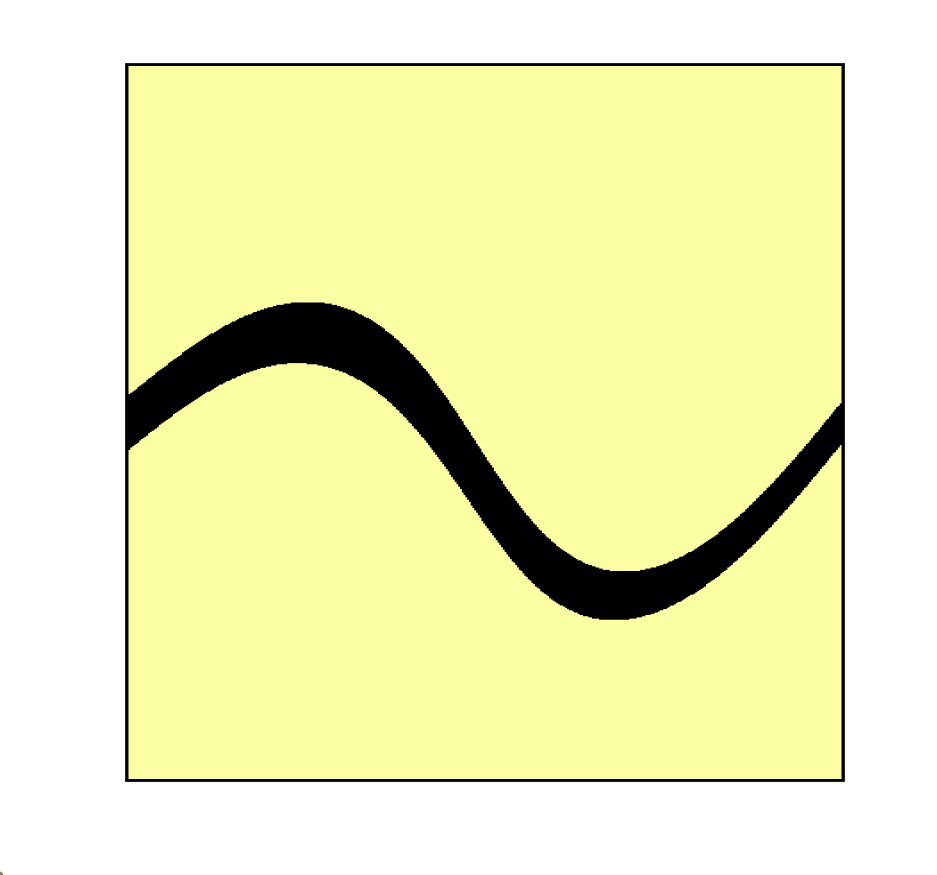}
\includegraphics[width=0.38\columnwidth]{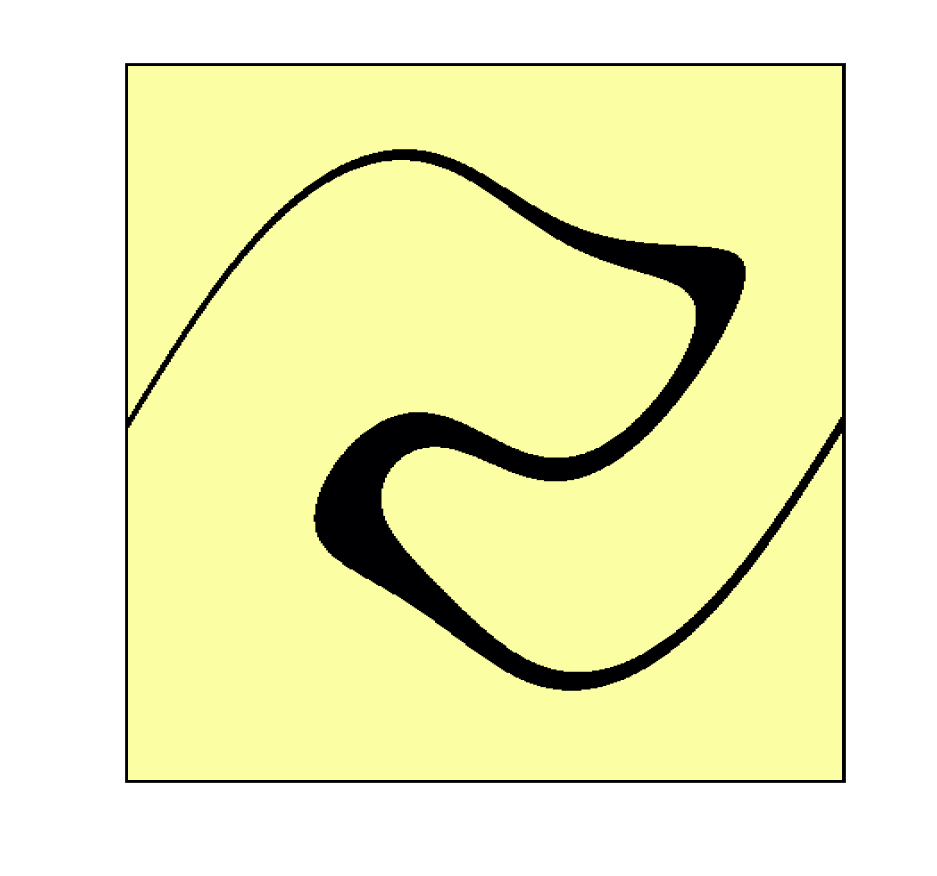}
\includegraphics[width=0.38\columnwidth]{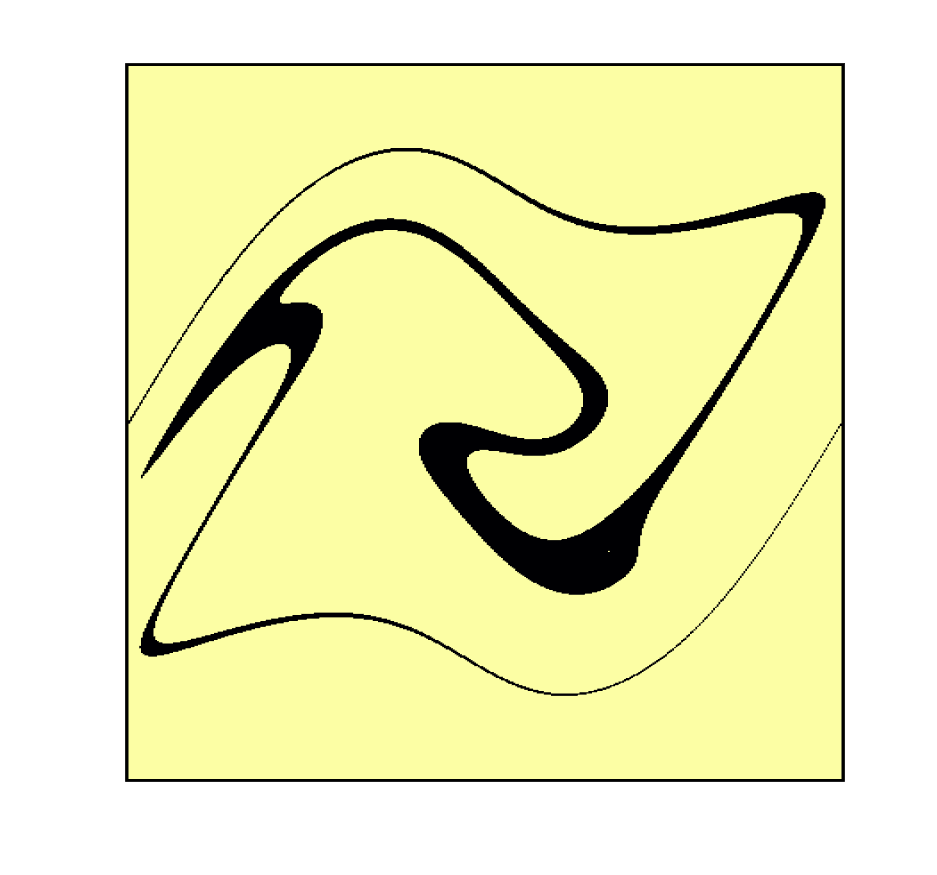}
\includegraphics[width=0.38\columnwidth]{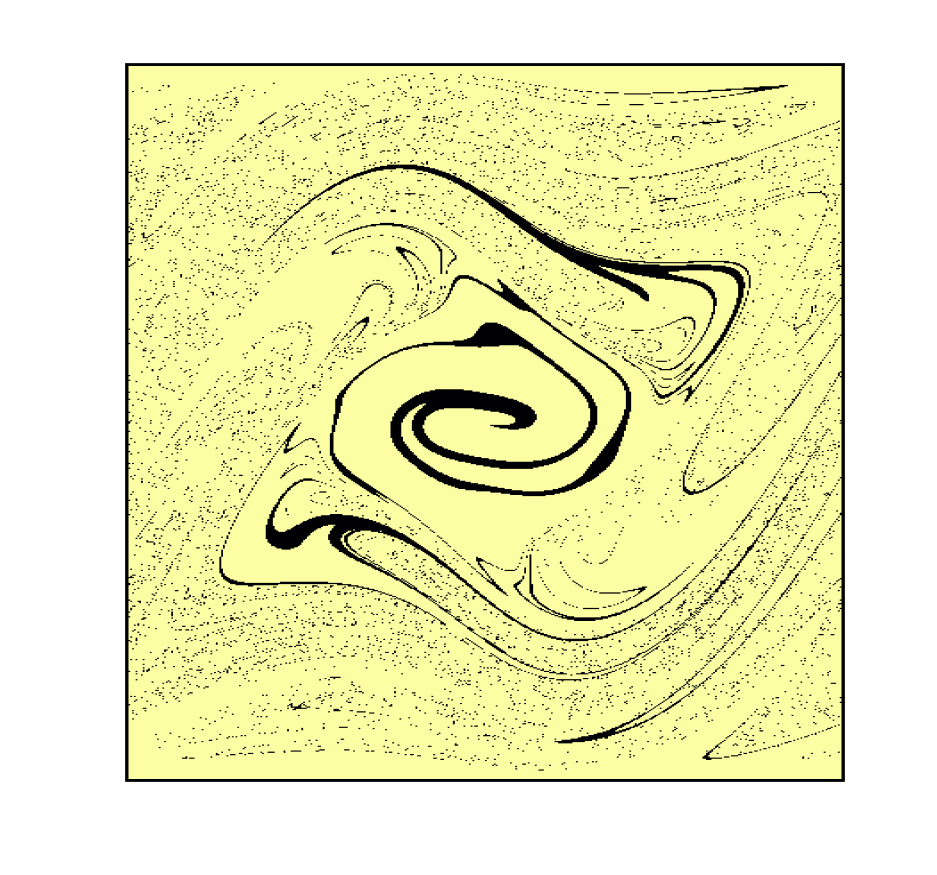}
\includegraphics[width=0.38\columnwidth]{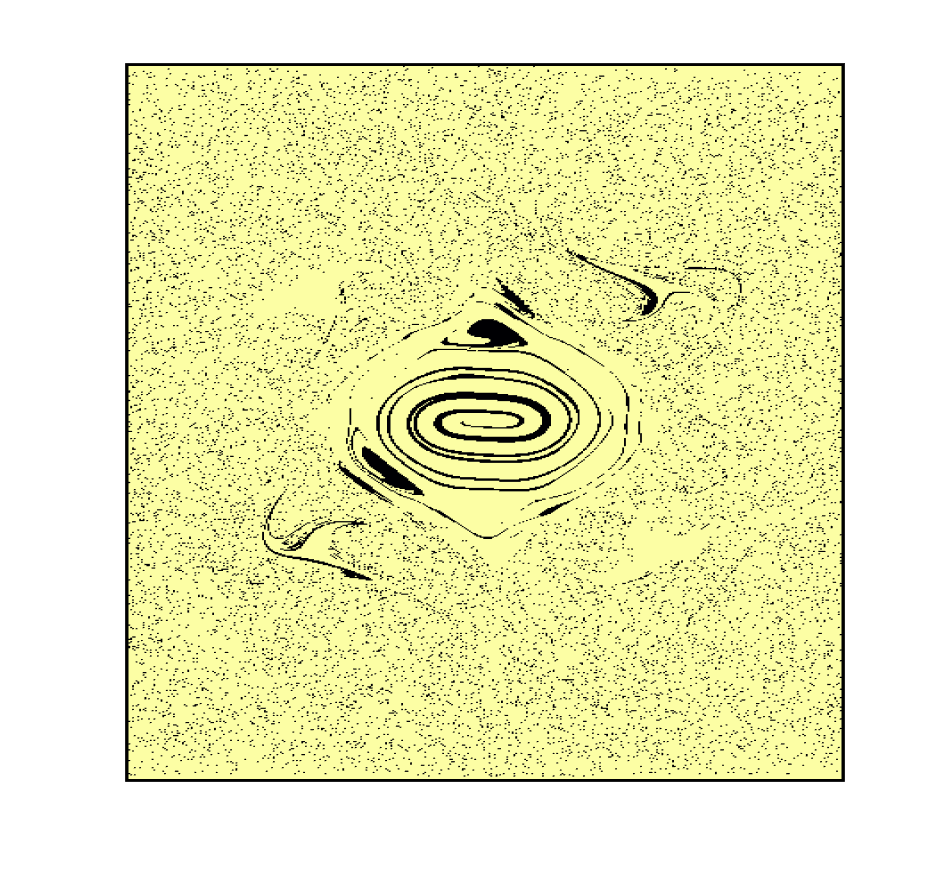}\\
\includegraphics[width=0.38\columnwidth]{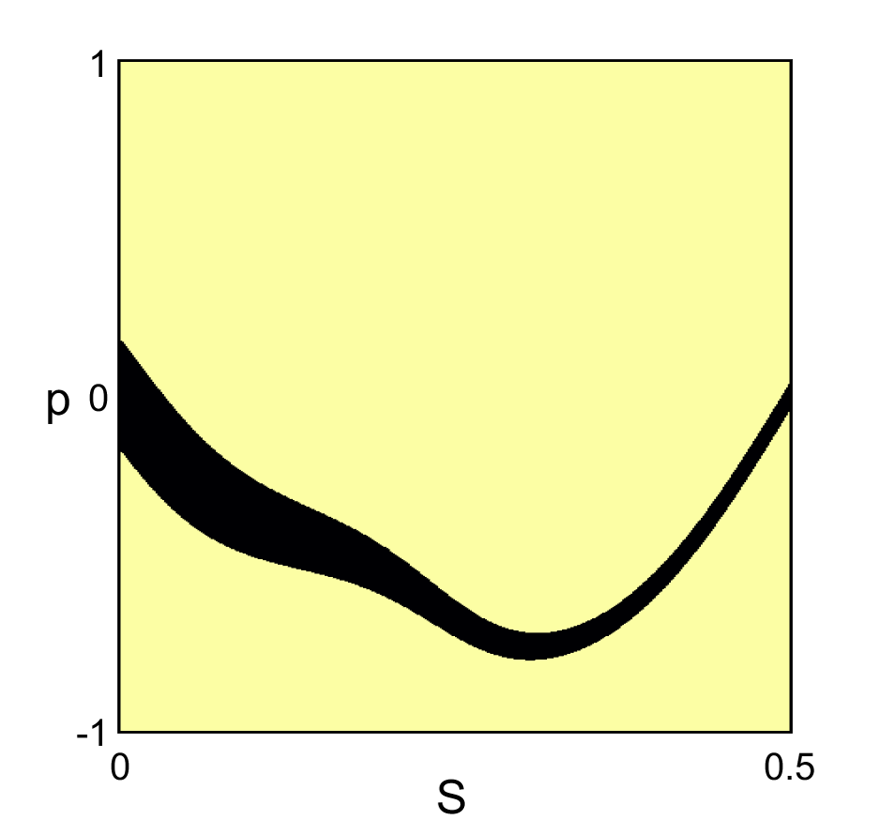}
\includegraphics[width=0.38\columnwidth]{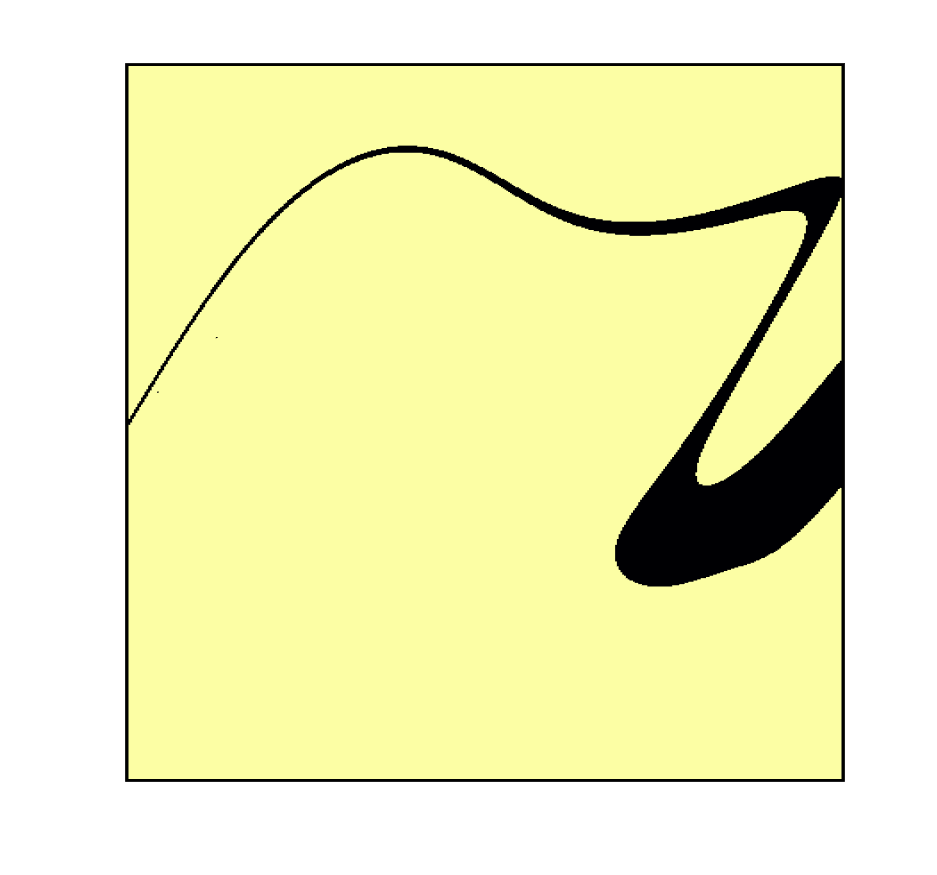}
\includegraphics[width=0.38\columnwidth]{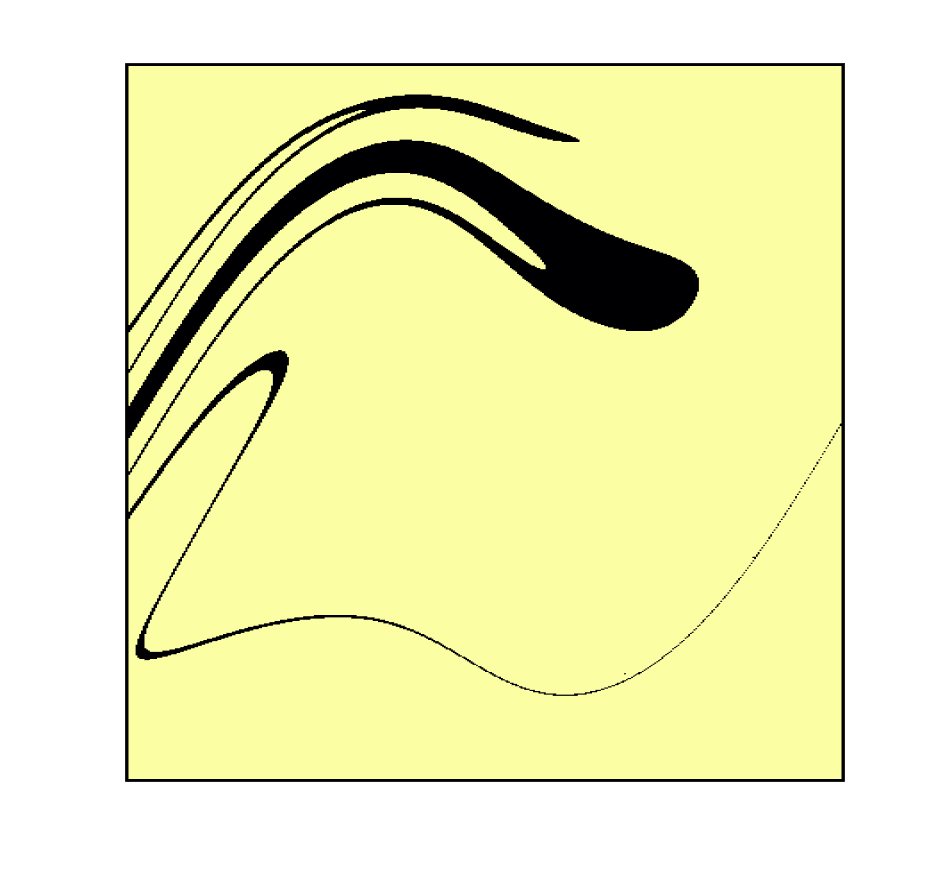}
\includegraphics[width=0.38\columnwidth]{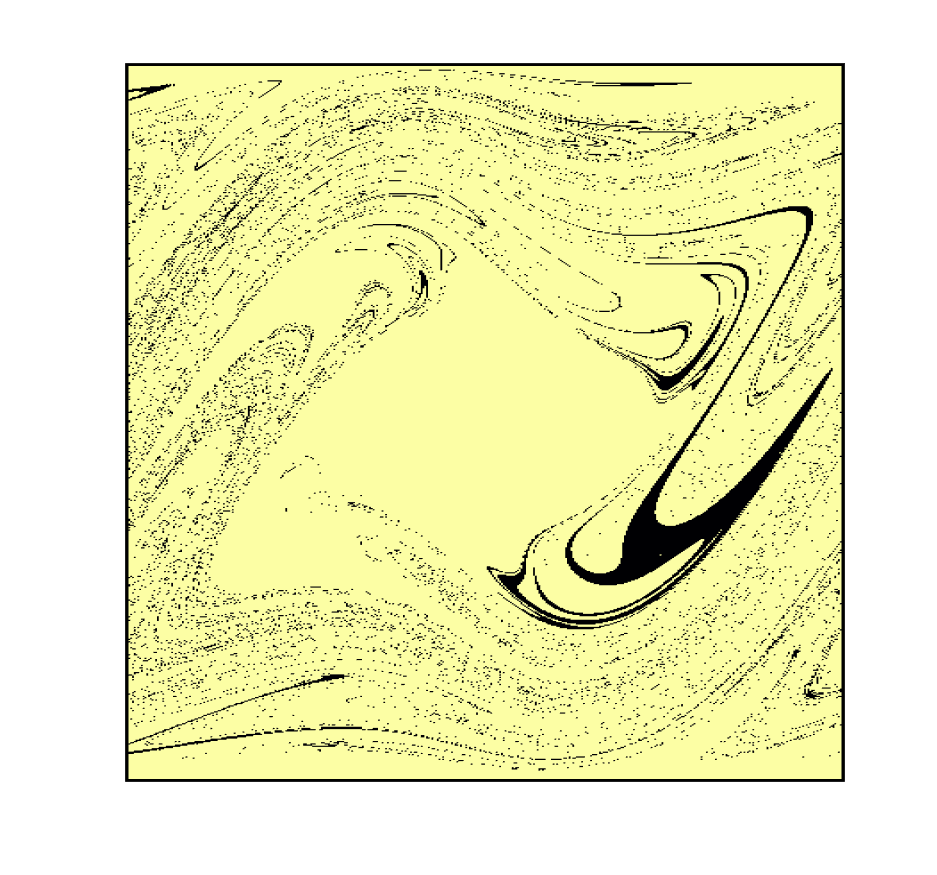}
\includegraphics[width=0.38\columnwidth]{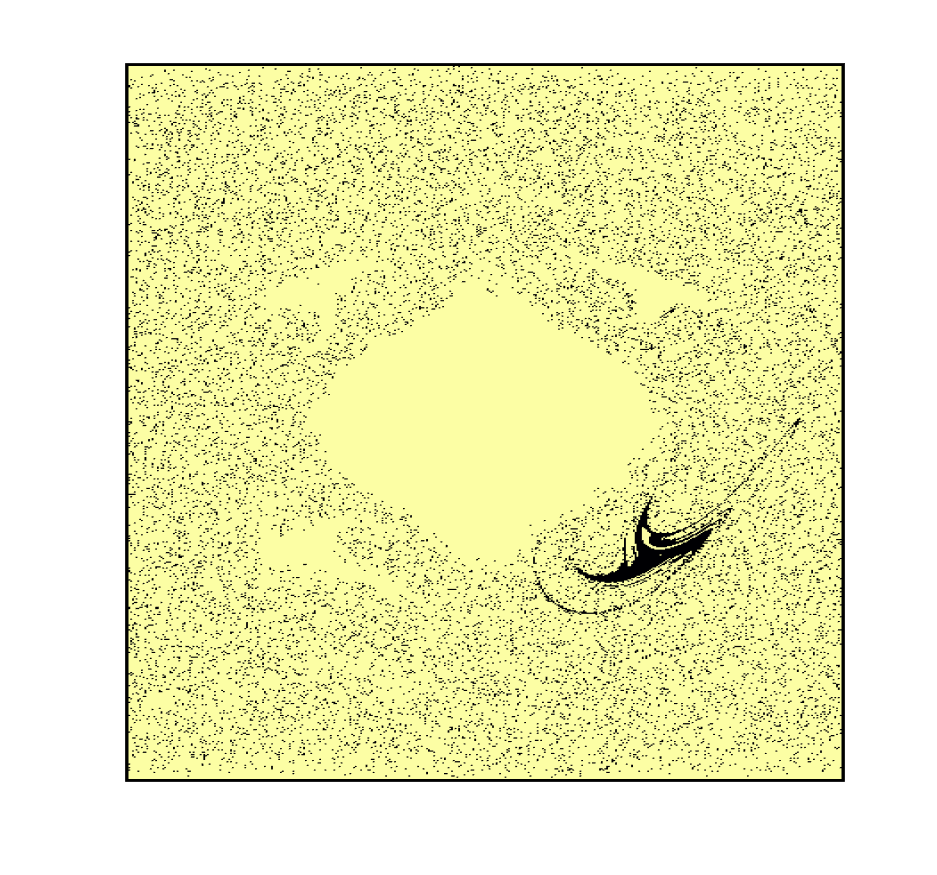}
\caption{The set of initial conditions for the oval billiard that intersects the absorbing region at each iteration for radius $R=0.1$ centred at values $d=0.2$ (top) and $d=0.9$ (bottom) to the right of the billiard centre, depicted over half of the phase space \(S\in[0,0.5]\). From left to right the iterations are $n=1,2,3,10$ and $30$.}
\label{diagram:offset_oval_Sn}
\end{figure*} 

In this appendix we provide additional examples of the circle, elliptic and oval billiard with an internal circular absorbing region of radius \(R = 0.1\) offset away from the billiard centre. Specifically, the centre of the absorbing region is displaced to the right of the billiard centre along the horizontal axis by $d$. A similar set-up, with a scatterer instead of an absorber has been investigated in \cite{da2015dynamics}. Here we show the sets \(S_{n}\) on half of the phase space \(S\in[0,0.5]\) of each billiard for two examples with \(d=0.2\) and \(d=0.9\), respectively. 

Figure \ref{diagram:offset_circle_Sn} shows the resulting sets ${\cal S}_n$ for $n=1,2,3,10$ and $30$, for the circle billiard, with \(d=0.2\) in the top row, and \(d=0.9\) in the bottom row. Unlike in the case where the absorbing region is at the billiard centre, the boundaries of the set ${\cal S}_1$ are no longer horizontal lines, and thus
the set is no longer invariant under the billiard map. Each individual point is still transported along horizontal lines in the phase space, and thus the sets ${\cal S}_n$ are confined to the same range of $p$ as the set ${\cal S}_1$, spreading into thinner and thinner regular filaments. For \(d=0.2\) the range of \(p\) values of \(S_{1}\) is a small proportion of the full $p$ range, leading to filaments filling a band within the phase space. For \(d=0.9\), on the other hand, where the absorbing region touches the billiard perimeter at \(S=0\), trajectories of all reflection angles intersect the absorbing region, and thus \(S_{1}\) spans all values of \(p\), and the filament structure of ${\cal S}_n$ fills out the whole phase space.

Figure \ref{diagram:offset_ellipse_Sn} shows the same sets \({\cal S}_n\) for the elliptic billiard. For the case where \(d=0.2\), the absorbing region is located between the focal points of the ellipse, and thus the set \({\cal S}_1\) encloses mostly box trajectories, just as in the symmetric case \(d=0\). A difference to the symmetric case is a slight asymmetry of the strip \({\cal S}_1\) which is slightly thinner towards $S=0.5$ than around $S=0$. Importantly, it does not cover the elliptic two-cycle. Thus, while most of the trajectories contributing to the iterated sets \({\cal S}_n\) are box trajectories, just as in the symmetric case, resulting in a whorl structure within the separatrix, the lack of contribution from the elliptic two-cycle at the centre of the phase space, leads to a sparser structure compared to the symmetric case. For the case \(d=0.9\) the absorbing region is displaced outside of and does not intersect the focal points of the billiard. As a result we observe contributions from only the loop trajectories in the \({\cal S}_{n}\) diagrams. This can be seen most prominently for \({\cal S}_{30}\), which is zero for the region contained by the separatrix, but develops a dense line structure around this region. 

Finally, Figure \ref{diagram:offset_oval_Sn} displays the equivalent plots for the oval billiard. We see that for \(d=0.2\), some of the trajectories corresponding to the regular islands that did not intersect the symmetrically placed absorbing region now intersect. In the iterated sets \({\cal S}_n\), the regular islands lack definition, however, as the corresponding separatrices are not contained within \({\cal S}_1\). The elliptic fixed point and the sticky trajectories also intersect the patch, however the sticky trajectories in particular are sparsely filled. For the case \(d=0.9\), none of the regular trajectories are contained within \({\cal S}_1\), and in fact only one localised region corresponding to sticky trajectories intersects the absorbing region on a given iteration. In both cases \(d=0.2\) and \(d=0.9\) for the oval billiard, we see that for later iterations the chaotic sea is well populated.

 These visual representations of the sets \({\cal S}_{n}\) are useful tools to obtain first insights into the structures emerging from a placement of the absorbing region that breaks the symmetry. They can be further built upon to develop intensity landscapes, which would be a natural extension to be addressed in future research.

\bibliography{bib}

\end{document}